\documentclass[twocolumn,superscriptaddress,amsmath,amssymb,aps,prx]{revtex4-2}

\usepackage{wasysym}
\usepackage{tikz}
\usetikzlibrary{decorations.markings, arrows,arrows.meta}
\usetikzlibrary{external}
\usepackage{relsize}
\tikzset{fontscale/.style = {font=\relsize{#1}}}

\usepackage{placeins} 
\usepackage{amsthm,amssymb}
\usepackage{nicematrix}
\usepackage{diagbox}
\usepackage[unicode=true,bookmarks=true,bookmarksnumbered=false,bookmarksopen=false,breaklinks=false,backref=false,colorlinks=true]{hyperref}
\hypersetup{urlcolor=blue,linkcolor=violet,citecolor=blue}
\usepackage{natbib}
\usepackage{dcolumn}
\usepackage{bm}
\usepackage{oplotsymbl}
\usepackage{graphicx}
\graphicspath{ {./images/} }
\usepackage{physics}
\usepackage[english]{babel}
\usepackage{xcolor}
\usepackage[normalem]{ulem}
\usepackage{mathtools}
\usepackage{tabularx}
\usepackage{array}
\usepackage{booktabs}
\usepackage{sansmath}
\usepackage{bm}
\usepackage{yhmath}
\usepackage{ragged2e}
\usepackage[nameinlink,capitalize]{cleveref}
\usepackage{times}
\usepackage{orcidlink}
\usepackage{svg}

\usepackage[caption=false]{subfig}
\definecolor{col_green}{RGB}{38,162,105}
\definecolor{col_poly}{RGB}{255,159,72}
\definecolor{col_dim}{RGB}{38,162,105}
\definecolor{col_mon}{RGB}{255,163,72}
\definecolor{col_link}{RGB}{140,140,140}
\definecolor{col_membrane}{RGB}{38,162,105}
\newcommand\scl{0.20}
\newcommand{\plaqA}{ \hspace{0.5mm} \begin{tikzpicture}[scale=0.20,baseline=0.1mm]
 \draw[line width=0.20mm, pink] (0,0) -- (1,0);
\draw[line width=0.4mm, col_link] (1,0) -- ++(90:1);
  \draw[line width=0.25mm, pink] (1,0)++(90:1) -- (90:1);
  \draw[line width=0.4mm, col_link] (0,0) -- ++(90:1);
 \filldraw[col_dim] (0,0) circle (6pt);
  \filldraw[col_dim] (1,0) circle (6pt);
    \filldraw[col_dim] (1,0) ++(90:1) circle (6pt);
    \filldraw[col_dim] (90:1) circle (6pt);
 \end{tikzpicture} \hspace{0.5mm}}

 \newcommand{\plaqB}{  \hspace{0.5mm}\begin{tikzpicture}[scale=0.20,baseline=0.1mm]
 \draw[line width=0.40mm, col_link] (0,0) -- (1,0);
\draw[line width=0.25mm, pink] (1,0) -- ++(90:1);
  \draw[line width=0.4mm, col_link] (1,0)++(90:1) -- (90:1);
  \draw[line width=0.25mm, pink] (0,0) -- ++(90:1);
 \filldraw[col_dim] (0,0) circle (6pt);
  \filldraw[col_dim] (1,0) circle (6pt);
    \filldraw[col_dim] (1,0) ++(90:1) circle (6pt);
    \filldraw[col_dim] (90:1) circle (6pt);
 \end{tikzpicture} \hspace{0.5mm}}

 \newcommand{\octA}{\hspace{0.5mm} \begin{tikzpicture}[scale=0.15,baseline=-0.8mm]
        \draw[line width=0.40mm, col_link] (0.5, 0.5) -- (1.207, 1.207)  (2.207, 1.207) -- (2.9142, 0.5)  (2.9142, -0.5) -- (2.2071, -1.2071)  (1.2071, -1.2071) -- (0.5, -0.5) ;
        \draw[line width=0.20mm, pink] (1.207, 1.207) -- (2.207, 1.207)  (2.9142, 0.5) -- (2.9142, -0.5)  (2.2071, -1.2071) -- (1.2071, -1.2071)  (0.5, -0.5) -- (0.5, 0.5);
         \filldraw[col_dim] (0.5, 0.5) circle (6pt);
         \filldraw[col_dim] (1.207, 1.207) circle (6pt);
         \filldraw[col_dim] (2.207, 1.207) circle (6pt);
         \filldraw[col_dim] (2.9142, 0.5) circle (6pt);
         \filldraw[col_dim] (2.9142, -0.5) circle (6pt);
         \filldraw[col_dim] (2.2071, -1.2071) circle (6pt);
         \filldraw[col_dim] (1.2071, -1.2071) circle (6pt);
         \filldraw[col_dim] (0.5, -0.5) circle (6pt);
\end{tikzpicture} \hspace{0.5mm}}

 \newcommand{\octB}{ \hspace{0.5mm} \begin{tikzpicture}[scale=0.15,baseline=-0.8mm]
        \draw[line width=0.20mm, pink] (0.5, 0.5) -- (1.207, 1.207)  (2.207, 1.207) -- (2.9142, 0.5)  (2.9142, -0.5) -- (2.2071, -1.2071)  (1.2071, -1.2071) -- (0.5, -0.5) ;
        \draw[line width=0.40mm, col_link] (1.207, 1.207) -- (2.207, 1.207)  (2.9142, 0.5) -- (2.9142, -0.5)  (2.2071, -1.2071) -- (1.2071, -1.2071)  (0.5, -0.5) -- (0.5, 0.5);
         \filldraw[col_dim] (0.5, 0.5) circle (6pt);
         \filldraw[col_dim] (1.207, 1.207) circle (6pt);
         \filldraw[col_dim] (2.207, 1.207) circle (6pt);
         \filldraw[col_dim] (2.9142, 0.5) circle (6pt);
         \filldraw[col_dim] (2.9142, -0.5) circle (6pt);
         \filldraw[col_dim] (2.2071, -1.2071) circle (6pt);
         \filldraw[col_dim] (1.2071, -1.2071) circle (6pt);
         \filldraw[col_dim] (0.5, -0.5) circle (6pt);
\end{tikzpicture} \hspace{0.5mm}}

\renewcommand\scl{0.10}

\newcommand{\orangesq}{  \begin{tikzpicture}[scale=\scl]
 \draw[fill=col_poly, line width=0.2mm, semitransparent] (1,0) -- (0,1) -- (-1,0) -- (0,-1) -- cycle;
 \end{tikzpicture} \hspace{0.5mm} }

\newcommand{\orangetrA}{ \hspace{0.5mm} \begin{tikzpicture}[scale=0.18]

 \draw[fill=col_poly, line width=0.2mm, semitransparent] (0,0) -- (1,0) -- (0,1) -- cycle;
 \end{tikzpicture} 
 \hspace{0.5mm} }

 \newcommand{\orangetrB}{ \hspace{0.5mm}  \begin{tikzpicture}[scale=0.14]
 \draw[fill=col_poly, line width=0.2mm, semitransparent] (-1,0) -- (1,0) -- (0,1) -- cycle;
 \end{tikzpicture}\hspace{0.5mm}  }

 \newcommand{\orangetrC}{ \hspace{0.5mm}  \begin{tikzpicture}[scale=0.08]
 \draw[fill=col_poly, line width=0.2mm, semitransparent] (1,0) -- (0,2.414) -- (-1,0) -- cycle;
 \end{tikzpicture}\hspace{0.5mm}  }

 \newcommand{\orangetrD}{ \hspace{0.5mm}  \begin{tikzpicture}[scale=0.25]
 \draw[fill=col_poly, line width=0.2mm, semitransparent] (0,0) -- (1,0) -- (-0.707,0.707) -- cycle;
 \end{tikzpicture}\hspace{0.5mm}  }

\newtheorem{thm}{Theorem}[section]

\newtheorem{lem}[thm]{Lemma}

\renewcommand{\Re}{\operatorname{\mathrm{Re}}}
\newcommand{\im}{\operatorname{\mathrm{Im}}}
\newcommand{\re}{\operatorname{\mathrm{Re}}}
\newcommand\numberthis{\stepcounter{equation}\tag{\theequation}}
\newcommand{\E}{\mathbb{E}}
\renewcommand{\P}{\mathbb{P}}
\newcommand{\N}{\mathbb{N}}
\newcommand{\Z}{\mathbb{Z}}
\newcommand{\R}{\mathbb{R}}
\newcommand{\C}{\mathbb{C}}

\newcommand{\Kl}{\mathbb{K}} 
\newcommand{\Ki}{\mathcal{K}} 
\newcommand{\Kzz}{K^\zz}
\newcommand{\Kzo}{K^\zo}
\newcommand{\Koz}{K^\oz}
\newcommand{\Koo}{K^\oo}
\newcommand{\Kab}{K^\ab}
\newcommand{\zz}{{00}}
\newcommand{\zo}{{01}}
\newcommand{\oz}{{10}}
\newcommand{\oo}{{11}}
\newcommand{\ab}{{\alpha\beta}}

\newcommand{\dconfig}{\mathcal C}

\newcommand{\Pf}{\operatorname{Pf}}

\begin{document}

\title{Breakdown of the thermodynamic limit in quantum spin and dimer models}

\author{Jeet Shah\,\orcidlink{0000-0001-5873-8129}}
\thanks{Jeet Shah and Laura Shou contributed equally.}
\affiliation{Joint Quantum Institute, Department of Physics, University of Maryland,
College Park, MD 20742, USA}
\affiliation{Joint Center for Quantum Information and Computer Science, NIST/University of Maryland,
College Park, MD 20742, USA}

\author{Laura Shou}
\thanks{Jeet Shah and Laura Shou contributed equally.}
\affiliation{Joint Quantum Institute, Department of Physics, University of Maryland,
College Park, MD 20742, USA}

\author{Jeremy Shuler}
\affiliation{Joint Quantum Institute, Department of Physics, University of Maryland,
College Park, MD 20742, USA}

\author{Victor Galitski}
\affiliation{Joint Quantum Institute, Department of Physics, University of Maryland,
College Park, MD 20742, USA}

\date{\today}

\begin{abstract}

The thermodynamic limit is foundational to statistical mechanics, underlying our understanding of many-body phases. It assumes that, as the system size grows infinitely at fixed density of particles, unambiguous macroscopic phases emerge that are independent of the system's boundary shape. We present explicit quantum spin and dimer Hamiltonians whose ground states violate this principle. Our construction relies on the previous mathematical work on classical dimers on the Aztec diamond and the square-octagon fortress, where geometry-dependent phase behaviors are observed in the infinite-size limit. We reverse engineer quantum spin Hamiltonians on the square and the square-octagon lattices whose ground states at the Rokhsar--Kivelson points are described by classical dimer coverings. On diamond-shaped domains, we find macroscopic boundary regions exhibiting distinct quantum phases from those on square-shaped domains. We study the nature of these phases by calculating the dimer-dimer and vison correlators and adapt Kasteleyn matrix based analytical and numerical methods for computing the vison correlator, which are significantly more efficient than standard Monte Carlo techniques. Our results show that the square-octagon lattice supports a single gapped short-range entangled phase, with exponentially decaying dimer correlators and a constant vison correlator. When the same model is considered on a diamond-shaped domain, an additional ordered phase emerges near the corners, where the dimers are in a staggered pattern.

\end{abstract}

\maketitle
\def\thefootnote{*}\footnotetext{These authors contributed equally to this work}\def\thefootnote{\arabic{footnote}}

\tableofcontents

\section{Introduction}
\label{sec:introduction}

A foundational assumption of statistical mechanics is that, in the thermodynamic limit, the shape of the system's boundary does not influence its bulk properties.
This is consistent with everyday experience: a cubic piece of iron is metallic throughout the bulk, as is a spherical piece of iron.
The negligible influence of the boundaries allows us to define a macroscopic bulk phase with uniform properties throughout and choose boundary conditions as convenient. 

In this work, we present two quantum spin models with local interactions that map to quantum dimer models, for which this assumption fails---i.e., the thermodynamic limit breaks down, and the shape of the system boundaries have a profound effect on the bulk phases.
Dimers are degrees of freedom that occupy pairs of neighboring sites, with the constraint that no two dimers share a site.
In the context of graph theory, dimer coverings are known as perfect matchings.
Quantum dimer models were first introduced by Rokhsar and Kivelson in Ref.~\cite{rokhsar1988superconductivity}, where they studied such a model on the square lattice.
Their Hamiltonian consisted of off-diagonal ring-exchange terms and diagonal Rokhsar--Kivelson (RK) potential energy terms.
At the RK point, where these terms have equal strength, the ground state is exactly solvable, and the dimer–dimer connected correlator exhibits power-law decay, implying that the model is at a critical point in the phase diagram. 
Remarkably, we find that for the same quantum dimer model when considered on the square lattice with boundaries at $\pm 45^\circ$ to the edges, different regions in the bulk have drastically different behavior.
Such a square lattice with diamond shaped ($\diamond$) boundaries is called the ``Aztec diamond" \cite{elkies1992alternating1}. 
\Cref{fig:az-random-config} shows an Aztec diamond along with one of its dimer coverings chosen randomly from the set of all dimer coverings.
This model is our first example of the breakdown of the thermodynamic limit.
\begin{figure}[tbp]
  \centering
\begin{tikzpicture}
\def\height{.9} 
\def\xs{8.25} 
\node at (0,0) {\includegraphics[height=\height\columnwidth]{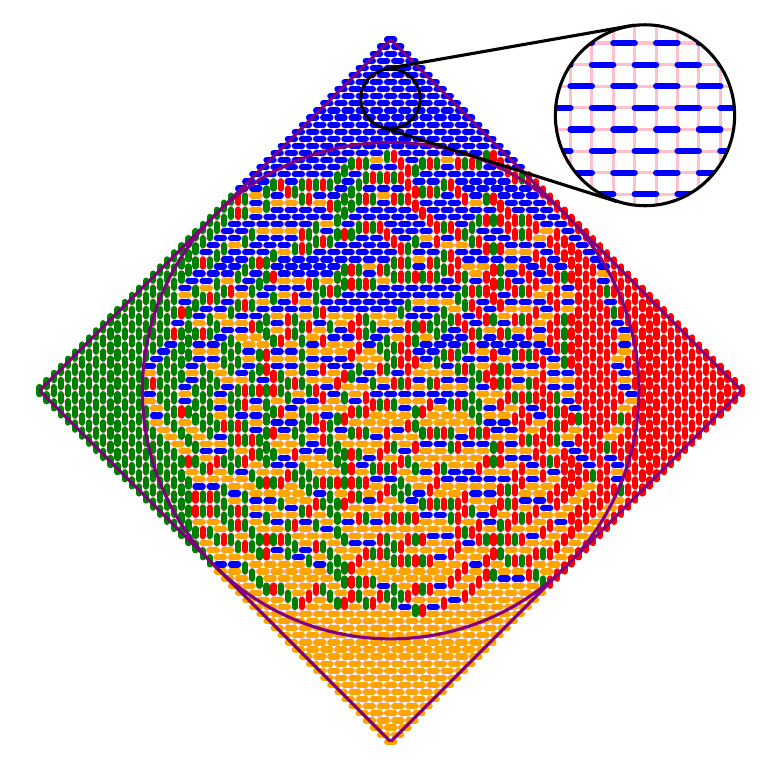}};
\def\lcoords{(-3.8,3.5)};
\foreach \xloc in {0}{ 
\begin{scope}[xshift={\xloc},{Latex[length=2mm]}-,fontscale=-.25]
\draw (-.2,3) -- (-1,3.5) node[left] {Frozen region};
\draw (-1,-1)--(-2.7,-2) node[below] {Critical region};
\end{scope}
}
\begin{scope}[{Latex[length=2mm]}-] 
\draw (.2,-2.45) to [out=270,in=180] (1,-3) node[right,fontscale=-.25] {Arctic circle};
\end{scope}
\end{tikzpicture}

  \caption{A random dimer covering of an Aztec diamond, generated via Monte Carlo simulation.
  The Aztec diamond is a bipartite graph, with vertices partitioned into two sublattices---commonly referred to as ``black" and ``white" vertices.
  Red and green edges represent vertical dimers, while blue and orange edges represent horizontal dimers; the two colors for each orientation indicate whether the black vertex lies on the top/bottom (for vertical) or left/right (for horizontal) end of the dimer.
  The purple circle indicates the \textit{Arctic circle}. 
  Outside the Arctic circle, the system is \textit{frozen}---dimers align in a staggered configuration (as shown in the inset) with probability tending to one as the diamond size tends to infinity.  
    Inside the Arctic circle, dimers can have random orientations.
    This configuration was generated by starting from an initial configuration with all horizontal dimers and performing $10^{11}$ ring-flip moves. 
  }
  \label{fig:az-random-config} 
\end{figure}
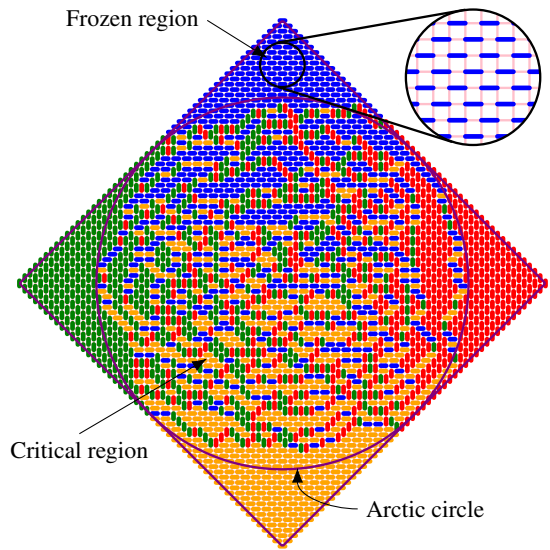
\begin{figure}[tbp]
\centering
     \includegraphics[width=0.7\columnwidth]{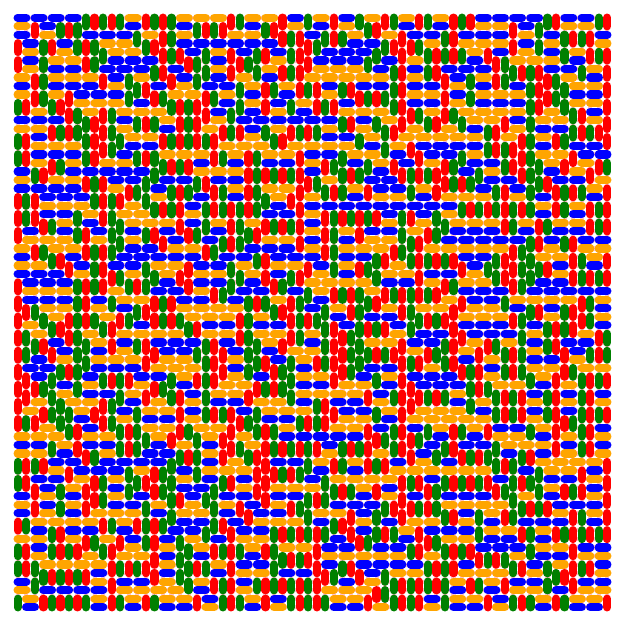}
  \caption{A random dimer configuration on the square lattice with square-shaped boundaries, generated via Monte Carlo sampling. Different colors indicate dimer orientations.}
  \label{fig:sq-random-config}
\end{figure}
One can immediately see phase separation in this figure: regions near the corners display a staggered dimer configuration, while the dimers within a circle inscribed in the diamond, known as the ``Arctic circle" \cite{jockusch1998random}, exhibit no apparent order.
This phenomenon in uniformly sampled random dimer coverings of the Aztec diamond is well known in the mathematical literature as the ``Arctic circle theorem"~\cite{elkies1992alternating1,elkies1992alternating2,kenyon1997local,jockusch1998random,cohn1996local,propp2003generalized,kenyon2009lectures}, and was proved by W. Jockusch, J. Propp, and P. Shor in Ref.~\cite{jockusch1998random}.
The Arctic circle phenomenon implies that for the RK dimer model on the Aztec diamond, the corners are in a \textit{frozen} phase, while the region inside the ``Arctic circle" will be critical, having power-law decaying dimer correlations.
This provides a rigorous example of spatial separation between quantum phases induced by the shape of the boundary.
Notably, the frozen regions are macroscopic relative to the system as a whole, implying that the phase separation is not localized to the boundaries.
In contrast, the quantum dimer model on the square lattice with rectangular boundaries, as proposed by Rokhsar and Kivelson in Ref.~\cite{rokhsar1988superconductivity}, does not exhibit such a phase separation.
\Cref{fig:sq-random-config} shows a randomly chosen dimer configuration of the square lattice on a square-shaped domain,
where it is apparent that no frozen regions appear.
Importantly, we keep the Hamiltonian the same for both the Aztec diamond and rectangular boundaries in the quantum dimer model.
Because differently shaped pieces of the same model display drastically different bulk behavior, the thermodynamic limit is not well defined.
As the system size tends to infinity, one must also specify the shape of the system and the location of the origin relative to that shape.

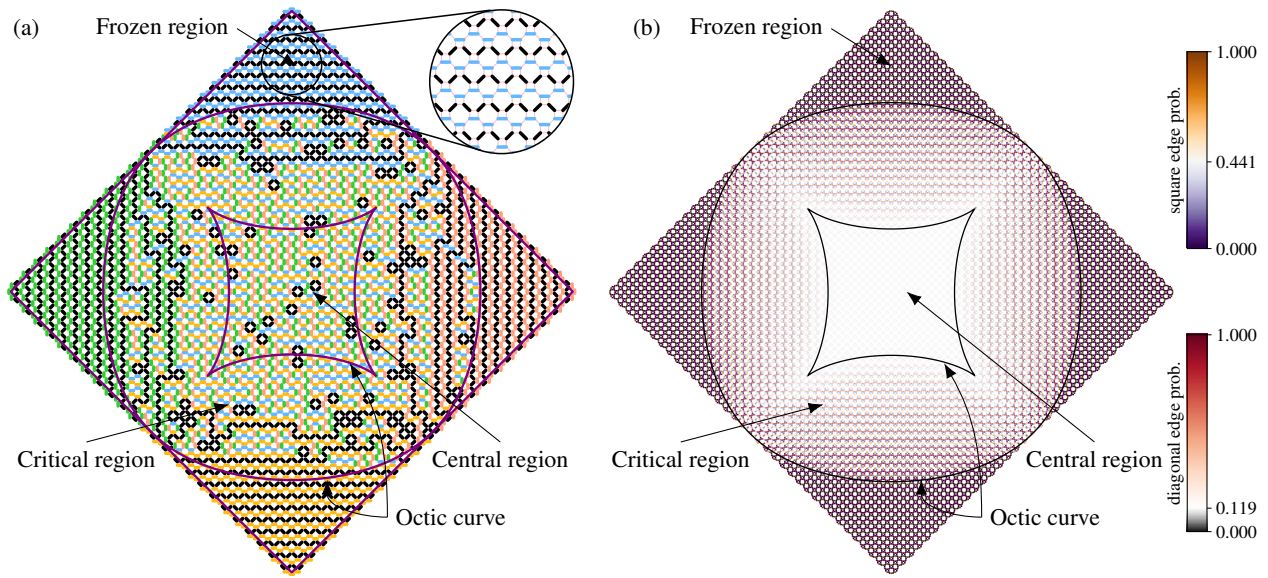
\begin{figure*}[htbp]
  \centering
\begin{tikzpicture}
\def\xs{8.25} 
\node at (0,0) {\includegraphics[height=.95\columnwidth]{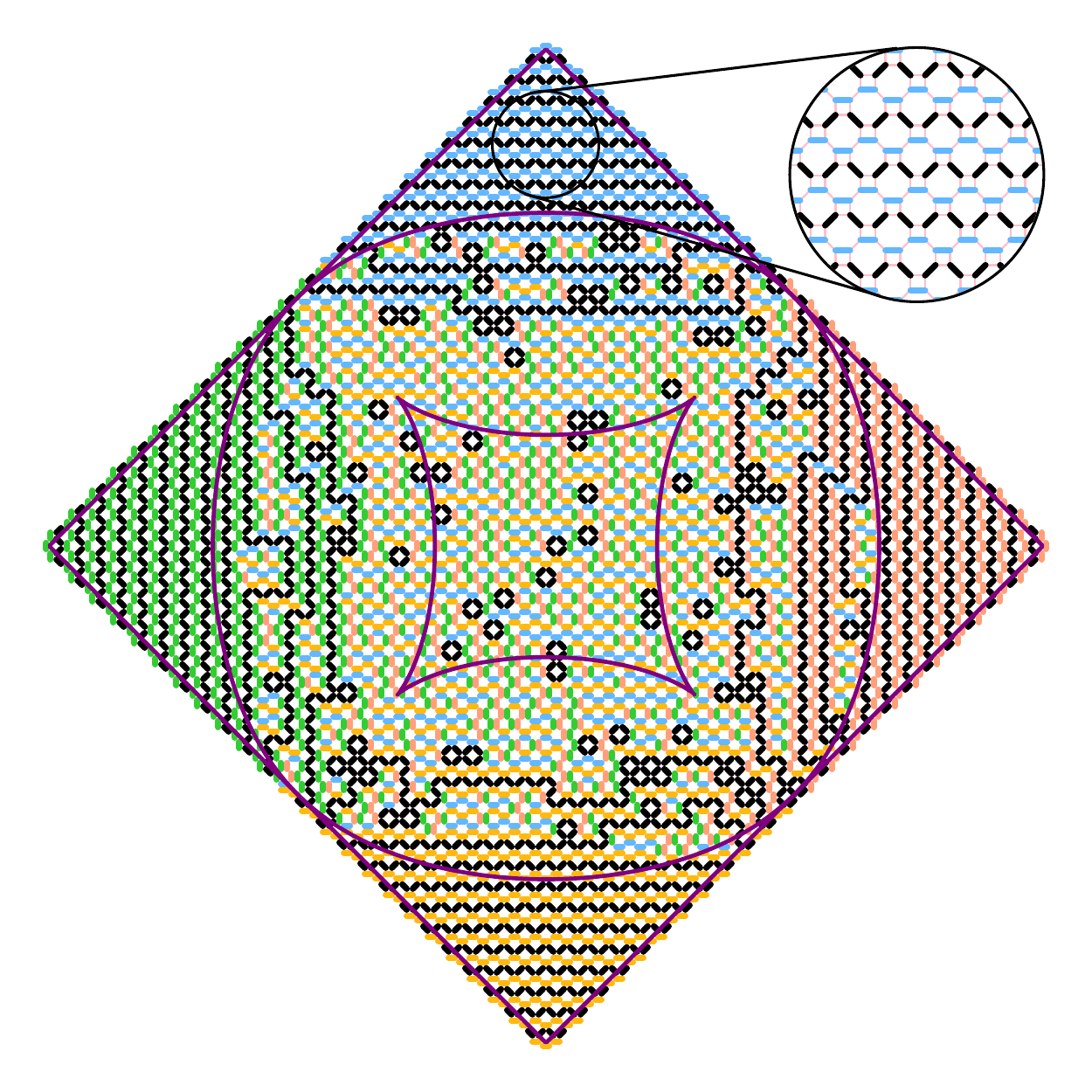}};$\;\;\;$
\node at (\xs,0) {\includegraphics[height=.9\columnwidth]{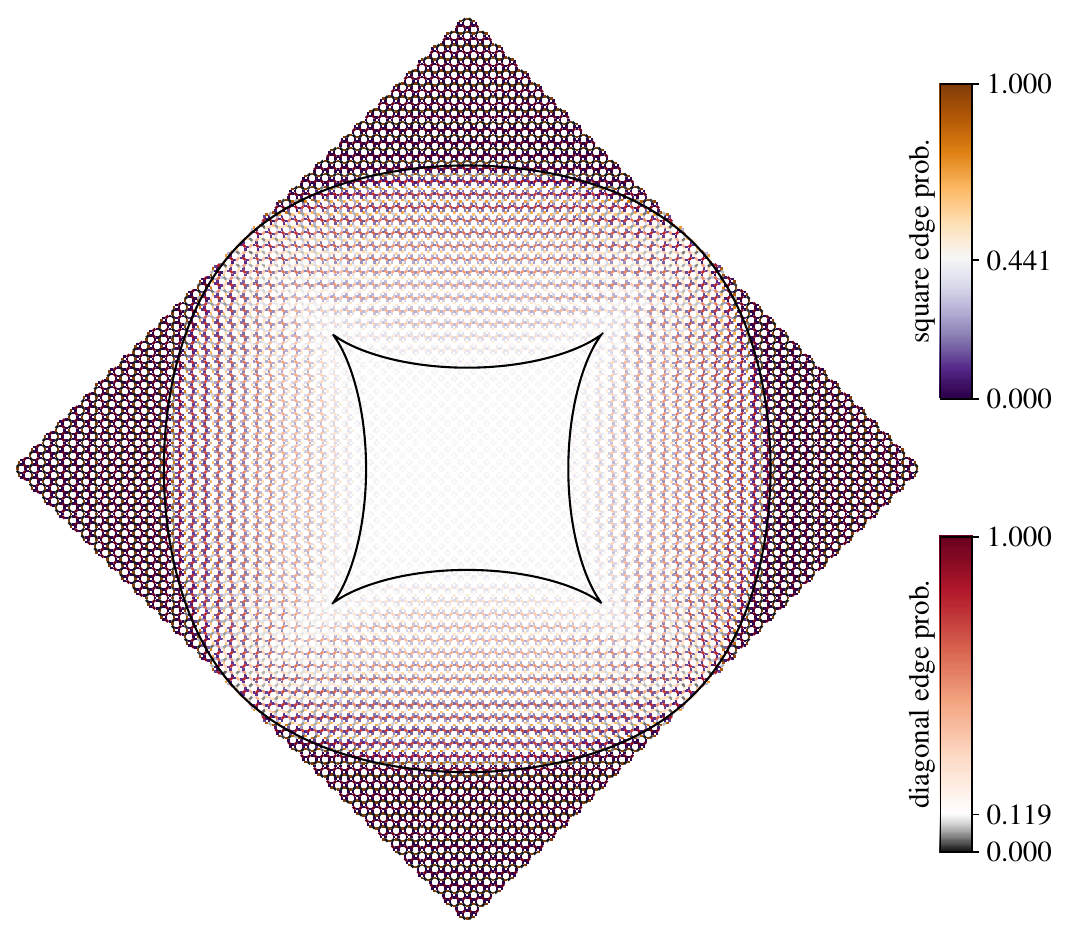}};
\def\lcoords{(-3.8,3.5)};
\node at \lcoords {(a)};
\node[xshift=8.25cm] at \lcoords {(b)};
\foreach \xloc in {0,{\xs cm-.35cm}}{ 
\begin{scope}[xshift={\xloc},{Latex[length=2mm]}-,fontscale=-.25]
\draw (-.2,3) -- (-1,3.5) node[left] {Frozen region};
\draw (-1.1,-1.5)--(-3,-2) node[below] {Critical region};
\draw (0,0)--(2.5,-2) node[below] {Central region};
\draw (.2,-2.5) to [out=270,in=180] (1,-3) node[right] {Octic curve};
\draw (.5,-.95) to [out=300,in=90] (1,-3);
\end{scope}
}
\end{tikzpicture}
  \caption{(a) A random dimer covering of the square-octagon fortress, generated via Monte Carlo simulation. 
  Edges colored in red, blue, green, and orange represent horizontal and vertical dimers, while black links represent diagonal dimers.
  The octic curve separates the graph into three distinct regions.
  The regions near the four corners are called the \textit{frozen regions}, where dimers exhibit an ordered configuration as shown in the inset.
  The \textit{critical region}, located between the oval and the star-shaped curves, shows power-law decay in the dimer-dimer connected correlator.
  The innermost \textit{central region} features exponentially decaying dimer-dimer connected correlator.
  (b) A color plot showing the probability with which dimers occupy each edge.
  The color scale is set such that white corresponds to a probability of $0.441$ on horizontal and vertical edges, and $0.119$ on diagonal edges.
  These values are the analytically calculated dimer probabilities for the infinite square-octagon lattice with square boundaries, as derived in \cref{appsub:dimerprob}.
The center region appears white, indicating probabilities close to the infinite-limit rectangular boundary values, while the frozen corner regions appear dark brown, purple, red, or black, indicating dimer probabilities close to zero or one.
The critical region between the center and the frozen corners shows a mix of colors that vary with position.
All probabilities are calculated using exact fraction-free LU decomposition, as described in \cref{subsec:vison}. 
  }
  \label{fig:so-random-config} 
\end{figure*}

Dimer models have long attracted interest from both physicists and mathematicians, albeit for different reasons.
In physics, early work \cite{kasteleyn1961statistics,temperley1961dimer,fisher1961stastical,montroll1963correlations} studied the relationship between correlation functions in classical dimer models and those in Ising models.
Since then, quantum dimer models have been studied because they can exhibit exotic states of matter such as topologically ordered quantum spin liquids which have fractionalized excitations \cite{moessner2011introduction,balents2002fractionalization,savary2017quantum,senthil2000z2gauge,moessner2001resonating,fendley2002classical,misguich2002quantum,hermele2004pyrochlore}.

In mathematics, the combinatorics of dimer coverings on the Aztec diamond were first studied in Ref.~\cite{elkies1992alternating1}. 
Since then, there has been much work on weighted dimer coverings of the Aztec diamond \cite{chhita2014coupling,cjy2015asymptotic,chhita2016domino,difrancesco2014acoe,duits2021the,berggren2019correlation,berggren2023geometry,boutillier2025focks},  where the probability of a dimer covering is proportional to the product of weights on edges occupied by dimers. 
For different weights, shapes, or (bipartite) lattices, it is well-known that various different curves separating up to three distinct phases are possible \cite{kenyon2006dimers,difrancesco2014acoe}.
Additionally, there are deep connections between weighted dimer models and algebraic geometry:
for example, the curves separating distinct phases, the number of ``gaseous'' phase regions, and the limiting shape behaviors in the model, all have geometric identifications with properties of Riemann surfaces \cite{kenyon2006dimers,berggren2023geometry}.

Returning to the quantum dimer model on the Aztec diamond, we note that since the region inside the Arctic circle is a critical region, it is not guaranteed if the quantum phase separation only occurs at the fine-tuned RK point or is stable to perturbations.
It is possible that perturbations away form the RK point will drive the critical region into an ordered phase, resulting in a uniform phase without any phase separation.
This raises a natural question: can the breakdown of the thermodynamic limit and the spatial separation of quantum phases persist over an extended region of the phase diagram?
We show that this is likely possible by introducing our second model, defined on the square-octagon lattice (also known as the truncated square lattice) which consists of alternating squares and octagons.
In \cref{fig:so-random-config}(a), we show a random dimer covering on a diamond-shaped domain of the square-octagon lattice, known as the ``square-octagon fortress".
\begin{figure}[htbp]
  \centering
  \includegraphics[width=0.9\columnwidth]{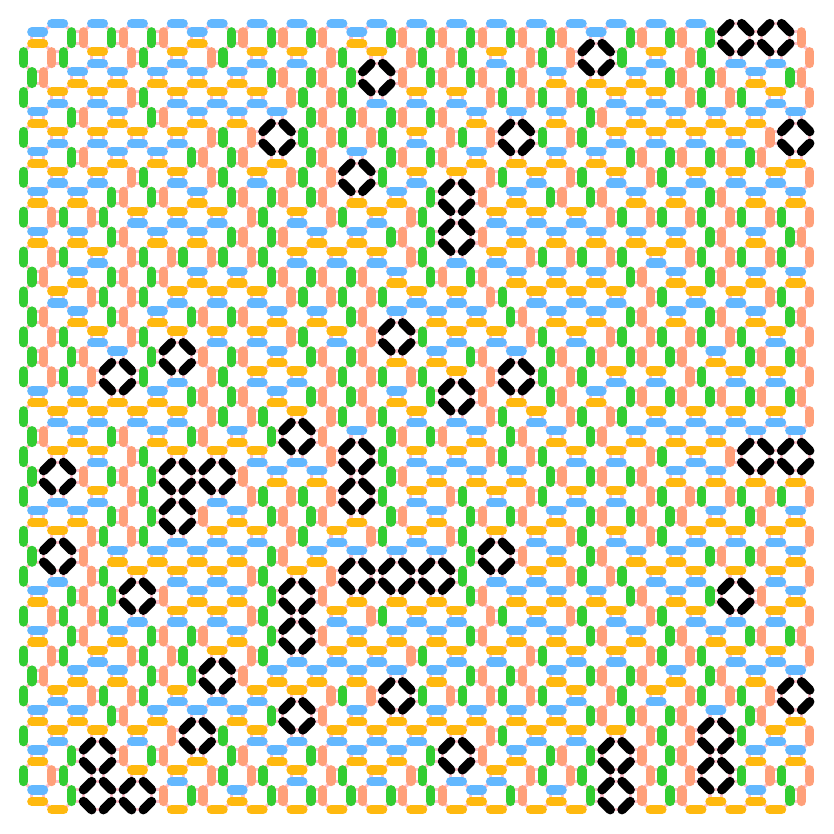}
  \caption{A random dimer configuration of the square-octagon lattice with rectangular boundaries. 
  Horizontal and vertical dimers are colored in orange, while diagonal dimers are colored in black.
  We see that all diagonal dimers appear as a part of some flippable octagon.
  This phase can be viewed as a background of flippable square plaquettes, interspersed with a finite density of black flippable octagons.}
  \label{fig:sor-random-config}
\end{figure}
From this figure, it is clear that, similar to the Aztec diamond, the corners of the square-octagon fortress exhibit an ordered dimer configuration \cite{propp2003generalized}.
The analogue of the ``Arctic circle" in this geometry is an octic curve~\cite{propp2003generalized,difrancesco2014acoe} that is the union of the oval and the star-shaped curves shown in \cref{fig:so-random-config}.
The ``Critical region" between these two curves is well-expected to have dimer-dimer connected correlators with power-law decay, while in the innermost region these correlators are well-expected to decay exponentially in the infinite-size limit, due to a relation with the dimer model on certain weighted Aztec diamonds
\cite{propp2003generalized,chhita2014coupling,difrancesco2014acoe,chhita2016domino}. 
We confirm the latter expectation numerically.
Although these regions are shown for a single dimer covering in the figure, the ground state of our quantum spin and dimer models at the RK point also possess these three regions.
These three regions are also clearly visible in the plot of dimer occupation probabilities for the RK wavefunction, shown in \cref{fig:so-random-config}(b).
In contrast, for the square-octagon lattice with square boundaries, a randomly
sampled dimer covering has neither the frozen nor critical regions, as seen from a typical dimer covering displayed in \cref{fig:sor-random-config}.
Perturbing away from the RK point of our dimer model, the critical region will likely cease to exist, possibly by turning into an ordered phase.
However, since the innermost region has exponentially decaying correlators, it is expected to remain stable under small perturbations.
Consequently, the ground state of the perturbed Hamiltonian will likely retain at least two macroscopic regions with distinct phases.
This demonstrates that the breakdown of the thermodynamic limit can persist beyond fine-tuned points and remain robust to perturbations.

We also investigate the phases of our dimer models on the square-octagon lattice with both rectangular and square-octagon fortress boundaries at their respective RK points.
Quantum dimer models are known to host exotic phases, such as quantum spin liquids---topologically ordered and highly entangled phases that are described by deconfined phases of gauge theories~\cite{savary2017quantum,moessner2001shortranged}.
For example, the RK quantum dimer model on the triangular lattice is known to lie in a $\mathbb{Z}_2$ quantum spin liquid phase~\cite{moessner2001resonating,misguich2008quantum,ioselevich2002groundstate}.
A $\mathbb{Z}_2$ quantum spin liquid phase supports two types of gapped excitations: electric charges and visons~\cite{senthil2000z2gauge}.
As a result, all correlators, including both the dimer-dimer connected correlator and the nonlocal vison correlator must decay exponentially with distance.
The vison, defined via a string operator, is a topological excitation that is a relative semion with respect to the electric charge: braiding one around the other produces a statistical phase of $-1$.
The presence of gapped visons implies fractionalization, and thus the existence of a quantum spin liquid~\cite{senthil2000z2gauge}, since gapped visons are not present in the ground state, causing the electric charges to be deconfined~\cite{balents2002fractionalization}.
Therefore, calculating the vison correlator is crucial for diagnosing if the dimer model is in a quantum spin liquid phase or not.
While classical Monte Carlo methods are often used to estimate the vison correlators~\cite{shah2025quantumpenrose,ioselevich2002groundstate,balents2002fractionalization}, they only provide approximate values due to finite sampling.
In this work, we describe an exact method to calculate the vison correlator using the Kasteleyn matrix~\cite{kasteleyn1961statistics,temperley1961dimer,fisher1961stastical} and its inverse.
This approach applies to the RK wavefunction on any planar lattice.
This method allows us to calculate the vison correlator easily to arbitrary precision (here we go up to $10^{-21}$), while Monte Carlo calculations would only give an accuracy of $10^{-6}$ in reasonable times. 
Furthermore, along certain paths of the square-octagon lattice with rectangular boundaries, we utilize the Toeplitz structure of the inverse Kasteleyn matrices to determine the vison correlator analytically.  

We calculate the dimer-dimer connected and vison correlators in the central region of the square-octagon fortress, as well as in the square-octagon lattice with rectangular boundaries, to determine whether these regions realize a quantum spin liquid phase.
In both geometries, we find that the dimer-dimer connected correlator decays exponentially, while the vison correlator approaches a constant along certain paths.
Moreover, the asymptotic values of the vison correlators appear to be essentially the same in both cases.
This rules out the possibility of a $\mathbb{Z}_2$ quantum spin liquid phase in these regions, indicating that the system is instead in an ordered phase.
We identify the nature of the order by studying the conditional probabilities of diagonal dimers and conclude that the system is in a short range entangled state similar to the Ruby family of Ref.~\cite{balasubramanian2022classical}.

The structure of the paper is as follows.
In \cref{sec:kasteleyn}, we begin by reviewing the Kasteleyn method to calculate the partition function and dimer probabilities. 
We then describe an exact method to calculate vison correlators for a general planar graph.
Next, in \cref{sec:aztec}, we provide our first example of the breakdown of the thermodynamic limit on the Aztec diamond. 
We define a quantum spin Hamiltonian that maps to a quantum dimer model on both the square lattice with rectangular boundaries and the Aztec diamond.
\Cref{sec:square-octagon-rect} focuses on the square-octagon lattice with rectangular boundaries.
We define the corresponding spin and low-energy effective dimer Hamiltonians and calculate dimer-dimer correlators in the infinite lattice limit using the asymptotic form of the Kasteleyn matrix inverse.
Using the method of \cref{subsec:vison}, we calculate the vison correlator and find that it approaches a constant along certain paths, implying that the system is in an ordered phase.
We end the section by commenting on the nature of the ordered phase.
In \cref{sec:sqoct-fortress}, we provide our second example of the breakdown of the thermodynamic limit---this time likely stable to perturbations---by studying the quantum dimer model on the square-octagon fortress.
We define the spin and dimer Hamiltonians as in the other sections, discuss the connection between the fortress and weighted Aztec diamonds, and calculate the dimer-dimer and vison correlators.
We find that the central region behaves exactly like the square-octagon lattice with rectangular boundaries considered in \cref{sec:square-octagon-rect}.
We again comment on the nature of the order in the different regions.
Finally, in \cref{sec:discusssion}, we summarize our results and outline several future directions.

\section{Exact correlators via Kasteleyn method}\label{sec:kasteleyn}

In this section, we briefly review Kasteleyn's method for counting the total number of dimer coverings of a planar graph, and describe its standard application to calculating dimer-dimer correlators. 
For further references, see Kasteleyn's original papers \cite{kasteleyn1961statistics,kasteleyn1963dimer,kasteleyn1967graph} and the overviews in Refs.~\cite{kenyon2009lectures, cimasoni2014geometry, gorin2021lectures}.
We then present an efficient, lesser-known method for calculating the vison correlator \emph{exactly}, which will be useful for both numerical and analytical calculations in \cref{sec:aztec,sec:square-octagon-rect,sec:sqoct-fortress}.

\subsection{Kasteleyn method overview}
\label{sec:kasteleyn-method-overview}

\begin{figure}[tb]
\begin{tikzpicture}
\begin{scope}[decoration={
    markings,
    mark=at position 0.5 with {\arrow[scale=1.1]{Stealth}}}] 
\foreach \y in {0,1}{
\foreach \x in {0,1,2,3}{
\draw[postaction={decorate}] (\x,\y)--(\x+1,\y);
}
\foreach \ux in {1,3}{
\draw[postaction={decorate}] (\ux,\y)--(\ux,\y+1);
}
\foreach \dx in {2}{
\draw[postaction={decorate}] (\dx,\y)--(\dx,\y-1);
}
}
\draw[postaction={decorate}] (1,-1)--(1,0);
\draw[postaction={decorate},xshift=2cm] (1,-1)--(1,0);
\draw[postaction={decorate}] (2,2)--(2,1);
\def\rad{.08}
\foreach \wc in {(1,0),(3,0),(2,1)}{
\draw[fill=white] \wc circle (\rad);
}
\foreach \bc in {(2,0),(1,1),(3,1)}{
\draw[fill=black] \bc circle (\rad);
}
\end{scope}
\end{tikzpicture}
\caption{A Kasteleyn orientation on the square lattice: all horizontal edges are oriented to the right, while vertical edges alternate in direction from column to column.}\label{fig:square-orientation}
\end{figure}
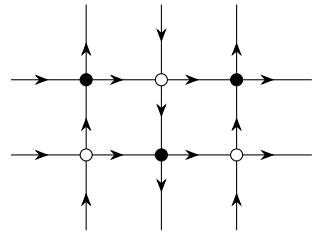

We start by counting the number of dimer coverings of a bipartite planar graph $G$ with $2N$ vertices.
A dimer covering, also known as a perfect matching of a graph, is a set of edges such that each vertex is connected to exactly one edge in the set.
Since $G$ is bipartite, we can color its vertices with two colors, say black and white, creating two disjoint sets $B=\{b_1,\ldots,b_N\}$ and $W=\{w_1,\ldots,w_N\}$, where a vertex $b\in B$ connects only to vertices $w\in W$, and vice versa. 
Due to the bipartite structure, all graph information is captured in a reduced, half-size $N\times N$ adjacency matrix $A$, with rows  indexed by vertices in $B$ and columns indexed by vertices in $W$.
This matrix is defined as $A(b_i,w_j)=1$ if vertices $b_i\in B$ and $w_j\in W$ are connected by an edge, and 0 otherwise. The determinant of $A$ expands as
\begin{align}\label{eqn:detA}
\det A=\sum_{\sigma\in S_{N}}\operatorname{sgn}(\sigma)A(b_1,w_{\sigma(1)})\cdots A(b_{N},w_{\sigma(N)}),
\end{align}
where $S_N$ is the set of all permutations of $N$ elements.
From this expansion, we see that there is a bijection between the nonzero terms in the sum and the dimer coverings of $G$. 
However, this determinant need not count the total number of dimer coverings, due to the sign $\operatorname{sgn}(\sigma)$ which can cause cancellations. 
Kasteleyn \cite{kasteleyn1961statistics,kasteleyn1963dimer} showed how to construct a \emph{signed} adjacency matrix $K$, now called a Kasteleyn matrix, such that all nonzero terms in the sum for $\det K$ have the same sign.
This gives the result:
\begin{align}
\label{eqn:detk}
\text{Number of dimer coverings}&=|\det K|.
\end{align}
The construction of a Kasteleyn matrix is always possible for planar graphs, and can be extended to count dimer coverings on a torus using four determinants \cite{kasteleyn1961statistics} which we briefly review in \cref{app:torus}.
Specifically, in the planar case, the Kasteleyn matrix $K$ is constructed from a \emph{Kasteleyn orientation}, which is an assignment of directions to each edge such that every face has an odd number of edges oriented clockwise around its boundary \cite{kasteleyn1963dimer,kasteleyn1967graph,fendley2002classical,cimasoni2014geometry}.
For each directed edge $(b\to w)$, we assign it a positive ($+1$) orientation if it agrees with the Kasteleyn orientation and a negative ($-1$) orientation if it disagrees.
An example Kasteleyn orientation for the square lattice is shown in \cref{fig:square-orientation}.
The Kasteleyn matrix $K$ is then defined as:
\begin{align}\label{eqn:K-def}
K(b,w)&=\begin{cases}
1,&(b\to w)\text{ has positive orientation}\\
-1,&(b\to w)\text{ has negative orientation}\\
0,&(b,w)\text{ is not an edge}
\end{cases}.
\end{align}
More generally, for non-bipartite graphs, one uses the same method but with a full-size $2N\times 2N$ Kasteleyn matrix $\Kl$ and its Pfaffian (see e.g. \cite{kasteleyn1961statistics,cimasoni2014geometry} or \cref{subsec:large} for a brief review).
However, since this paper focuses on the Aztec diamond and square-octagon graphs, which are bipartite, we will primarily work with the smaller,  ``half-size'' Kasteleyn matrices $K$. 

Kasteleyn's method gives a formula for the probability that a particular set of dimers appear, expressed in terms of the inverse of the Kasteleyn matrix $K$. 
In the context of bipartite lattices, this is often referred to as Kenyon's formula after Ref.~\cite{kenyon1997local}. 
The number of dimer coverings where a dimer occupies each edge in a set of edges $E$, can be counted using Kasteleyn's method by noting that enforcing the presence of a dimer on an edge $e=(b,w)$ is effectively equivalent to removing the vertices $b$ and $w$ (along with all their incident edges) from the graph. 
This implies that the number of dimer coverings with dimers on all edges in $E$ is given by the determinant of a submatrix of $K$, which, by Jacobi's minor theorem, can be written in terms of the matrix elements of $K$ and $K^{-1}$ corresponding to the vertices in $E$.

\vspace{1mm}
\noindent Kenyon's formula~\cite{kenyon1997local}:
\textit{
Let $K$ be a Kasteleyn matrix for a bipartite graph $G$.
Given a set of edges $E=\{e_1,\ldots,e_k\}$, with $e_i=(b_i,w_i)$, the probability that the edges in $E$ occur in a dimer covering is
\begin{multline}\label{eqn:kenyon}
\P[e_1,\ldots,e_k\text{ all occur in a dimer covering}]\\
=\left[\prod_{i=1}^kK(b_i,w_i)\right]\det\left[K^{-1}(w_i,b_j)\right]_{1\le i,j\le k},
\end{multline}
where the determinant is over the submatrix of $K^{-1}$ indexed by rows $w_i$ and columns $b_j$, $(w_i,b_j)_{1\le i,j\le k}$.
}

Kenyon's formula is commonly used to calculate an important quantity---the dimer-dimer connected correlator, defined as
\begin{align}
C_{\mu\nu}^d=\langle d_\mu d_\nu\rangle-\langle d_\mu\rangle\langle d_\nu\rangle,
\end{align}
where $\mu$ and $\nu$ label edges of the graph, and $d_{\mu} = 1$ if a dimer is present on edge $\mu$, and $d_{\mu} = 0$ otherwise.
Letting $\mu=(b_1,w_1)$ and $\nu=(b_2,w_2)$, \cref{eqn:kenyon} implies
\begin{align}\label{eqn:Kdimer}
C_{\mu\nu}^d&=-K(b_1,w_1)K(b_2,w_2)K^{-1}(w_2,b_1)K^{-1}(w_1,b_2).
\end{align}
Therefore, knowing the values of $K^{-1}$ on the relevant vertices allows one to calculate the dimer-dimer correlator.
More generally, \cref{eqn:kenyon} shows that the matrix $L$ defined by $L(w_i,b_j)=K(b_i,w_i)K^{-1}(w_i,b_j)$, serves as the correlation kernel for the determinantal point process formed by the edges~\cite{kenyon2009lectures,kenyon1997local}.

\subsection{Vison correlator via Kasteleyn matrices}\label{subsec:vison}
In this section, we explain how to calculate the vison correlator both efficiently and exactly.
It is well known that dimer models can be written as $\mathbb{Z}_2$ gauge theories~\cite{moessner2001shortranged,balents2002fractionalization,savary2017quantum}.
The electric charges of these gauge theories correspond to sites where the dimer constraint is violated while the flux excitations of these theories are known as visons.
The vison correlator~\cite{senthil2001fractionalizationPRL,senthil2001fractionalizationPRB,senthil2000z2gauge} is a nonlocal string correlator that serves as a key diagnostic for identifying phases in quantum dimer models~\footnote{In terms of concepts from the mathematics literature, the vison correlator is related to a form of the height function modulo two.}.
If the vison correlator decays exponentially with the distance between the string endpoints, the visons are gapped and can be viewed as excitations of the dimer model, implying the presence of fractionalized quasiparticles and that the system is a quantum spin liquid.

The type of calculation presented here traces back to the work of Onsager in the 1950s, Kasteleyn \cite{kasteleyn1961statistics}, and Montroll, Potts, and Ward \cite{montroll1963correlations} concerning spin correlations.
However, in the context of visons, this method is often overlooked in favor of approximate approaches such as random sampling or Monte Carlo methods.

We focus on planar bipartite graphs with $2N$ vertices, where we can work with the half-size Kasteleyn matrices $K$, although the same method (using the full-size Kasteleyn matrices and Pfaffians) applies to non-bipartite planar graphs.
The case of graphs on a torus is also similar; since we will use this for analytic results, we describe the necessary adjustments in \cref{app:torus}.

For a fixed path $p$ between faces $f_1$ and $f_2$, let $E=\{e_1,\ldots,e_\ell\}$ be the set of edges crossed by $p$. Given a dimer configuration $\dconfig$, let $\eta_E=\eta_E(\dconfig)$ denote the number of dimers present on the edges in $E$. The vison correlator between $f_1$ and $f_2$---which is independent of the choice of path $p$ as long as the paths are deformable into each other---is
\begin{align}\label{eqn:vison-kasteleyn}
\langle(-1)^{\eta_E}\rangle &= \frac{\sum_{\{\text{configs. }\dconfig\}}(-1)^{\eta_E(\dconfig)}}{\sum_{\{\text{configs. }\dconfig\}}1}
=\frac{\det \tilde{K}}{\det K},
\end{align}
where $\tilde K$ is defined as
\begin{align}
\tilde K(b,w)&=\begin{cases}-K(b,w),&(b,w)\text{ is an edge in }E\\
K(b,w),&\text{otherwise}
\end{cases}.
\end{align}
The determinant $\det\tilde{K}$ counts the signed sum $\sum_{\{\text{configs. }\dconfig\}}(-1)^{\eta_E(\dconfig)}$ for the same reason that $\det K$ counts the total number of configurations: the Kasteleyn orientation of $K$ was chosen precisely so that every nonzero term $\operatorname{sgn}(\sigma) K(b_1,w_{\sigma(1)})\cdots  K(b_N,w_{\sigma(N)})$ in the determinant expansion [cf. \cref{eqn:detA}] has the same sign.
By flipping the sign in $\tilde K$ along every edge $(b,w)$ in $E$, a given term $\operatorname{sgn}(\sigma)\tilde K(b_1,w_{\sigma(1)})\cdots \tilde K(b_N,w_{\sigma(N)})$ picks up an additional negative sign if and only if the configuration contains an odd number of dimers on edges in $E$. Thus, $\det\tilde{K}$ is equal (up to an overall sign) to $\sum_{\{\text{configs. }\dconfig\}}(-1)^{\eta_E(\dconfig)}$. \Cref{eqn:vison-kasteleyn} can be simplified further. 
Since $\tilde{K}=K-2K'$, where 
\begin{align}
\label{eqn:kprime-definition}
K'(b,w)&=\begin{cases}K(b,w),&(b,w)\text{ is an edge in }E\\
0,&\text{otherwise}
\end{cases},
\end{align}
we can view $K'$ as a block matrix with only one nonzero block.
This gives: 
\begin{align*}
\langle(-1)^{\eta_E}\rangle=\frac{\det \tilde K}{\det K}&=\det(I-2K'K^{-1})\\
&=\det\left(I_{\ell}-2(K')_E(K^{-1})_{E}\right),\numberthis\label{eqn:vison-fast}
\end{align*}
where $(K^{-1})_E$ and $(K')_E$ are the submatrices corresponding to the vertices in $E$.
Because this is only an $\ell\times\ell$ matrix determinant (for the bipartite case; in the non-bipartite case, it is the square root of a $2\ell\times2\ell$ determinant), it becomes very convenient to calculate, provided we can reliably extract the required $\ell\times\ell$ submatrix of $K^{-1}$. 
In the case of an infinite periodic lattice (with periodic boundary conditions), we can use the limiting formula for the inverse Kasteleyn matrix, expressed as contour integrals, to obtain an exact (though potentially complicated) analytic expression for the vison correlator.
We will apply this approach to analytically determine the asymptotic behavior of the vison correlator along a simple path in the square-octagon lattice in \cref{subsec:sqoct-vison}.
For other cases, we can use sparse LU decomposition to efficiently determine the required submatrix of $K^{-1}$ numerically.
However, for certain graphs, the Kasteleyn matrix $K$ can be ill-conditioned, meaning it has a large condition number $\kappa=\|K\|\|K^{-1}\|$.
We find this to be the case for the Aztec diamond and the square-octagon fortress, as detailed in \cref{sec:numdetails}.
In such cases, standard floating-point LU factorization is unreliable for numerically computing the required submatrix and can produce highly inaccurate results.
To avoid this issue, we implement sparse fraction-free LU factorization~\cite{lee1995fraction, nakos1997fraction, wilson1997determinant}, which relies entirely on exact integer arithmetic.
We will refer to fraction-free sparse LU as FFLU for short.

In summary, the numerical procedure for calculating vison correlators along a string $E$ is as follows:
\begin{enumerate}
    \item Construct the Kasteleyn matrix $K$ with integer values and calculate its fraction-free sparse LU decomposition~\cite{lee1995fraction}.
    \item Use the LU factors to perform fraction-free forward and back substitution~\cite{nakos1997fraction} to calculate the submatrix $(K^{-1})_E$, by solving $Kx=\delta_{v}$ for each vertex $v\in B$ contained in $E$, while maintaining the exact rational form of the values. 
    \item Calculate the determinant in \cref{eqn:vison-fast} using floating-point numbers with sufficient precision to account for the condition number $\kappa$ of the matrix $I_\ell-2(K')_E(K^{-1})_E$.
\end{enumerate}
If the graph is not bipartite, only minor adjustments are needed: in the second step above, one solves $Kx=\delta_v$ for every vertex in $E$, and the matrices in \cref{eqn:vison-fast}---$I_{2\ell}$, $(K')_E$, and $(K^{-1})_E$---are of size $2\ell\times2\ell$, with the final result having the square root of the determinant due to using Pfaffians.

We note that the fraction-free sparse LU method described above is sufficiently efficient that all numerical results presented in this paper, including correlators as small as $<10^{-20}$, were calculated using plain Python, with each plot generated in under a few hours on a desktop computer.
In contrast, calculating the correlators up to a precision of $10^{-5}$ for the square-octagon lattice dimer model using Monte Carlo calculations shown in \cref{sec:square-octagon-rect} required approximately one day on 100 computing cores.
Moreover, Monte Carlo convergence for the square-octagon fortress is observed to be slow~\cite{bhakta2016markov,bhakta2017sampling}.
As a result, calculating correlators on the fortresses via Monte Carlo is extremely computationally expensive, and we therefore do not calculate them using Monte Carlo.
This comparison highlights the remarkable efficiency and power of the Kasteleyn matrix-based method for calculating vison correlators.

\section{Phase separation example I: Aztec diamond}
\label{sec:aztec}

\subsection{Square lattice with rectangular boundaries}

In this section, we present the spin Hamiltonian that, in the low-energy limit, maps onto a quantum dimer model on a square lattice with rectangular boundaries. 
This construction is inspired by the Hamiltonians in Refs.~\cite{balents2002fractionalization,balasubramanian2022classical,shah2025quantumpenrose}.
We also review some of the known results about dimer models on the square lattice with rectangular boundaries.
\begin{figure}[htbp]
  \centering
  \includegraphics[width=0.6\columnwidth]{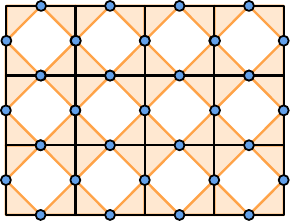}
  \caption{Square lattice with rectangular boundaries, where edges are shown in black.
  Blue dots represent spin-1/2 sites. 
  Spins belonging to the same plaquette are connected by orange edges, forming a lattice of corner-sharing polygons, which are shaded in orange.}
  \label{fig:square-spin}
\end{figure}

Consider an $n \times m$ rectangular section of the square lattice, with vertices labeled by pairs $(i,j)$ where $ 1 \leq i \leq n$ and $1 \leq j \leq m$, as shown in \cref{fig:square-spin}.
Place spin-1/2 degrees of freedom on the links of the lattice---these are indicated by the blue dots in the figure.
Next, construct the line graph of the square lattice: in this graph, each vertex corresponds to a spin, and two vertices are connected if the associated edges of the original square lattice share a common vertex.
This line graph can be interpreted as a graph of corner-sharing polygons, where the polygons in the bulk are squares while those on the boundaries are triangles, as illustrated by the orange polygons in \cref{fig:square-spin}.
This corner-sharing graph provides a natural graph to  define our Hamiltonian, which has two pieces:
\begin{align}
\label{eqn:hsq}
    H_{\text{sq}} = H_{\text{sq},0} + H_{\text{sq},1}~,
\end{align}
where $H_{\text{sq},0}$ is designed to impose the dimer constraint energetically and $H_{\text{sq},1}$ is designed to generate the ring-exchange and the Rokhsar--Kivelson potential~\cite{rokhsar1988superconductivity}. 
The classical part of the Hamiltonian, $H_{\text{sq},0}$ is given by:
\begin{align}
\label{eqn:hsq-0}
    H_{\text{sq},0} = U \sum_{p \in \left\{ \orangesq, \orangetrB, \orangetrA \right\}} \left( S^z_{p} - h_p \right)^2,
\end{align}
where $S^z_{\orangesq}$ denotes the sum of the $S^z \in \{ +1/2, -1/2 \}$ over the four spins located on the vertices of an orange-shaded square $\orangesq$.
Similarly, $S^z_{\orangetrA}$ and $S^z_{\orangetrB}$ represent the sum of $S^z$  over the two and three spins on the boundaries of triangles similar to $\orangetrA$ and $\orangetrB$, respectively.
We define $h_p \equiv q_p/2 - 1$, where $q_p$ is the number of spins on the boundary of $p$. 
Thus, we have $h_{\orangesq} = 1, h_{\orangetrB} = 1/2$, and $h_{\orangetrA} =  0$. 
Note that in the sum in \cref{eqn:hsq-0}, we also include orientations of $\orangetrB$ and $\orangetrA$, obtained by a rotation  by $90^\circ, 180^\circ $, and $270^\circ$, which are present in \cref{fig:square-spin}.

The ground state manifold of $H_{\text{sq},0}$ maps to dimer coverings of the square lattice.
Since all terms of $H_{\text{sq},0}$ commute with each other, the ground state manifold can be obtained by minimizing each term in the sum individually. 
We will see below that this simultaneous minimization can be done, implying that $H_{\text{sq},0}$ is also frustration free.
Minimizing the term corresponding to orange squares $\orangesq$ gives the constraint $S^z_{\orangesq} = 1$ for each $\orangesq$, implying that three of the spins have $S^z = +1/2$ and one spin has $S^z = -1/2$ on every $\orangesq$.
Mapping $S^z_{\mu} = \pm 1/2$ to the absence/presence of a dimer on link $\mu$, we see that $\left( S^z_{\orangesq} - 1 \right)^2$ is minimized when exactly one dimer touches the site at the center of $\orangesq$ giving us the hard-core dimer constraint.
It is easy to check that  minimizing $\left( S^z_{p}  \right)^2$ for $p \in \{ \orangetrB, \orangetrA \}$ imposes the hard-core dimer constraint on the boundary sites of the square lattice.

We now describe $H_{\text{sq},1}$, which is designed to map to the ring-exchange and the Rokhsar--Kivelson potential terms~\cite{rokhsar1988superconductivity} when interpreted in terms of dimers.
$H_{\text{sq},1}$ is defined as:
\begin{align}
\label{eqn:hsq-1}
    H_{\text{sq},1} = \sum_{\square}  \left(V - J \prod_{\nu =1}^{4} 2S^x_\nu \right) \left[ \sum_{\eta \in \{\pm\}} \prod_{\mu = 1}^{4} \left( S^z_{\mu} + \eta \frac{(-1)^\mu}{2} \right) \right]
\end{align}
where the outer sum runs over all plaquettes $\square$ of the square lattice, and $S_1, S_2, S_3, S_4$ denote the four spins located on the edges of a square plaquette, taken in a clockwise order.
The term in square brackets acts (up to a constant multiple) as a projection operator that selects a flippable plaquette.
A plaquette is said to be flippable if it has dimers on a pair of opposite edges.
In terms of spin configurations, a plaquette is flippable if $(S^z_1, S^z_2, S^z_3, S^z_4)$ is $ (1,-1,1,-1)/2$ or $ (-1,1,-1,1)/2$.
The term proportional to $J$ flips all four spins on a plaquette, provided the plaquette is flippable, thereby generating the ring-exchange process.
The terms proportional to $V$ correspond to the RK potential~\cite{rokhsar1988superconductivity}.

In the limit $U \gg |V|, |J|$, the low-energy states are those that correspond to valid dimer coverings of the square lattice. 
Within this subspace, $H_{\text{sq},1}$ acts as the RK quantum dimer model~\cite{rokhsar1988superconductivity}:
\begin{align}\label{eqn:hsq1}
\begin{split}    
    H_{\text{sq},1}^{\text{dimer}} = \sum_{\square} & -J \Big[ \ket{\plaqB}\bra{\plaqA} + \ket{\plaqA}\bra{\plaqB} \Big] \\
    &+ V \Big[ \ket{\plaqB}\bra{\plaqB} + \ket{\plaqA}\bra{\plaqA} \Big],
\end{split}
\end{align}
where  $\ket{\plaqA}$  and $\ket{\plaqB}$ denote the two flippable dimer configurations on a square plaquette.
For $J=V$, this dimer Hamiltonian can be rewritten as a sum of (non-commuting) projectors:
\begin{align}
\label{eqn:hsq1-dimer}
    H_{\text{sq},1}^{\text{dimer}} = J \sum_{\square}  \Big[ \ket{\plaqB} - \ket{\plaqA} \Big]\Big[ \bra{\plaqB} - \bra{\plaqA} \Big].
\end{align}
For $J>0$, the ground state of this Hamiltonian is the uniform superposition of all dimer configurations---known as the RK wavefunction---since it is annihilated by every projector in~\cref{eqn:hsq1-dimer}.
\begin{equation}
    \ket{\Psi_{\text{RK}}^{\text{sq}}} = \sum_{\mathcal{C}} \ket{\mathcal{C}}
\end{equation}
where the sum runs over all dimer coverings $\mathcal{C}$ of the square lattice.

Certain well-known properties of the RK wavefunction $\ket{\Psi_{\text{RK}}}$ follow from known results about the classical dimer model on the square lattice with rectangular boundaries.
These properties can can be derived using Kasteleyn matrix methods~\cite{kasteleyn1967graph,kasteleyn1961statistics,temperley1961dimer,fisher1961stastical}, which we reviewed in~\cref{sec:kasteleyn}.
In particular, the long-distance behavior of the dimer-dimer connected correlator can be computed analytically using~\cref{eqn:Kdimer}. 
It is known that $C_{\mu \nu}^d$ decays as a power law at long distances: in general, $C_{\mu \nu}^d \propto 1/R_{\mu \nu}^2$,
where $R_{\mu\nu}$ denotes the distance between the edges $\mu$ and $\nu$~\cite{fisher1961stastical,fisher1963statistical}. (Depending on the relative orientation of the dimers and the vector joining $\mu$ and $\nu$, the decay of the dimer-dimer connected correlator may follow a higher power~\cite{fisher1963statistical}.)
Since $d_\mu = S^z_{\mu} + 1/2$, the dimer-dimer connected correlator is equal to the spin-spin connected correlator in the $S^z$ basis.

The power-law decay of the dimer-dimer correlator implies that the quantum dimer model, and hence the spin Hamiltonian in \cref{eqn:hsq}, is at a critical point in the phase diagram.
Indeed, the RK point $J=V$ lies at a quantum critical point separating different ordered phases~\cite{yan2021widely,banerjee2016finite,moessner2011introduction}.

To summarize, in the limit $U\rightarrow \infty$, the spin Hamiltonian in~\cref{eqn:hsq} maps onto a quantum dimer model on the square lattice with rectangular boundaries. 
At the RK point, the system resides at a critical point characterized power-law dimer-dimer connected correlations, separating distinct ordered phases.

\subsection{Aztec diamond}

In this section, we consider the spin and dimer models defined on the Aztec diamond.
Although this Hamiltonian differs from that on the square lattice with rectangular boundaries in \cref{eqn:hsq} only by boundary terms, the ground state at the RK point exhibits drastically different properties.
This provides our first example of the breakdown of the thermodynamic limit.

\begin{figure}[htbp]
  \centering
  \includegraphics[width=0.7\columnwidth]{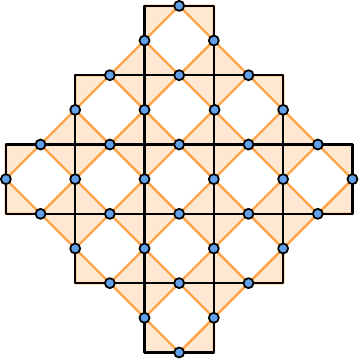}
  \caption{The black edges define an Aztec diamond. 
  Spin-1/2 degrees of freedom (blue dots) are located at the midpoints of these edges, forming a square lattice with square boundaries.
  As in \cref{fig:square-spin}, we construct the line graph of the Aztec diamond, resulting in a graph of corner-sharing polygons, shaded in orange.
  In this construction, the spins are naturally viewed as residing on the vertices of the resulting graph.
  }
  \label{fig:aztec-spin}
\end{figure}

The Aztec diamond is a finite region of the square lattice, with boundaries oriented at $\pm 45^\circ$ relative to the lattice basis vectors, such that each corner is one square wide, as shown in~\cref{fig:aztec-spin}.
As with the square lattice with rectangular boundaries, we place the spin-1/2s on the edges of the Aztec diamond and construct the line graph of the Aztec diamond.
This gives us a graph composed of corner-sharing polygons, as shown in \cref{fig:aztec-spin}.
The spin Hamiltonian on the Aztec diamond is closely related to that in~\cref{eqn:hsq} and is defined as:
\begin{equation}
\label{eqn:haz}
    H_{\text{az}} = H_{\text{az},0} + H_{\text{az},1}~,
\end{equation}
where
\begin{equation}
\label{eqn:haz0}
     H_{\text{az},0} = U \sum_{p \in \left\{ \orangesq,  \orangetrA \right\}} \left( S^z_{p} - h_p \right)^2.
\end{equation}
Here, the sum runs over all plaquettes of the shape $\orangesq$ and the various orientations of $\orangetrA$.
Symbolically, $H_{\text{az},1}$ is given by the same expression as $ H_{\text{sq},1}$ in \cref{eqn:hsq-1}, except that the sum is now taken over the plaquettes of the Aztec diamond rather than the rectangle.
In the limit $U\gg |V|,|J|$, the ground-state manifold of $H_{\text{az},0}$ consists of dimer coverings of the Aztec diamond, and in terms of dimers, $H_{\text{az},1}$ corresponds to the RK Hamiltonian.
As before, for $V=J$, the RK Hamiltonian becomes a sum of projectors, and a ground state is the RK wavefunction $\ket{\Psi_{\text{RK}}^{\text{az}}}$ on the Aztec diamond.
Notably, $H_{\text{az},0}$ differs from $H_{\text{sq},0}$ only at terms which are at the boundaries of the two graphs.

To understand the properties of the RK wavefunction on the Aztec diamond, we first calculate the dimer density $\langle d_\mu \rangle$ using the Kasteleyn method described in~\cref{sec:kasteleyn-method-overview}, specifically applying~\cref{eqn:kenyon} with $k=1$.
In agreement with the infinite-limit asymptotics proved in Ref.~\cite{cohn1996local},
the behavior stands in a stark contrast to that of the square lattice with rectangular boundaries, where the dimer density remains constant in the bulk, $\langle d_\mu \rangle = 1/4$ for all edges.
As briefly discussed in~\cref{sec:introduction}, the Aztec diamond features ``frozen" corners: certain links are occupied, while others are not occupied, with a probability approaching one in the infinite system-size limit~\cite{jockusch1998random,cohn1996local}.
In a typical configuration, these frozen corners exhibit a staggered dimer pattern, as illustrated in \cref{fig:az-random-config} and its inset.
Inside the circular region shown in the same figure, however, the dimers are not frozen.
This central area, labeled as the ``Critical region" is separated from the frozen corners by a boundary known in the mathematical literature as the ``Arctic circle"~\cite{jockusch1998random,cohn1996local}, since it separates the frozen and the ``liquid" regions.
It is also known that, inside the Arctic circle, the dimer-dimer connected correlator decays as $1/(\text{distance})^2$~\cite{cohn1996local,kenyon1997local,kenyon2006dimers, chhita2014coupling,chhita2016domino} similar to the behavior in the square lattice with rectangular boundaries.

The above results imply that, in the spin system, the corner regions are in an ordered phase, while the region inside the Arctic circle is in a critical phase.
Importantly, the frozen regions occupy a finite fraction of the Aztec diamond (specifically $1-\pi/4\approx 21\%$)indicating that this is not merely a boundary effect. 
To the best of our knowledge, this quantum phenomenon---where a change in the shape of the boundaries gives rise to distinct spatial regions residing in different quantum phases---has not been reported in the literature.
We refer to this effect as \textit{quantum phase separation}.
This phenomenon bears resemblance to classical phase coexistence, such as the simultaneous presence of liquid and solid phases in systems undergoing a first-order transition, 
although in our case the phase separation is induced purely by the geometry of the boundaries.
Additionally, we note that the dimers in the top and the bottom corners are ordered horizontally, while those in the left and right corners are oriented vertically.

So far we have explained the coexistence of frozen and critical (or liquid) regions in the ground state at the special RK point $V=J$ of our Hamiltonian~\cref{eqn:haz}. 
A natural questions arises: what happens when we move away from this point?
Since the region inside the Arctic circle is critical, it remains unclear whether quantum phase separation is unique to the RK point or persists over an extended region of the phase diagram.
It is conceivable that even an infinitesimal deviation from the RK point causes the critical region to transition into an ordered phase, similar to the square lattice with rectangular boundaries.
This could result in a uniform ordered phase across the system with no spatial phase separation.
If this is the case, quantum phase separation would occur only at a single fine-tuned point in the phase diagram.
However, if the system exhibited phase separation between two (or more) regions with exponentially decaying correlations (i.e., gapped phases), one would expect stability under small perturbations away from the RK point, since the gap cannot close abruptly.
In~\cref{sec:sqoct-fortress}, we present an example of such a system on the square-octagon fortress, featuring three distinct regions: two with exponentially decaying (connected) dimer correlations and one with critical correlations.
Before turning to that, however, we first study a model on a simpler graph: the square-octagon lattice with rectangular boundaries.

\section{Square-octagon lattice with rectangular boundaries}
\label{sec:square-octagon-rect}

\subsection{Hamiltonian}
\label{sec:ham-sor}

\begin{figure}[tbp]
  \centering
  \includegraphics[width=0.8\columnwidth]{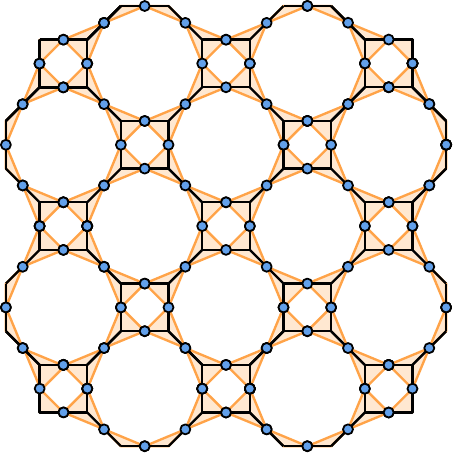}
  \caption{The black lines form a square-octagon lattice with a square boundary.
  Blue dots represent the spin-1/2 degrees of freedom, located at the vertices of the line graph of the square-octagon lattice.
  The edges of this line graph are marked in orange.
  This line graph can be viewed as a graph of corner-sharing triangles, which are shaded in orange.}
  \label{fig:so-rect}
\end{figure}

We now describe the square-octagon lattice (also known as the ``truncated square tiling'') and introduce the spin Hamiltonian that maps onto a quantum dimer model defined on this lattice.

The square-octagon lattice, composed of alternating squares surrounded by octagons, is shown in~\cref{fig:so-rect}.
In this section, we focus on square-shaped regions of this lattice like the one displayed in \cref{fig:so-rect} and we simply refer to such sections as the square-octagon lattice.
Analogous to \cref{sec:aztec}, we construct its line graph to obtain a graph of corner-sharing triangles, on which we define a spin Hamiltonian.
In this line graph, each vertex represents an edge of original lattice, and two vertices are connected if and only if their corresponding edges in the original lattice share a common endpoint. 
The Hamiltonian, analogous to~\cref{eqn:hsq,eqn:hsq-0,eqn:hsq-1}, for the square-octagon lattice is given by:
\begin{align}
\label{eqn:hsor}
    H_{\text{sos}} = H_{\text{sos},0} + H_{\text{sos},1},
\end{align}
where $H_{\text{sos},0}$ energetically enforces the dimer constraint, while $H_{\text{sos},1}$ generates the RK Hamiltonian.
Note that ``sos" here refers to ``square-octagon.
We define $H_{\text{sos},0}$ as:
\begin{align}
\label{eqn:hsor-0}
    H_{\text{sos},0} = U \sum_{p \in \left\{\orangetrC, \orangetrA, \orangetrD \right\}} \left( S^z_{p} - h_p \right)^2,
\end{align}
where $S^z_{p}$ is the sum of the $S^z$ over all spins on triangle $p$, and $h_p \equiv 1 -  q_p/2$, as in~\cref{sec:aztec}.
Explicitly, $h_{\orangetrC} = -1/2$, while $h_{\orangetrD} = h_{\orangetrA} = 0$.
We define $H_{\text{sos},1}$ as:
\begin{align}
\label{eqn:hsor-1}
    H_{\text{sos},1} &= \sum_{\square} \left( V_4 - J_4 \prod_{\nu = 1}^4 2 S^x_{\nu} \right) \left[ \sum_{\eta \in {\pm} } \prod_{\mu =1}^4 \left( S^z_{\mu} + \eta \frac{(-1)^\mu}{2} \right) \right] \nonumber \\
    & +\sum_{\octagon} \left( V_8 - J_8 \prod_{\nu = 1}^8 2 S^x_{\nu} \right) \left[ \sum_{\eta \in {\pm} } \prod_{\mu =1}^8 \left( S^z_{\mu} + \eta \frac{(-1)^\mu}{2} \right) \right],
\end{align}
with the first sum running over all square plaquettes and the second over all octagonal plaquettes.
Here, $S^z_1, S^z_2, S^z_3,$ and $S^z_4$ in the first sum refer to the spins along a square plaquette (ordered clockwise), and
 $S^z_1, S^z_2, \ldots , S^z_8$ in the second sum refer to the spins along an octagonal plaquette (also ordered clockwise).

In the limit $U \gg |V_4|, |V_8|, |J_4|, |J_8|$, $H_{\text{sor,0}}$ imposes the dimer constraint is energetically, and  the low-energy manifold consists of perfect dimer coverings of the lattice.
Within this manifold, $H_{\text{sor,1}}$ acts as the RK Hamiltonian:
\begin{align}
\begin{split} 
\label{eqn:hsos1-dimer}
    H_{\text{sos},1}^{\text{dimer}}  = & \sum_{\square} - J_{4} \Big[ \ket{\plaqB}\bra{\plaqA} + \ket{\plaqA}\bra{\plaqB} \Big] \\
    & \quad\quad\quad +  V_{4} \Big[ \ket{\plaqB}\bra{\plaqB} + \ket{\plaqA}\bra{\plaqA} \Big] \\
    & + \sum_{\octagon} -J_{8} \Big[ \ket{\octA}\bra{\octB} + \ket{\octB}\bra{\octA} \Big] \\
    & \quad \quad\quad\quad + V_{8} \Big[ \ket{\octB}\bra{\octB} + \ket{\octA}\bra{\octA} \Big]
\end{split}
\end{align}
For $V_4 = J_4 > 0$, $V_8 = J_8 > 0$, and arbitrary $V_4/V_8$, $H_{\text{sos,1}}^{\text{dimer}}$ becomes a sum of (non-commuting) projectors, and the uniform superposition of all dimer coverings, $\ket{\Psi_{\text{RK}}^{\text{sos}}}$, is the ground state.
It is known that the ring-exchange moves on the square-octagon lattice are ergodic~\cite{bhakta2016markov,bhakta2017sampling}, implying that all dimer coverings belong to a single connected sector. 
As a result, the ground state of \cref{eqn:hsos1-dimer} at the RK point is unique.

To understand the properties of the RK wavefunction on the square-octagon lattice, we will compute the dimer–dimer and vison correlators using Kasteleyn-like methods.

\subsection{Inverse Kasteleyn matrix asymptotics}

In this section, we compute the asymptotic forms of the partition function and the inverse Kasteleyn matrix in the infinite-size limit for the square-octagon lattice with rectangular boundaries and periodic boundary conditions (PBC).
These asymptotics are later used in~\cref{subsec:dimer-dimer-sos,subsec:vison-sos} to evaluate the dimer-dimer and vison correlators, respectively.
This follows standard methods, as described for example in Refs.~\cite{cohn2001variational, kenyon2006dimers}.
Here, we describe only the key definitions, the fundamental domain (unit-cell) diagram, and final results, deferring the full derivation to~\cref{app:inf}.

\begin{figure}[tbp]
\centering
\begin{tikzpicture}[scale=0.85]
\def\rad{.08}
\begin{scope}[decoration={
    markings,
    mark=at position 0.5 with {\arrow[scale=1.1]{Stealth}}}
    ] 
\foreach \col in {0,2,4}{
\foreach \row in {0,-1}{
\draw[postaction={decorate},yshift={\row cm}] (\col-1,{(2*\row+1)*mod(\col,4)/2})--(\col,{(2*\row+1)*mod(\col,4)/2});
}
}
\draw[postaction={decorate}](0,-1)--++(0,1); 
\draw[postaction={decorate}](1,2)--++(0,-1);
\draw[postaction={decorate}](2,1)--++(0,1);
\draw[postaction={decorate}](1,-2)--(1,-3);
\draw[postaction={decorate}](2,-3)--(2,-2);
\draw[postaction={decorate}](3,0)--(3,-1);
\draw[postaction={decorate}](0,0)--(1,1) ;
\draw[postaction={decorate}](0,-1)--(1,-2) ;
\draw[postaction={decorate}](2,1)--(3,0) ;
\draw[postaction={decorate}](2,-2)--(3,-1) ;
\end{scope}

\draw[fill=black] (0,0) circle (\rad);
\draw[fill=black] (2,1) circle (\rad);
\draw[fill=black] (1,-2) circle (\rad);
\draw[fill=black] (3,-1) circle (\rad);
\draw[fill=white] (1,1) circle (\rad);
\draw[fill=white] (3,0) circle (\rad);
\draw[fill=white] (0,-1) circle (\rad);
\draw[fill=white] (2,-2) circle (\rad);

\begin{scope}[yshift=.9mm]
    \node[above] at (0,0) {$v_1$};
    \node[left] at (1,1) {$v_6$};
    \node[right] at (2,1) {$v_2$};
    \node[above] at (3,0) {$v_7$};
\end{scope}
\begin{scope}[yshift=-.9mm]
    \node[below] at (3,-1) {$v_3$};
    \node[right] at (2,-2) {$v_8$};
    \node[left] at (1,-2) {$v_4$};
    \node[below] at (0,-1) {$v_5$};
\end{scope}

\end{tikzpicture}
\caption{Square-octagon fundamental domain with eight vertices $v_1,\ldots,v_8$. 
The Kasteleyn orientation, indicated by arrows on the edges, is inherited from the square lattice: by ``flattening'' the octagon---lowering vertices $v_6$ and $v_2$ to align with $v_1$, and raising $v_4$ and $v_8$ to align with $v_5$---the square-octagon lattice is realized as a subgraph of the square lattice (see also \cref{fig:sqoct-jk}).}\label{fig:sqoct-fc}
\end{figure}
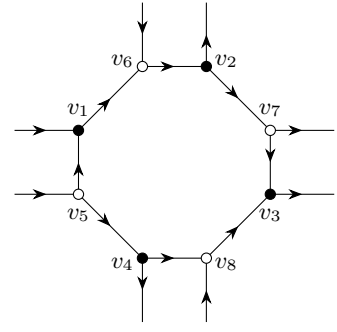

\Cref{fig:sqoct-fc} shows a fundamental domain of the square-octagon lattice, with eight vertices labeled $v_1,\ldots,v_8$, and a Kasteleyn orientation indicated by the arrows on each edge.
We construct the graph $G_n$ by taking $n$ horizontal and $n$ vertical translates of the fundamental domain. 
For analytical convenience, we impose periodic boundary conditions, though the resulting infinite-limit asymptotics agree with the finite-size, open-boundary numerical calculations.
We label vertex coordinates in $G_n$ as triples $|x,y,s\rangle$, where $(x,y)\in\{0,\ldots,n-1\}^2$ identifies the translated copy of the fundamental cell, and $s\in\{1,\ldots,8\}$ specifies the vertex $v_1,\ldots,v_8$ within the cell.

Correlators of diagonal operators in the RK wavefunction $\ket{\Psi_{\text{RK}}^{\text{sos}}}$ are equal to the correlators of a classical dimer model on the square-octagon lattice. 
Since all dimer coverings occur in $\ket{\Psi_{\text{RK}}^{\text{sos}}}$ with equal probabilities, the partition function of the classical dimer model $Z_n$, is simply
\begin{align}
Z_n=\#\{\text{dimer configurations of }G_n\}.
\end{align}
Because the graph $G_n$ is embedded on a torus, counting the number of dimer coverings requires determining four specific Kasteleyn matrices: $\Kzz_n,\Kzo_n,\Koz_n$, and $\Koo_n$ \cite{kasteleyn1961statistics}. 
The partition function can then be expressed as
\begin{align}\label{eqn:Z-torus}
Z_n=\frac{1}{2}\left(-Z^\zz_n+Z^\zo_n+Z^\oz_n+Z^\oo_n\right),
\end{align}
where $Z^\ab_n=\det K^\ab_n$.
As described for general graphs in \cite[\S3]{kenyon2006dimers}, the Kasteleyn matrices $K_n^{\ab}$ can be block-diagonalized via a Fourier transform.
This allows using a Riemann sum approximation to evaluate quantities such as the limiting free energy per unit cell, $\lim_{n\to\infty}n^{-2}\log Z_n$ as a double contour integral.
The limiting entries of $\left(\Kab_n\right)^{-1}$, for $\ab=\zz,\zo,\oz,\oo$, which are needed to analytically calculate the dimer and vison correlators, are also determined similarly.
From the block-diagonalization of $\Kab_n$, we obtain (see \cref{app:inf}) that, in the limit $n\to\infty$, the matrix entries $\langle 0,0,s|\Kab_n|x,y,t\rangle$ for any of the $\ab$ converge to
\begin{widetext}
\begin{align}\label{eqn:k-inv} 
\langle 0,0,s|\Ki^{-1}|x,y,t\rangle:=\begin{dcases}
\int_0^{2\pi}\int_0^{2\pi}\frac{Q_{st}(e^{i k_x},e^{i k_y})e^{-i (k_x x + k_y y)}}{-5+2\cos{k_x}+2\cos{k_y}} \, \frac{d k_x}{2\pi}\frac{d k_y}{2\pi},&s,t\text{ different color vertices}\\
 0,&s,t\text{ same color vertices}
\end{dcases},
\end{align}
where $Q_{st}(z,w)=-Q_{ts}(z,w)$ is a rational function given by the entries in the table
\begin{align}\label{eqn:numerator}
\begin{array}{c|cccc} 
\hline\hline
\hbox{\diagbox[height=5mm]{$s$}{$t$}}&1&2&3&4\\\hline
5&2-w^{-1} & 1-z^{-1}w^{-1} & 1-2 z^{-1} & -z^{-1}-w+1 \\ 
6&z+w-1 & 2-z^{-1} & 1-wz^{-1} & 1-2 w \\ 
7&2 z-1 & z+w^{-1}-1 & 2-w & 1-zw \\ 
8&1-zw^{-1} & 1-2 w^{-1} & -z^{-1}-w^{-1}+1 & z-2 \\\hline\hline
\end{array}~,
\end{align}
\end{widetext}
whose rows are indexed by $s=5,6,7,8$ and columns by $t=1,2,3,4$. (As seen in \cref{fig:sqoct-fc}, the vertices $v_s$ and $v_t$ can be split into different colors with $s\in\{5,6,7,8\}$ and $t\in\{1,2,3,4\}$.)
By translational invariance, this also determined the matrix entries for any pair of vertices $\ket{w,z,s}$ and $\ket{x,y,t}$.
\Cref{eqn:k-inv} exemplifies the more general fact that the  matrix entries of the limiting $\Ki^{-1}$ are always given by contour integrals which give Fourier coefficients of one of finitely many trigonometric polynomials~\cite[\S4]{kenyon2006dimers}.
The integrals in \cref{eqn:k-inv} can be reduced to a single integral using the residue theorem, as detailed in~\cref{app:inf}. 

It is well known that the total number of dimer coverings can be expressed as the partition function of a free-fermion hopping model defined on the same lattice as the dimer model~\cite{fendley2002classical}, 
providing a physics-friendly interpretation of \cref{eqn:k-inv}. 
With periodic boundary conditions, the system exhibits translational invariance, allowing diagonalization of the free-fermion model in momentum space.
For each momentum mode, one obtains an $8\times 8$ ``Hamiltonian", whose eigenvalues will give eight band dispersions.
The Green's function in the momentum space is obtained by inverting this momentum dependent Hamiltonian.
Doing so will lead to the appearance of the determinant of the Hamiltonian in the denominator, which is a product of the eight band dispersions given by the denominator of the integrand in \cref{eqn:k-inv}: $-5 + 2 \cos{k_x} + 2 \cos{k_y}$.
The real-space Green's function, $\langle 0,0,s| \mathcal{K}^{-1} |x, y, t\rangle$, is obtained by performing a Fourier transform of the momentum space Green's function.
The factor $Q_{st}(e^{ik_x}, e^{ik_y})$ in \cref{eqn:k-inv} arises from the sublattice structure.

Having determined the inverse Kasteleyn matrix elements in the infinite-system limit, we now use them to calculate the dimer-dimer and the vison correlators in the following sections.

\begin{figure}[htbp]
\includegraphics[width=.4\textwidth]{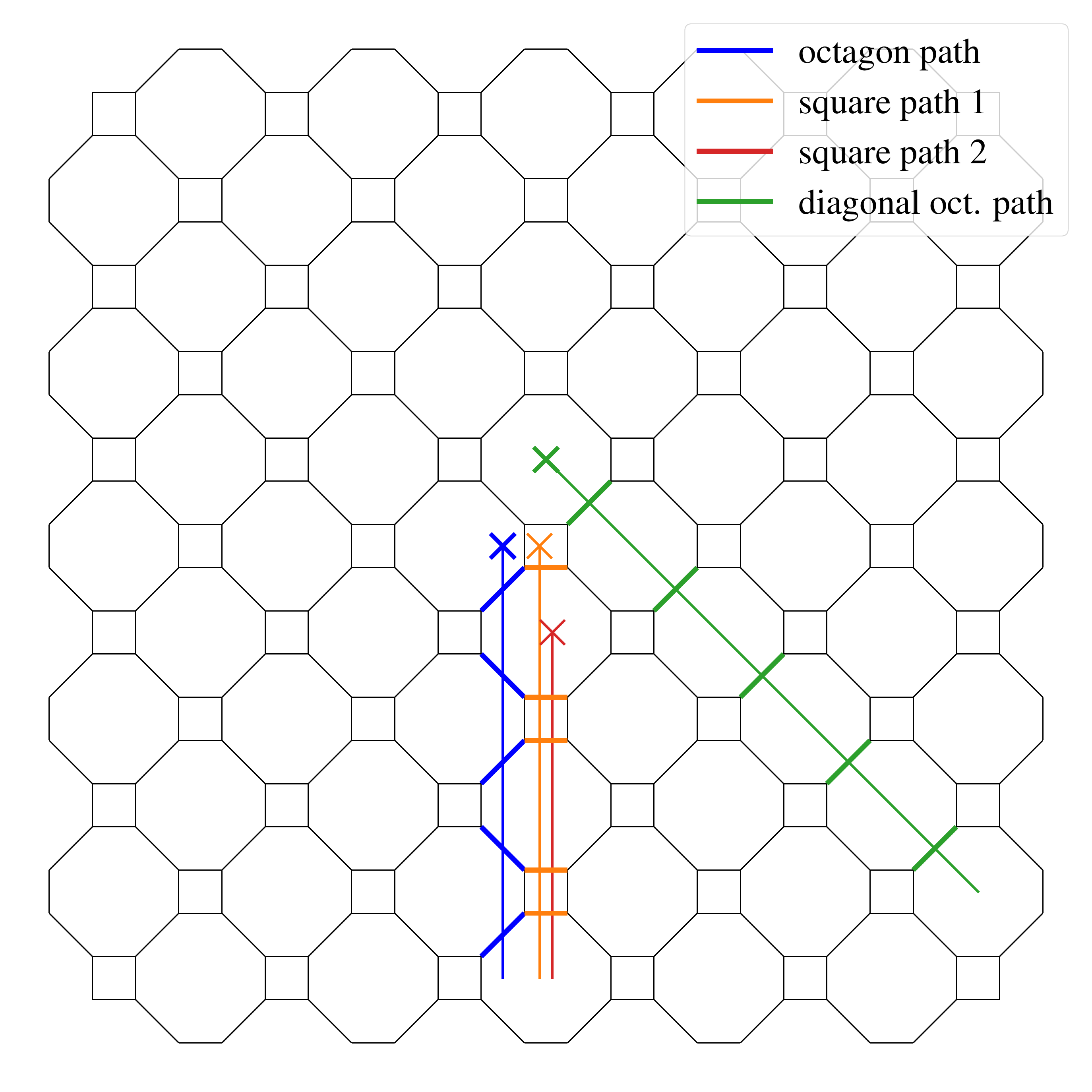}
\caption{Diagram of the four paths considered on the square-octagon lattice, drawn on a lattice with radius $R=6$. Here, ``radius" refers to the number of edges crossed by a line starting from the center square and going up vertically. 
The two square paths, shown in similar colors (orange and red), differ only in their starting point: one begins on a square face, and the other on an octagon face.
We refer to the longer of the two as ``square path 1''.
The ``octagon path'' and ``diagonal octagon path'' go through octagon faces only, with the former referring to horizontal or vertical paths and the latter to a diagonal path. These paths cross only diagonal edges which border the octagon faces.
}
\label{fig:sqoct-paths}
\end{figure}

\subsection{Dimer-dimer connected correlator}
\label{subsec:dimer-dimer-sos}

In this section, we calculate the dimer-dimer connected correlator $C_{\mu\nu}^d$ along the four paths shown in \cref{fig:sqoct-paths}.
A general argument from Ref.~\cite{kenyon2006dimers} shows that exponential vs power-law decay of the dimer-dimer connected correlator is determined by the \textit{characteristic polynomial} $P(z,w)$ of the lattice, which for the square-octagon lattice is given by $P(z,w)=-5+z+z^{-1}+w + w^{-1}$.
We explain this interpretation below.
The dimer-dimer connected correlator is given in terms of the inverse Kasteleyn matrix according to \cref{eqn:Kdimer}.
In the infinite-size limit, we replace the matrix elements of the finite inverse Kasteleyn matrix in \cref{eqn:Kdimer} with those of the infinite inverse Kasteleyn matrix $\mathcal{K}^{-1}$ defined in \cref{eqn:k-inv} (see \cref{app:torus} for details).
The denominator of the integrand in \cref{eqn:k-inv} is the characteristic polynomial $P(z,w)$, with $z=e^{ik_x}$ and $w=e^{ik_y}$.
Since for $k_x, k_y \in [0, 2\pi)$, the square-octagon lattice
$P(e^{ik_x}, e^{ik_y})$ does not have any zeros, and $Q_{st}(e^{ik_x}, e^{ik_y})$ is analytic, the integrand of \cref{eqn:k-inv} is an analytic function of $k_x$ and $k_y$. 
Thus its Fourier transform written in \cref{eqn:k-inv} will decay exponentially in $|x|$ and $|y|$.

This argument translated in the language of the free fermion model goes as follows: the product of band dispersion of the free fermions is given by $P(e^{ik_x}, e^{ik_y}) = -5 + 2 \cos{k_x} + 2\cos{k_y}$.
Since this function has a gap (it never vanishes for real $k_x, k_y$), all the bands of the free fermion model are gapped and the real-space Green's function and thus the dimer-dimer connected correlator will decay exponentially with distance.

We can also obtain the exact decay rates and correlation lengths using the following standard result:
\begin{lem}\label{lem:fourier}
If $f$ is $2\pi$-periodic and real analytic with analytic extension to a neighborhood of the strip $\{z\in\C:|\im z|\le\kappa\}$ for some $\kappa>0$, then the Fourier coefficients $\hat{f}(\xi)=\frac{1}{2\pi}\int_0^{2\pi}f(x)e^{-i\xi x}\,dx$ satisfy
\begin{align}
|\hat{f}(\xi)|\le C e^{-\kappa|\xi|},
\end{align}
for some constant $C=C(f)$. Conversely, if the Fourier coefficients have the above exponential decay, then $f$ has analytic extension to  the strip $\{z\in\C:|\im z|<\kappa\}$.
\end{lem}
Since the double integrals in \cref{eqn:k-inv} can be reduced to single integrals [see e.g. \cref{eqn:single-int}], the one-dimensional case stated above is enough. More generally, a $d$-dimensional version can be formulated, see for example, Ref.~\cite[Lemma 5.6]{broer2011dynamical}.

Applying the single integral expression \cref{eqn:single-int} to \cref{eqn:k-inv},
we see that the resulting integrand will always have denominator $\sqrt{(5-2\cos\theta)^2-4}$.
When extended to the complex plane, the nearest zero of this denominator to the real line occurs when $\cos(z)=3/2$, i.e., at $z=\arccos(3/2)=-\log[(3-\sqrt{5})/2]i$.
Now, applying the lemma, we find that 
\begin{align}\label{eqn:Kinv-expdecay}
|\langle0,0,s|\Ki^{-1}|x,y,t\rangle|&\le C\left(\frac{3-\sqrt{5}}{2}\right)^{\max(|x|,|y|)(1-\varepsilon)},
\end{align}
for any $\varepsilon>0$ and a (possibly different) constant $C$. 

For numerical comparison, it is convenient to index distances using the ``edge distance'' $\ell$, measured by counting the number of edges traversed, rather than by fundamental cell translations.
Since the four paths we consider pass through two edges per fundamental cell, this introduces a factor of $1/2$ in the exponent.
By the formula for the dimer-dimer connected correlator \cref{eqn:Kdimer}, the bound \cref{eqn:Kinv-expdecay} implies an upper bound on the dimer-dimer correlation length $\xi$ with respect to the edge distance $\ell$:
\begin{equation}\label{eqn:dcorr-ub}
\xi\le -\log\left(\frac{3-\sqrt{5}}{2}\right)^{-1}
=1.039043\ldots.
\end{equation}
In fact, along a horizontal or vertical line, indexing the dimers in the path by $\ell\in\N$ and applying \cref{eqn:Kdimer} and the converse statement \cref{lem:fourier} gives the asymptotic correlation length to actually be $\xi=-\log\left(\frac{3-\sqrt{5}}{2}\right)^{-1}$.

\begin{figure}[htb]
\includegraphics[width=.47\textwidth]{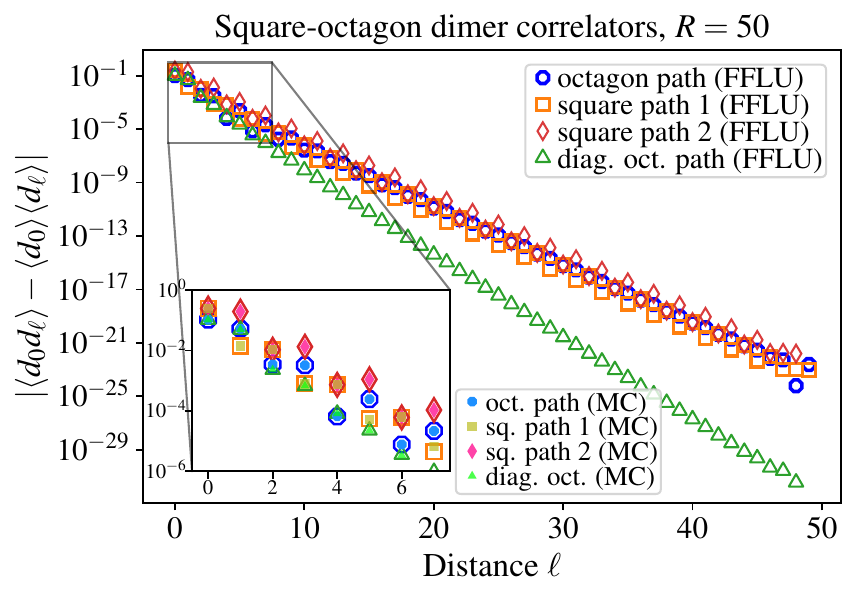} 
\caption{
Log plot of the dimer-dimer connected correlator along the four paths, calculated using the fraction-free LU (FFLU) method and compared against Monte-Carlo (MC) data (inset). 
The lattice has radius $R=50$, meaning there are $50$ edges along a horizontal or vertical path from the center to the boundary, corresponding to a graph with 19,996 vertices.
All dimer correlators show exponential decay.
The best-fit inverse correlation lengths (given by the negative  slopes of the log plot) lie in the range $0.978-0.991$ for the first three paths (calculated over the range $\ell=30$ to $46$), and $1.405$ for the fourth path in the same $\ell$ range.
These inverse correlation lengths are calculated with respect to the edge distance $\ell$, and are to be compared with the infinite limit lower bound $-\log\left(\frac{3-\sqrt{5}}{2}\right)=0.962424$.}\label{fig:dimerdimer}
\end{figure}

We compare the analytically obtained exponential decay in~\cref{eqn:dcorr-ub} to the numerical results for finite-size systems shown in \cref{fig:dimerdimer}.
For direct comparison with the best-fit slopes, we note that the reciprocal
$-\log\left(\frac{3-\sqrt{5}}{2}\right)=0.962424\ldots$ provides a lower bound on the inverse correlation length.
The numerical best-fit slopes for the octagon and square paths are all in the range $0.978-0.991$, which show close agreement with the above value, as expected since the bound is attained along these paths in the infinite size limit.

Let us briefly contrast the above with the \emph{square} lattice, where the corresponding denominator is the characteristic polynomial $P(z,w)=4+z+z^{-1}+w+w^{-1}$, which \emph{does} have a zero on the unit circles at $z=w=-1$.
As shown in Ref.~\cite{cohn2001variational} (and more generally in Ref.~\cite[\S4.2.2]{kenyon2006dimers}), this implies that the analogue of the integral in~\cref{eqn:k-inv} decays according to a power-law, leading to power-law decay of the dimer correlator. 
In terms of the associated free-fermion model on the square lattice, the band dispersion $4+2 \cos{k_x} + 2 \cos{k_y}$ is gapless at $(k_x, k_y) = (\pi, \pi)$ in the Brillouin zone, which explains power-law decay of the dimer-dimer correlator.
A similar phenomenon occurs on the hexagonal lattice, which is also critical.
More generally, since the characteristic polynomial $P(z,w)$ can be quickly computed form the fundamental cell, one can readily check whether a periodic lattice exhibits exponential or power-law decay of correlations~\cite{kenyon2006dimers}.
For example, the characteristic polynomials for a number of different lattices were calculated in Ref.~\cite{wu2006dimers}.

\subsection{Vison correlator}\label{subsec:sqoct-vison}
\label{subsec:vison-sos}

From the exponential decay of the dimer-dimer correlator, we infer that the quantum dimer model shown in \cref{eqn:hsos1-dimer} is likely not at a critical point, unlike the square lattice RK dimer model.
To determine the phase of the dimer model, we review the connection between gauge theories and dimer models.

It is well known that quantum dimer models can be effectively described by gauge theories~\cite{moessner2001shortranged,balents2002fractionalization}.
Gauge theories can either be in a deconfined phase, corresponding to the dimer model being in a quantum spin liquid phase, or in a confined phase, corresponding to an ordered phase such as a valence bond solid (VBS).
In particular, the dimer models studied in this paper can be mapped to $\mathbb{Z}_2$ gauge theories, which feature both electric and flux (vison) excitations.
The dimer constraint at each site plays the role of a Gauss law in the gauge theory, while electric charges correspond to violations of this constraint (i.e., sites with the incorrect number of dimers).
Visons are flux excitations that always appear in pairs, and the vison operator associated with a given path creates a pair of visons at its endpoints.
If the $\mathbb{Z}_2$ gauge theory is in the deconfined (gapped) phase, both the dimer–dimer and vison correlators decay exponentially with distance, and the visons behave as well-defined particle-like excitations.
Moreover, the exponential decay of the vison correlator to zero as the string length increases necessarily implies that the charges are fractionalized~\cite{senthil2000z2gauge}.
However, if the vison correlator approaches a nonzero constant, it indicates that the gauge theory is in the confined phase and the dimer model is in an ordered phase.
This behavior can be directly checked, for example, in the square lattice RK dimer model~\cite{rokhsar1988superconductivity}.
For RK potentials much larger than the ring-exchange coupling, the system exhibits columnar or staggered order, depending on the sign of the RK potential.
When the ring-exchange coupling is exactly zero, it is straightforward to see that the vison correlator is a constant in the staggered and the columnar configurations.
Even when the ring-exchange coupling is small (compared to the RK potential), the system remains ordered, and the vison correlator still approaches a nonzero constant.

\begin{figure*}[htb]
\begin{tikzpicture}
\def\hth{1.65in};
\def\rshift{5.9};
\node at (0,0) {\includegraphics[height=\hth]{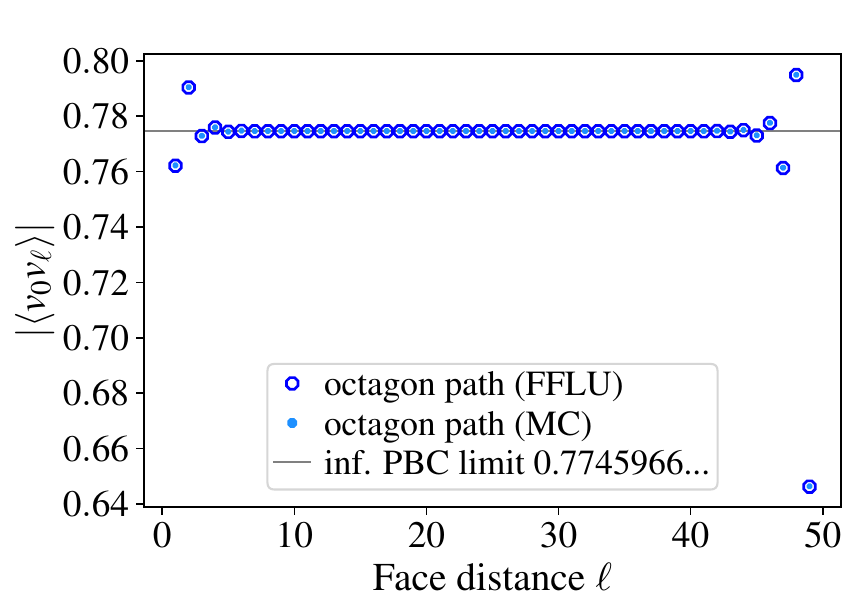}};
\node at (\rshift,0) {\includegraphics[height=\hth]{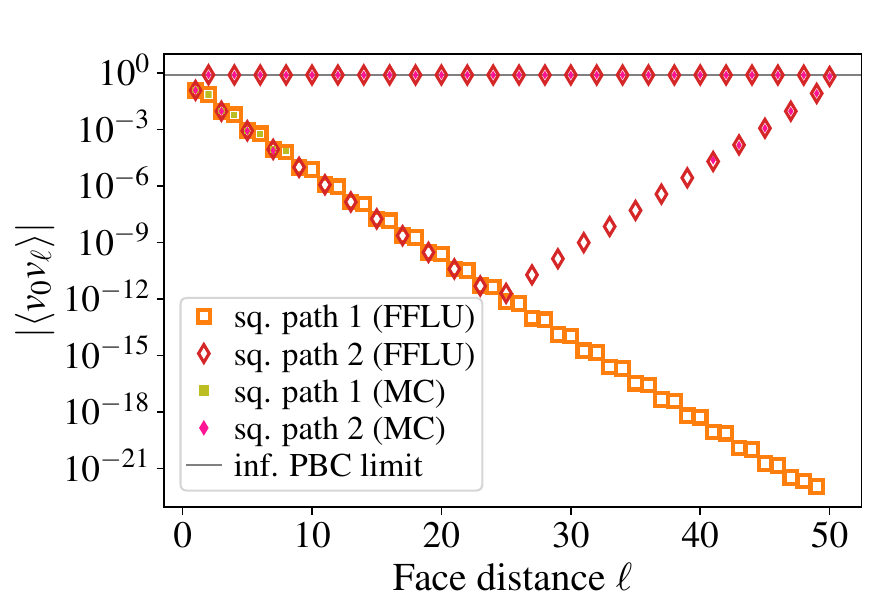}};
\node at (2*\rshift,0) {\includegraphics[height=\hth]{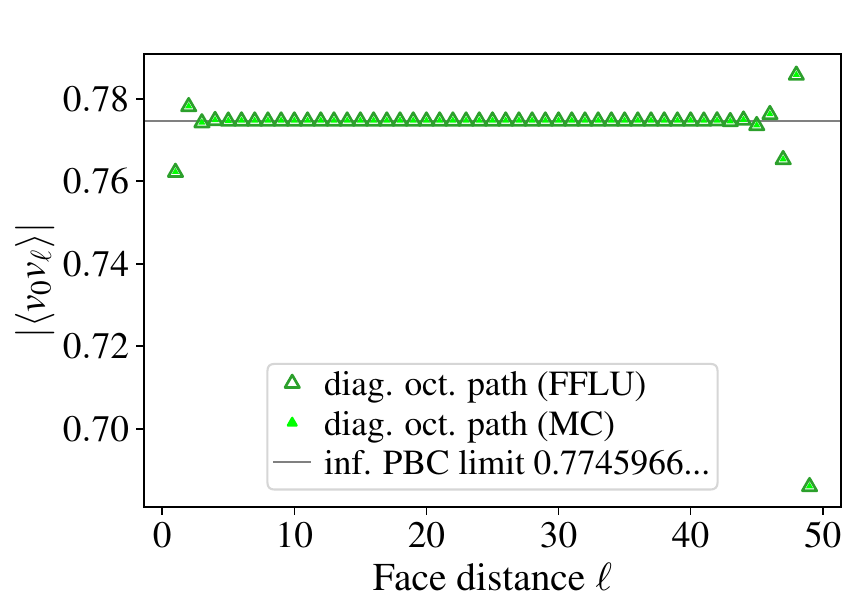}};
\def\xcoord{2.3};
\def\ycoord{2.0};
\begin{scope}[fontscale=-.5]
\node at (-\xcoord-0.3,\ycoord) {(a)};
\node[xshift={\rshift cm}] at (-\xcoord -0.15,\ycoord) {(b)};
\node[xshift={2*\rshift cm}] at (-\xcoord +0.05,\ycoord) {(c)};
\end{scope}
\end{tikzpicture}
\caption{
Vison correlator along different paths in the square-octagon lattice with open rectangular boundary conditions, calculated using fraction-free LU (FFLU) and Monte-Carlo (MC) methods.
All calculations were carried out on a lattice with a radius $R = 50$.
The x-axis in these plots is the number of edges crossed by the paths, denoted $\ell$.
(a) Vison correlator along the ``octagon path" from \cref{fig:sqoct-paths}, showing close agreement among FFLU, Monte-Carlo, and the analytically calculated infinite-size limit~\cref{eqn:sqoct-vison-limit0} indicated by the black horizontal line. 
(b) Vison correlators along the two square paths\footnote{For this plot, square path 2 is taken from the same starting face as in \cref{fig:sqoct-paths}, but continues to the top edge instead of the bottom, making it slightly longer.}, which show a mixture of constant and exponential decay behavior. 
The Monte Carlo data is plotted for values exceeding $2\times 10^{-5}$, where it agrees well with the FFLU results.
For the ``square path 1", represented by orange square markers, the vison correlator decays exponentially with string length $l$.
Fitting a straight line to the log plot over the range $\ell = 20$ to $40$, we obtain an inverse correlation length of $0.9955$ from the negative of the slope.
(c) Vison correlator along the ``diagonal octagon path'', again showing close agreement among FFLU, Monte Carlo, and the analytic infinite-size limit.
We note that for the octagon path shown in blue, similar plots can be efficiently generated for much larger radii (e.g. $R=100,200,500$) using sparse floating-point LU decomposition, all within a few minutes on a standard laptop.
}\label{fig:sqoct-vison}
\end{figure*}

Using~\cref{eqn:vison-fast}, we numerically calculate the vison correlator along several paths on a square-shaped section of the square-octagon lattice, similar to the one shown in~\cref{fig:sqoct-paths}. 
The figure also shows the four paths along which we calculate the vison correlator.
We fix one endpoint near the center of the graph and vary the other to generate strings of different lengths.
The resulting vison correlators are plotted in~\cref{fig:sqoct-vison}.
We find that the vison correlator behaves differently depending on the path.
Specifically, it remains constant along the ``octagon" and ``diagonal octagon" paths, decays exponentially along ``square path 1", and shows mixed behavior along ``square path 2": decaying exponentially if the endpoint lies on a square plaquette, but remaining constant if it ends on an octagonal plaquette.
Since the vison correlator depends only on the endpoints—not the specific path—the constant value along the octagon path implies it is also constant along ``square path 2" when both endpoints lie on octagons.
Since the vison correlator approaches a constant along certain paths, the system is not in a $\mathbb{Z}_2$ spin liquid phase and is instead in an ordered phase.

We now analytically prove that the vison correlator converges to a constant along the octagon path in the infinite-size square-octagon lattice with periodic boundary conditions.
Our final result is that the vison correlator along the octagon path approaches a constant as the path length goes to infinity:
\begin{align}\label{eqn:sqoct-vison-limit0}
\lim_{\ell\to\infty}\left|\langle v_0v_\ell\rangle\right|= 0.774596669\ldots.
\end{align}
To show this, we use vertex coordinates $(x,y,t)$, where $(x,y)\in\Z^2$ denotes the shift of the fundamental cell and $t\in\{1,\ldots,8\}$ labels the vertex within the cell, following the numbering in \cref{fig:sqoct-fc}.
The octagon path can be taken as the path crossing the $v_4v_5$ and $v_1v_6$ edges of fundamental cells with $(x,y)$ coordinates $(0,y)$ for $y=0,1,2,\ldots$.
The other vertical or horizontal octagon paths are equivalent by translational and reflection symmetry.
Recall that to analytically calculate the vison correlator using \cref{eqn:vison-fast}, we require the submatrices $(\mathcal{K}^{-1})_{E}$ and $(\mathcal{K}^{'})_{E}$.
To determine these, we first calculate the relevant matrix elements of $\mathcal{K}^{-1}$ using the infinite-size limit expression given in \cref{eqn:k-inv,eqn:numerator}, along with the single-integral simplification in \cref{eqn:single-int}.
The required matrix elements are:
\begin{align}
\begin{aligned}
\langle 0,0,5|\Ki^{-1}|0,y,4\rangle &= - \langle 0,0,6| \Ki^{-1} |0,y,1\rangle \\
&=- \int_0^{2\pi} e^{-i\theta y} F_1(\theta) \, \frac{d\theta}{2\pi} ,\\
\langle 0,0,5| \Ki^{-1} |0,y,1 \rangle &= -\langle 0,0,6| \Ki^{-1} | 0,y+1,4\rangle \\
&= - \int_0^{2\pi} e^{-i\theta y} F_2(\theta) \, \frac{d\theta}{2\pi} , \\
\end{aligned}\label{eqn:octpath-int}
\end{align}
where $F_1(\theta)$ and $F_2(\theta)$ are defined as: 
\begin{align}
\begin{aligned}
F_1(\theta) &\equiv \frac{-3-2e^{i\theta}+2\cos \theta+\sqrt{(5-2\cos \theta)^2-4}}{2\sqrt{(5-2\cos \theta)^2-4}},
\\ F_2(\theta) &\equiv \frac{2-e^{-i\theta}}{\sqrt{(5-2\cos \theta)^2-4}}.
\end{aligned}
\end{align}
Note that by translational invariance, $\langle 0,y_1,t|\Ki^{-1}|0,y_2,s\rangle=\langle 0,0,t|\Ki^{-1}|0,y_2-y_1,s\rangle$.
Thus, the four matrix elements above are sufficient to obtain all the required entries along the octagon path.
Thus we have,
\begin{align}
\begin{aligned}\label{eqn:kinv-octpath}
\langle0,y_1,5|\Ki^{-1}|0,y_2,4\rangle&=-\langle 0,y_1,6|\Ki^{-1}|0,y_2,1\rangle\\
&=-\hat{F}_1(y_2-y_1),\\
\langle 0,y_1,5|\Ki^{-1}|0,y_2,1\rangle&=-\langle0,y_1,6|\Ki^{-1}|0,y_2+1,4\rangle\\
&=-\hat{F}_2(y_2-y_1),
\end{aligned}
\end{align}
where $\hat{F}_{1, 2}$ denotes the Fourier transform of $F_{1, 2}$.
To obtain the vison correlator along the octagon path, we take the string of edges in \cref{eqn:vison-fast} to be $(v_4,v_5),(v_1,v_6)$, followed by their vertical translations.
The submatrix $(\Ki^{-1})_E$ is given by:
\begin{widetext}
\begin{align}
\label{eqn:k-inv-E-oct-path}
(\Ki^{-1})_E&=\begin{bNiceMatrix}[first-col,first-row]
   &0,4&0,1&1,4&1,1&\cdots&\ell-1,4&\ell-1,1\\
0,5& -\hat{F}_1(0)&-\hat{F}_2(0)&-\hat{F}_1(1)&-\hat{F}_2(1)&\cdots&-\hat{F}_1(\ell-1)&-\hat{F}_2(\ell-1)\\
0,6& \hat{F}_2(-1)&\hat{F}_1(0)&\hat{F}_2(0)&\hat{F}_1(1)&\cdots& \hat F_2(\ell-2) &\hat F_1(\ell-1)\\
1,5& -\hat{F}_1(-1)&-\hat{F}_2(-1)&-\hat{F}_1(0)&-\hat F_2(0) &\cdots & -\hat F_1(\ell-2)&-\hat F_2(\ell-2)\\
1,6&\hat{F}_2(-2)&\hat F_1(-1) &\hat F_2(-1) &-\hat F_1(0)&\cdots& \hat F_2(\ell-3)&\hat F_1(\ell-2)& \\
\vdots & \vdots &&&\vdots & &&\vdots 
\\
\ell-1,5& \hat F_2(-\ell)&\hat F_1(-\ell+1) & \hat F_2(-\ell+1) & \hat F_1(-\ell+2)&\cdots&\hat F_2(-1) & \hat F_1(0)
\end{bNiceMatrix},
\end{align}
\end{widetext}
where the row and column labeling $(y,t)$ is shorthand for vertex coordinates $(0,y,t)$.

The next quantity needed to calculate the vison correlator using \cref{eqn:vison-fast} is the submatrix $(\Ki')_E$.
Here, $\mathcal{K}'$ refers to the infinite-size limit of $K'$ as defined in \cref{eqn:kprime-definition}.
The submatrix $(\Ki')_E$ can be directly read from \cref{fig:sqoct-fc}; it is simply a diagonal matrix with alternating entries $-1,1,-1,1,\ldots$, in the same basis as $(\mathcal{K}^{-1})_{E}$ shown in~\cref{eqn:k-inv-E-oct-path}. Multiplying $(\Ki')_E$ against $(\Ki^{-1})_E$ flips the signs of every other row, giving a Toeplitz matrix $(\Ki')_E(\Ki^{-1})_E$.
This Toeplitz matrix is generated by a periodic function $F$, whose Fourier coefficients are given by interweaving those of $F_1$ and $F_2$, i.e.,
\begin{align*}
F(\theta)&=\sum_{j\in\Z}\hat{F}_1(j)e^{i2j\theta}+\sum_{j\in\Z}\hat{F}_2(j)e^{i2j\theta}e^{i\theta}\\
&=F_1(2\theta)+e^{i\theta}F_2(2\theta).\numberthis
\end{align*}
Thus,~\cref{eqn:vison-fast} [or more accurately \cref{eqn:vison-torus} in the infinite-size limit] gives the vison correlator as
\begin{align}
\langle(-1)^{\eta_E}\rangle&=\det(I_{\ell}-2T_{\ell}(F)),
\end{align}
where $T_{\ell}(F)$ is the $\ell\times\ell~$ Toeplitz matrix with matrix elements $[T_{\ell}(F)]_{ij}= \hat{F}(j-i)$, where the Fourier coefficients are given by $\hat{F}(j)=\frac{1}{2\pi}\int_0^{2\pi}F(\theta)e^{-ij\theta}\,d\theta$.
The limit of the determinant as $\ell\to\infty$ can be evaluated using the strong Szeg\H{o} limit theorem for Toeplitz matrices \cite{szego1952on,widom1974asymptotic,bottcher1990analysis,bottcher1995the,basor2019modified}.

\begin{thm}[Strong Szeg\H{o} limit theorem]
Let $f:\R/(2\pi\Z)\to\C$ be a smooth function which has no zeros and has winding number zero, and let $T_\ell(f)$ be the Toeplitz matrix with elements $[T_\ell(f)]_{ij} = \hat{f}(j-i)$, where $i,j \in \{0,1,2,\ldots, l-1\}$. 
If $\hat{c}_k=\frac{1}{2\pi}\int_0^{2\pi}e^{-i\theta k}\log f(\theta)\,d\theta$ are the Fourier coefficients of $\log F\in L^1$, then
\begin{align}
\lim_{\ell\to\infty}\frac{\det T_\ell(f)}{e^{(\ell+1)\hat{c}_0}}&=\exp\left(\sum_{k=1}^\infty k \hat{c}_{-k}\hat{c}_k\right).
\end{align}
\end{thm}

In our case, the symbol is
\begin{align}
f(\theta)=1-2F(\theta)&=\frac{3+2e^{2i\theta}+2e^{-i\theta}-4e^{i\theta}-2\cos(2\theta)}{\sqrt{(5-2\cos(2\theta))^2-4}}.
\end{align}
Since $\Re[1-2F(\theta)]>0$ for all $\theta$, the symbol is nonzero and has winding number zero, ensuring that the strong Szeg\H{o} limit theorem applies.
The Fourier coefficients $\hat{c}_{k}$ are given by:
\begin{align}
\hat{c}_k=\frac{1}{2\pi}\int_0^{2\pi}e^{-i\theta k} \log[1-2F(\theta)]\,d\theta.
\end{align}
We can next verify that $\hat{c}_0 = 0$ 
using symmetries of $\log[1-2F(\theta)]=\log|1-2F(\theta)|+i\operatorname{Arg}[1-2F(\theta)]$.
In particular, trigonometric manipulation shows that $\im[\log f(\theta)]=\im[\log f(-\theta)]$,  $\re[\log f(\theta)]=\re[\log f(\pi-\theta)]$, and $\re[\log f(\theta)]=\re[\log f(-\theta)]$, which together imply $\hat{c}_0=0$. 
The strong Szeg\H{o} limit theorem then gives the vison correlator in the $\ell\to\infty$ limit as the constant
\begin{align}\label{eqn:sqoct-vison-limitsum}
\lim_{\ell\to\infty}\left|\langle(-1)^{\eta_{E_\ell}}\rangle\right|&=\exp\left(\sum_{k=1}^\infty k \hat{c}_{-k}\hat{c}_k\right).
\end{align}

The Fourier coefficients $\hat{c}_k$ decay exponentially in $|k|$ since $\log[1-2F(\theta)]$ is analytic and $\Re[1-2F(\theta)]>0$. 
As a result, the sum in~\cref{eqn:sqoct-vison-limitsum} converges, and the limit is nonzero.
Evaluating $\hat{c}_k$ and summing over all $|k|\le 50$ gives
\begin{align}\label{eqn:sqoct-vison-limit}
\lim_{\ell\to\infty}\left|\langle(-1)^{\eta_{E_\ell}}\rangle\right|= 0.774596669\ldots,
\end{align}
which appears to be $\sqrt{60}/10 = 0.774596669\ldots$. 
We expect an exact evaluation may be possible, for example, using the methods of Ref.~\cite{basor2007asymptotics}, which evaluates a more challenging limit.
For the purpose of determining the phase, however, it suffices to note that the limit is a nonzero constant, as guaranteed by the convergence of the series in~\cref{eqn:sqoct-vison-limitsum}.

We note that calculating spin-spin correlations in the Ising model using Toeplitz determinants goes back to at least Ref.~\cite{montroll1963correlations}, and even an earlier letter in 1950 by Onsager (see for example the introduction in Ref.~\cite{bottcher1995the}).
In the vison case considered here, we were fortunate that $(\Ki^{-1})_E(\Ki')_E$ became exactly a Toeplitz matrix along this particular path.
More generally, including for other paths on the square-octagon lattice, $(\Ki^{-1})_E(\Ki')_E$ can take the form of a block Toeplitz matrix, to which one may attempt to apply the Szeg\H{o}--Widom generalization \cite{widom1974asymptotic,basor2007asymptotics}. 
However, in the block case, there is the potential for the limiting coefficient to be zero, in which case the Szeg\H{o}--Widom theorem need not give any information on the decay rate, though one can in some cases obtain corrections \cite{basor2019modified}. 
One instance of the precise calculations needed for the block Toeplitz determinants appears in Ref.~\cite{basor2007asymptotics}, which determined the classical
monomer-monomer correlations for the triangular lattice dimer model (see the determinant set-up of Ref.~\cite{fendley2002classical}).

We also note that the oscillating correlations as seen for square path 2 in \cref{fig:sqoct-vison} can be expressed as determinants of block Toeplitz matrices with alternating \emph{uneven} size blocks, for which evaluating the limit may be more difficult.

\subsection{Nature of the phase}
\label{subsec:phase-description}

\begin{figure}[htb]
\begin{tikzpicture}
\node at (0,0) {\includegraphics[height=3.5cm]{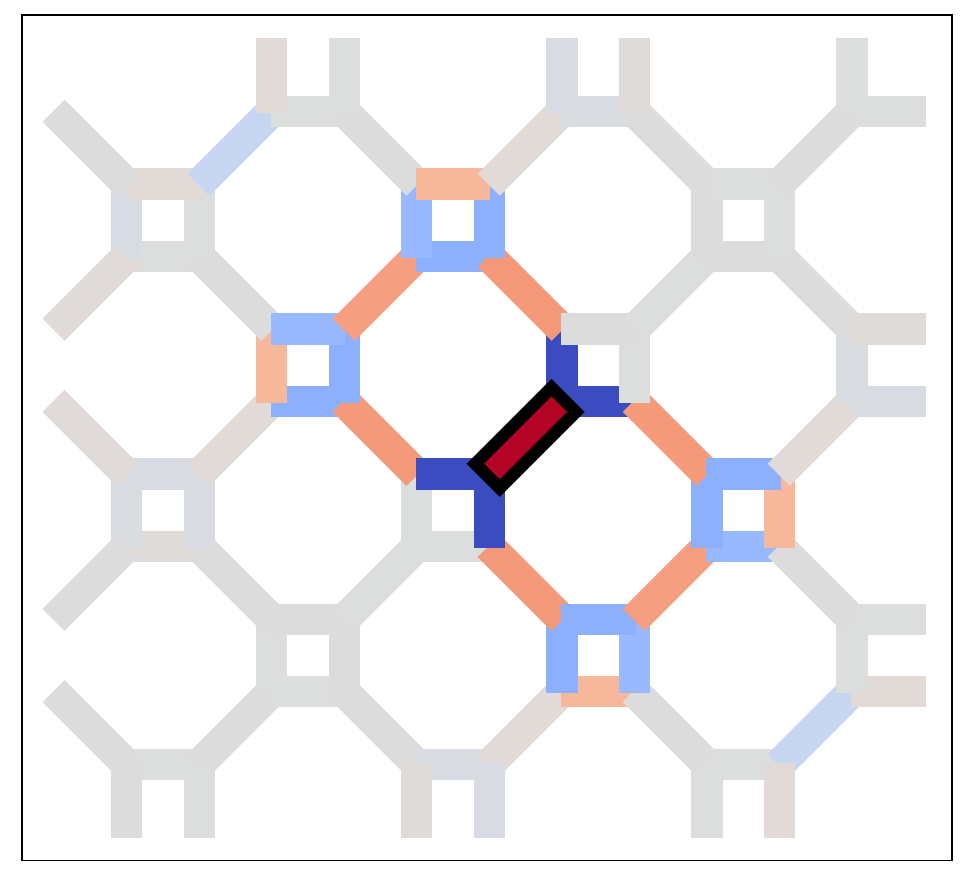}};
\node at (4.25,0) {\includegraphics[height=3.5cm]{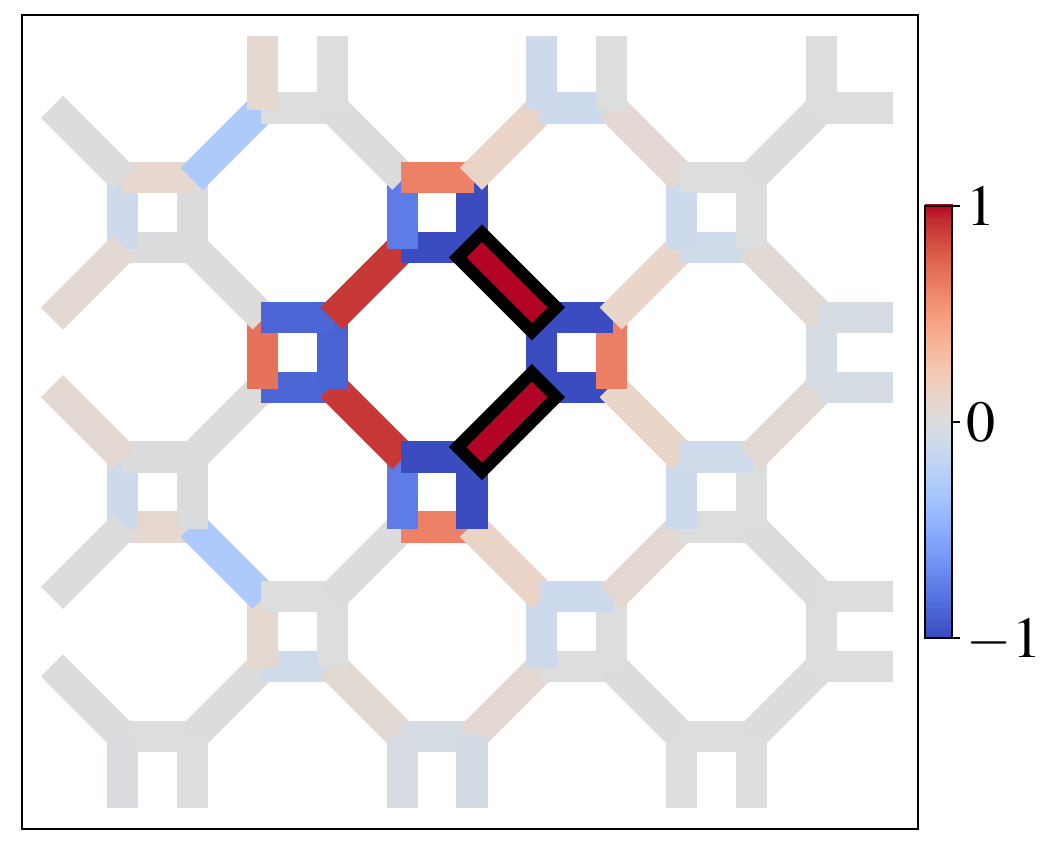}};
\def\lcoords{1.5};
\node at (-\lcoords,\lcoords) {(a)};
\node[xshift=4cm] at (-\lcoords,\lcoords) {(b)};
\end{tikzpicture}
\caption{Change in conditional probabilities, compared to the usual dimer probabilities, for the square-octagon lattice. 
The color scale corresponds to the scaled difference between the conditional probability and the non-conditional dimer density.
The value 1 (dark red) represents the largest possible increase, corresponding to a conditional probability of 1, while $-1$ (dark blue) represents the largest possible decrease, corresponding to a conditional probability of 0. The value 0, which is gray, represents no change in the conditional probability from the usual (non-conditional) probability of that dimer.
From the two plots, we see that a dimer placed on a diagonal edge is very likely to appear as part of a group of four dimers surrounding an octagon face. 
(a) Change in conditional probabilities given that a dimer occupies the outlined diagonal edge. 
(b) Change in conditional probabilities given that the two outlined dimers appear on those diagonal edges.  
This plot shows that the dimers tend to come in a single octagon face, rather than e.g. a 50\% chance of figure 8 shape which would be consistent with (a).
The plots are taken in a central region of a radius $R=50$ region.}\label{fig:cond-prob}
\end{figure}

We now interpret the phase of the square-octagon lattice dimer model. 
Although the dimer-dimer correlator decays exponentially in all directions we considered, the constant vison correlator along certain directions implies that the system is in an ordered phase, not a quantum spin liquid. 
To understand the nature of the order, we examine the plot of a randomly chosen dimer covering of the square-octagon lattice shown in~\cref{fig:sor-random-config}. 
We observe that the system does not exhibit one of the typical dimer-ordered phases, such as a staggered or columnar phase. 
However, this figure reveals key features of the ordering: most of the square plaquettes are flippable, and the diagonal (black) dimers appear only as part of flippable octagon(s).
To quantify this latter feature, we compute the conditional probabilities of finding a dimer on various edges, given that there is a dimer on a specific diagonal edge.
These conditional probabilities are plotted in \cref{fig:cond-prob}.
From this figure, one can see that when a diagonal dimer appears, it tends to come in as part of a group of four, all surrounding an octagon face. 
The exact asymptotic probability can be calculated using the infinite-limit values of $\Ki^{-1}$ from~\cref{eqn:k-inv}, combined with the dimer probability formula~\cref{eqn:kenyon}.
Given a diagonal edge $e$, let $P_1$ and $P_2$ be the two adjacent octagon plaquettes.
Then we can calculate
\begin{align}\label{eqn:pflip}
\P[P_1\text{ or }P_2\text{ is flippable}|\text{diag. dimer on}\; e] 
&=0.9892. 
\end{align}
Thus, diagonal dimers are accompanied by a flippable octagon face in nearly $99\%$ of the cases.

This gives us the following intuitive picture for the phase of the dimer model: there is a background of flippable square plaquettes, over which there is a finite density of flippable octagonal defects, shown in black in~\cref{fig:sor-random-config}.
Such an order is similar to the short range entangled states seen for the Ruby family of Ref.~\cite[\S5]{balasubramanian2022classical}, where different plaquettes are disconnected to from each other.
Although in our case on the square-octagon lattice, the square plaquettes are not disconnected, the large number of flippable square plaquettes in \cref{fig:sor-random-config} suggests that it has the same order.
Using this picture, we can give an intuitive estimate for the vison correlator along the octagon path and the diagonal octagon path shown in \cref{fig:sqoct-paths}, both of which intersect only diagonal edges.
Let us label the octagonal plaquettes along the path as $P_0, P_1, P_2, \ldots, P_\ell$.
Recall that the vison correlator along a path is given by $|p_{\text{odd}} - p_{\text{even}}|$, where $p_{\text{odd}}$ and $p_{\text{even}}$ are the probabilities that an odd and even number of dimers appear along the path, respectively.
Assuming that any diagonal dimer appears only as part of flippable octagon(s), the parity of the number of dimers along the interior octagons $P_1, P_2, \ldots, P_{\ell-1}$ will always be even.
Let $p_{\text{df}}$ denote the probability that an octagonal plaquette is flippable and has dimers along its diagonal edges (we refer to such plaquettes as ``diagonal-flippable").
The only way to have an odd number of dimers along the vison path is if $P_0$ is diagonal-flippable and $P_\ell$ is not, or vice versa.
Thus, 
\begin{equation}\label{eqn:podd}
    p_{\text{odd}} \approx 2 p_{\text{df}} (1-p_{\text{df}}),
\end{equation}
where we use that the edge configurations in $P_0$ and $P_\ell$ are nearly independent for large $\ell$, since the dimer-dimer correlator decays exponentially.
Thus, the vison correlator is $|1-2p_{\text{odd}}| \approx |1-4p_{\text{df}} ( 1- p_{\text{df}})|$.
Using Kasteleyn matrix methods, we calculate $p_{\text{df}} = 0.0590$, which gives a vison correlator of $0.7779$---in close agreement with the exact analytic result in~\cref{eqn:sqoct-vison-limit}.
Thus, we have shown that the vison correlators along the octagon and diagonal octagon paths are constant, under the assumption that diagonal dimers always appear as part of flippable octagon(s).

\begin{figure}[htbp]
\begin{tikzpicture}[scale=1.0]
\def\shft{2.8};
\def\xcoord{1.2};
\def\ycoord{2.4};
\node at (\xcoord,\ycoord) {(a)};
\node at (\xcoord + \shft,\ycoord) {(b)};
\node at (\xcoord + 2*\shft,\ycoord) {(c)};
\draw[gray,dashed] (0,0) -- (0.8,0.8) -- (1.6,0.8) -- (1.6,1.6) -- (0.8,1.6) -- (0,2.4);
\draw[gray,dashed] (2.4,2.4) -- (1.6,1.6);
\draw[gray,dashed] (2.4,0.0) -- (1.6,0.8);
\draw[black,thick] (0.8,1.6) -- (0.8,0.8);
\draw[black,thick] (1.6,0.8) -- (1.6,1.6);
\draw[fill=black] (0.8,0.8) circle (0.08);
\draw[fill=black] (1.6,1.6) circle (0.08);
\draw[fill=black] (0.8,1.6) circle (0.08);
\draw[fill=black] (1.6,0.8) circle (0.08);

\draw[gray,dashed] (2.8,0) -- (3.6,0.8) -- (4.4,0.8) -- (4.4,1.6) -- (3.6,1.6) -- (3.6,0.8);
\draw[gray,dashed] (5.2,2.4) -- (4.4,1.6);
\draw[gray,dashed] (5.2,0) -- (4.4,0.8);
\draw[thick] (2.8,0) -- (3.6,0.8);
\draw[thick] (3.6,1.6) -- (2.8,2.4);
\draw[black,thick] (4.4,0.8) -- (4.4,1.6);
\draw[fill=black] (3.6,0.8) circle (0.08);
\draw[fill=black] (4.4,1.6) circle (0.08);
\draw[fill=black] (3.6,1.6) circle (0.08);
\draw[fill=black] (4.4,0.8) circle (0.08);

\draw[gray,dashed] (5.6,0) -- (6.4,0.8) -- (7.2,0.8) -- (7.2,1.6) -- (6.4,1.6) -- (6.4,0.8);
\draw[thick] (8,2.4) -- (7.2,1.6);
\draw[thick] (6.4,1.6) -- (5.6,2.4);
\draw[thick] (7.2,0.8) -- (8,0);
\draw[thick] (5.6,0.0) -- (6.4,0.8);
\draw[fill=black] (6.4,0.8) circle (0.08);
\draw[fill=black] (7.2,1.6) circle (0.08);
\draw[fill=black] (6.4,1.6) circle (0.08);
\draw[fill=black] (7.2,0.8) circle (0.08);
\end{tikzpicture}
\caption{The three allowed configurations of dimers surrounding a square face (up to rotation and reflection).
Figures (a), (b), and (c) have zero, two, and four diagonal dimers, respectively, illustrating that the number of diagonal dimers around any square face is always even.
}
\label{fig:diagdimers-allowed}
\end{figure}
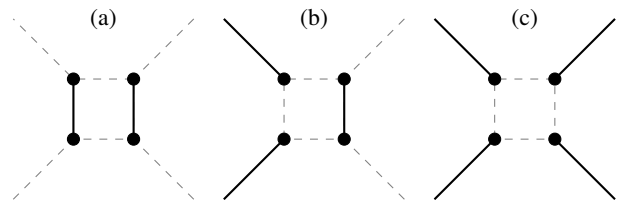

From~\cref{eqn:pflip},
we see that there is a small probability that the above assumption---that diagonal dimers always appear as part of flippable octagons---does not hold.
One might wonder whether such rare cases could lead to an exponential decay of the vison correlator, albeit with a large correlation length.
We argue below why this does not occur.
Essentially, when a diagonal dimer does not appear as part of a flippable octagon, it instead participates in a larger flippable loop.
To see this, we first note that only three arrangements of diagonal dimers are allowed in the square-octagon lattice by the dimer constraint: no diagonal dimers, dimers extending from two adjacent vertices, or dimers extending from all four vertices, as shown in~\cref{fig:diagdimers-allowed}.
If we consider the graph formed solely by the diagonal edges with dimers present---obtained by collapsing each incident square plaquette into a vertex---we see that every interior (non-boundary) vertex in the graph must have even degree. 
This requirement implies that the collection of diagonal dimers must either form closed loops or extended chains that start and end on the boundary. 

Now, looking at~\cref{fig:sor-random-config}, consider all places at the boundaries that present unique situations where an orthogonal dimer can be found, as we have established already that wherever a square connects to four diagonal edges, an even number of them must be dimers. However, the same arguments apply to the corner squares that connect to only three diagonal edges: if one or three of those edges is a dimer, then one vertex of the square will either be under- or overloaded. The bare horizontal edges that connect to two diagonal edges also have the same issue: one of these edges being occupied implies the other. This shows that it is impossible for a chain of diagonal dimers to begin or end even at the boundary, and therefore all diagonal dimers occur in loops.

In the fortress (see~\cref{fig:fortress}), meanwhile, there are spots where chains can terminate, those being the squares ``inset" into the boundary such that they lack a protruding corner. In that case, one of the diagonal edges along the very boundary can be occupied by a dimer, and the orthogonal edge connecting the two vertices of the square \textit{not} adjoining that dimer can be occupied without any problem (since the fourth vertex that would be underloaded is missing). Therefore, there is no need for a second diagonal dimer to emanate from the inset square.

It is likely that the probability of finding a flippable chain along a path/loop decays exponentially with the length of the path/loop.
In view of this, the earlier assumption that any diagonal dimer appears only as part of a flippable octagon is equivalent to assuming that the only diagonal flippable loops present are the minimal ones surrounding a single octagon face.
Now, considering any flippable path of diagonal edges, we can make the following argument: for a diagonal dimer (on edge $e$) in the bulk of the path to change the parity of the vison string, a diagonal-flippable loop containing $e$ must enclose one of the endpoints of the path.
Since the probability of a flippable path occurring likely decays at least exponentially with the length of the path, a diagonal dimer on edge $e$ alters the parity of the vison string only with a probability that itself decreases exponentially with the string length.
As a result, the vison correlator can still approach a constant in the infinite-length limit.
In fact, we expect that if we replace $p_\text{df}$ in \cref{eqn:podd} with the probability
\begin{align}
p_\text{o-loops}&\equiv\P[\text{odd \# of diagonal dimer loops around $P$}],
\end{align}
we would obtain an even closer approximation to the analytic vison correlator value in~\cref{eqn:sqoct-vison-limit}.
In summary, our intuitive picture of the dimer model captures certain aspects of its behavior, providing a useful understanding of its underlying structure.

\section{Phase separation example II: Square-octagon fortress}\label{sec:sqoct-fortress}

\subsection{Hamiltonian}
\label{sec:ham-fortress}

In this section, we define the square-octagon fortress graph and the corresponding spin Hamiltonian, whose low-energy limit maps to a quantum dimer model on the square-octagon fortress.
The steps are similar to those outlined in~\cref{sec:ham-sor}.

\begin{figure}[t]
  \centering
  \includegraphics[width=0.9\columnwidth]{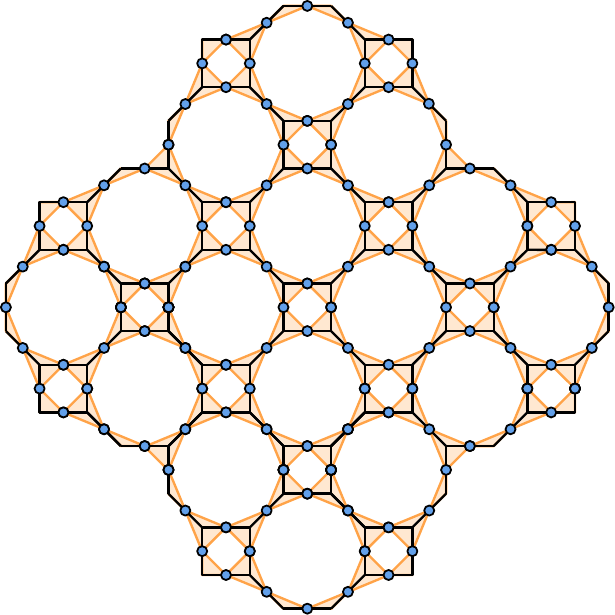}
  \caption{The figure displays a square-octagon fortress of radius $R=4$, defined by its black edges. 
  Blue dots indicate the positions of spin-1/2 degrees of freedom. 
  The orange edges represent the line graph of the fortress, forming a graph of corner-sharing polygons.}
  \label{fig:fortress}
\end{figure}

The square-octagon fortress is based on the square-octagon lattice described in~\cref{sec:square-octagon-rect}, but with boundaries rotated by $\pm 45^\circ$ relative to the square-boundary case.
\Cref{fig:fortress} shows an example of a square-octagon fortress with radius $R = 4$.
Here, the radius is defined as the number of edges crossed by a vertical path extending from the center to infinity.
This graph is interesting because it shows phase separation into three different regions---behavior that appears likely to persist not just at a single fine-tuned point in the phase diagram, but across an extended region.

We begin by explaining the spin and dimer Hamiltonians on the square-octagon fortress, simply called as fortress.
In the spin model, the spins are placed on the links of the fortress, as represented by the blue dots in~\cref{fig:fortress}.
We then construct the line graph of the fortress and interpret it as a graph of corner-sharing polygons, shaded in orange in the same figure.
This line graph helps us define the spin Hamiltonian.

The Hamiltonian of our model on the fortress is symbolically given by~\cref{eqn:hsor,eqn:hsor-0,eqn:hsor-1}, with the sums taken over polygons and plaquettes of the fortress, rather than the square-octagon lattice with rectangular boundaries.
In the limit $U \gg |V_4|, |V_8|,|J_4|,|J_8|$, we recover an RK quantum dimer model similar to~\cref{eqn:hsos1-dimer}, but now defined on the square-octagon fortress.
The ground state of the RK Hamiltonian is the RK wavefunction on the fortress, which is the uniform superposition of all dimer coverings.

Properties of uniformly sampled dimer coverings on fortress graphs are well studied in the mathematical literature, due to numerical results and their connection to certain weighted versions of the Aztec diamond dimer coverings~\cite{propp2003generalized, chhita2014coupling, difrancesco2014acoe, chhita2016domino}.
Below, we highlight some of these properties and discuss their implications for the phase of both the quantum dimer and spin models.

\subsection{The octic curve}
\label{sec:octic}

If one uniformly samples dimer coverings of the square-octagon fortress, then it is known, due to numerical results and a well-expected equivalence with a weighted Aztec diamond, that there are three distinct regions as shown in \cref{fig:so-random-config} 
\cite{propp2003generalized,chhita2014coupling,difrancesco2014acoe,chhita2016domino}. 
The regions near the corners are called the ``frozen" region similar to the frozen region of the dimer coverings of the Aztec diamond.
In the frozen region, a random dimer covering will have dimers in a staggered configuration with a probability approaching one in the limit of infinite system size. 
We can also see from \cref{fig:so-random-config}(b) that the average dimer occupation number is (approximately) 0 or 1 for different edges corresponding to the staggered pattern.
The innermost region in \cref{fig:so-random-config} is known as the ``gaseous" region in mathematical literature.
Finally, the region between the gaseous and the frozen regions is known as the ``liquid" region.

Using coordinates where the fortress boundaries are given by $|x| + |y| = 2$, the boundaries separating these three regions are given by the following ``octic (8th-order) curve'' \cite{propp2003generalized}:
\begin{multline}\label{eqn:octcurve}
400(x^8 + y^8) + 8025x^4 y^4 + 3400(x^2 y^6 + y^2 x^6)\\
+ 1000(x^6 + y^6) - 17250(x^4 y^2 + x^2 y^4)- 1431(x^4 + y^4)\\
 + 25812x^2 y^2 - 3402(x^2 + y^2) + 729 = 0.
\end{multline}
This curve is plotted in both parts of \cref{fig:so-random-config}, overlaid atop a random dimer configuration in (a) and atop a dimer density plot of the square-octagon fortress in (b).
There are two parts of this curve, and they split the fortress into the three distinct regions explained above. 

The octic curve formula \cref{eqn:octcurve} was first stated in \cite{propp2003generalized} for diabolo fortress tilings.
As explained further in~\cref{sec:correspondence}, there is also a two-periodic assignment of weights to the Aztec diamond, which makes it ``equivalent" to a weighted square-octagon fortress in the sense of sharing certain partition functions and behaviors, such as the Arctic curve and phases \cite{difrancesco2014acoe}, which can be obtained from them.
A general formula for the Arctic curve of a two-periodic Aztec diamond was given in Ref.~\cite[\S3.3]{difrancesco2014acoe} in terms of its face weights $a, b, c,$ and $d$, and specifiable entirely by a single parameter $\alpha = \left( \frac{4}{(\frac{a}{b} + \frac{b}{a})(\frac{c}{d} + \frac{d}{c})} \right)^2$.
For the unweighted fortress that we consider, the corresponding weighting is given by $a = b = c = 1$ and $d = 1/2$, so that $\alpha = 16/25$ and using that value in Ref.~\cite[Eq. (3.8)]{difrancesco2014acoe} gives~\cref{eqn:octcurve}.

\subsection{Dimer-dimer correlator across the phases}

\begin{figure}[htbp]
\includegraphics[width=.45\textwidth]{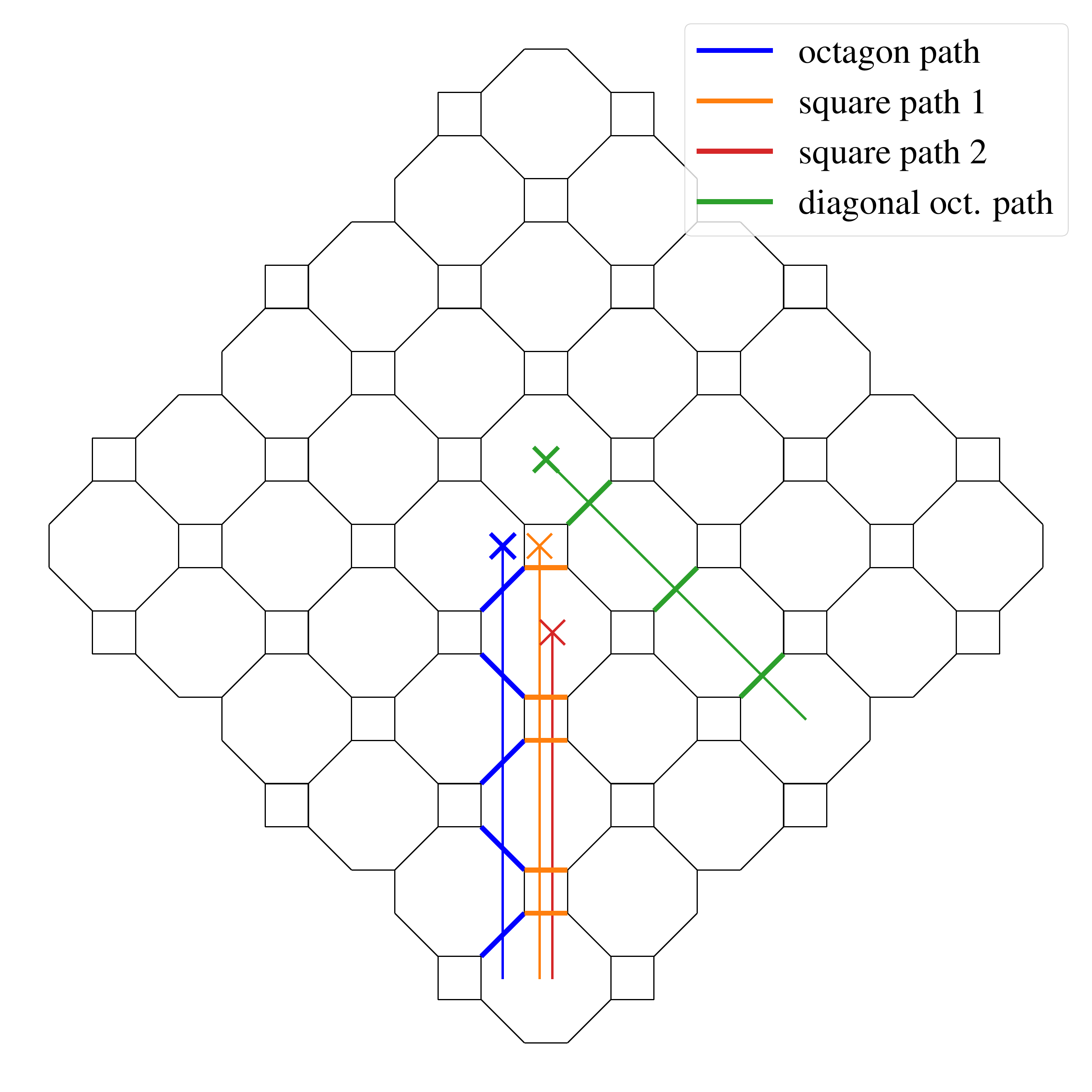}
\caption{Paths in the square-octagon fortress with radius $R=6$ along which we calculate the dimer-dimer and the vison correlators. 
}\label{fig:sqoct-fortress_paths}
\end{figure}

The dimer-dimer connected correlator shows distinct behavior in each region of the square-octagon fortress. 
Using the methods from~\cref{sec:kasteleyn}, we numerically calculate the dimer-dimer connected correlator and vison correlator along various paths  on a square-octagon fortress of radius $R = 100$, where the radius is defined as the number of edges along a horizontal or vertical path from the center to the boundary (including the boundary edge).
We evaluate these correlators along the four paths shown in~\cref{fig:sqoct-fortress_paths}, which are analogous to the paths considered in the square-octagon lattice with rectangular boundaries in~\cref{sec:sqoct-fortress}.

\begin{figure}[htbp]
\includegraphics[width=.45\textwidth]{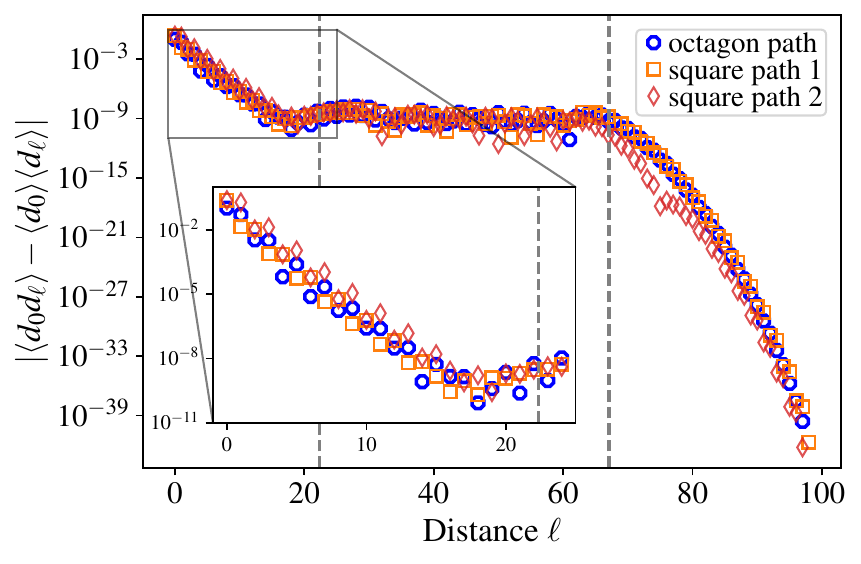}
\caption{Log plots of the dimer-dimer correlators along the octagon and the two square paths in the square-octagon fortress as a function of the edge distance $\ell$.
Here, recall $\ell$ is not the Euclidean or graph distance, but is the number of edges between the endpoints along the path.
These correlators were calculated using fraction-free LU decomposition.
The correlators show exponential decay in the center phase region ($\ell$ smaller than around 22 for this fortress), with negatives of the best fit slopes (inverse correlation lengths) between $1.0$ and $1.1$ as calculated for distances between $\ell=5$ to $16$.
This is in close agreement (within $\pm 0.05$) with the dimer correlator for the rectangular boundary square-octagon lattice shown in \cref{fig:dimerdimer}, as well as its corresponding analytic result $0.774597\ldots$.
The fortress for which the correlators are calculated has a radius $R=100$ and $40\,600$ vertices.
The vertical dashed gray lines show the limiting phase boundary locations according to \cref{eqn:octcurve}, which mark changes in the plotted correlators. 
}\label{fig:sqoctfortress-dimer}
\end{figure}

\begin{figure}[htbp]
\includegraphics[width=.38\textwidth]{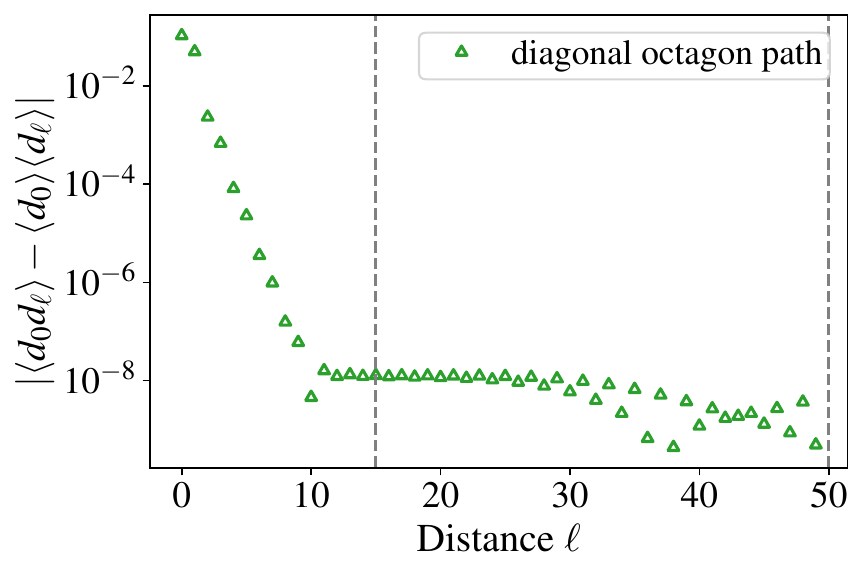}
\caption{Log plot of the dimer-dimer connected correlator along the diagonal octagon path as a function of the distance  $\ell$ between the endpoints.
Similar to \cref{fig:sqoctfortress-dimer}, the distance is not the Euclidean distance, rather it is the number of edges between the endpoints along the path.
We plot the correlator for the diagonal octagon path here separately from \cref{fig:sqoctfortress-dimer} 
since the path length and phase boundary locations differ (in particular, this path does not enter a frozen region). The negative of the best fit slope is $1.559$ as calculated from $\ell=3$ to 9, which is similar to that of the rectangular boundary square-octagon lattice shown in \cref{fig:dimerdimer}.}
\label{fig:sqoctfortress-dimer-green}
\end{figure}

In~\cref{fig:sqoctfortress-dimer}, we plot the dimer-dimer connected correlator along the octagon path, square path 1, and square path 2 as a function of the edge distance between the endpoints.
The correlator along the diagonal octagon path is shown separately in~\cref{fig:sqoctfortress-dimer-green}, since its path length and phase boundary locations differ. 
For each path, one endpoint is fixed within the innermost region of the fortress, as indicated by the crosses in~\cref{fig:sqoct-fortress_paths}. 
Using~\cref{eqn:octcurve} to identify the asymptotic phase boundaries, we observe that the dimer-dimer connected correlator exhibits distinct behaviors across different phases, consistent with expectations.
The correlators in the center phase region---covering distances $0$ to around $22$ along the octagon and square paths---exhibit similar behavior to that of the rectangular-boundary square-octagon lattice shown in~\cref{fig:dimerdimer}. 
In particular, the dimer-dimer connected correlator decays exponentially, with slopes (i.e., inverse correlation lengths) closely matching those observed in the infinite-size case.
In the intermediate critical region, the numerical plots are less conclusive; however, we expect the dimer-dimer connected correlator to exhibit power-law decay as is typical for critical regions.
In the frozen regions near the corners, the dimer-dimer connected correlator vanishes in the infinite-size limit. 

\subsection{Vison correlator across the phases}

\begin{figure*}[htbp]
\begin{tikzpicture}
\def\hfour{2.3in};
\def\hth{1.68in};
\def\rshift{5.9};
\node at (0,0) {\includegraphics[height=\hth]{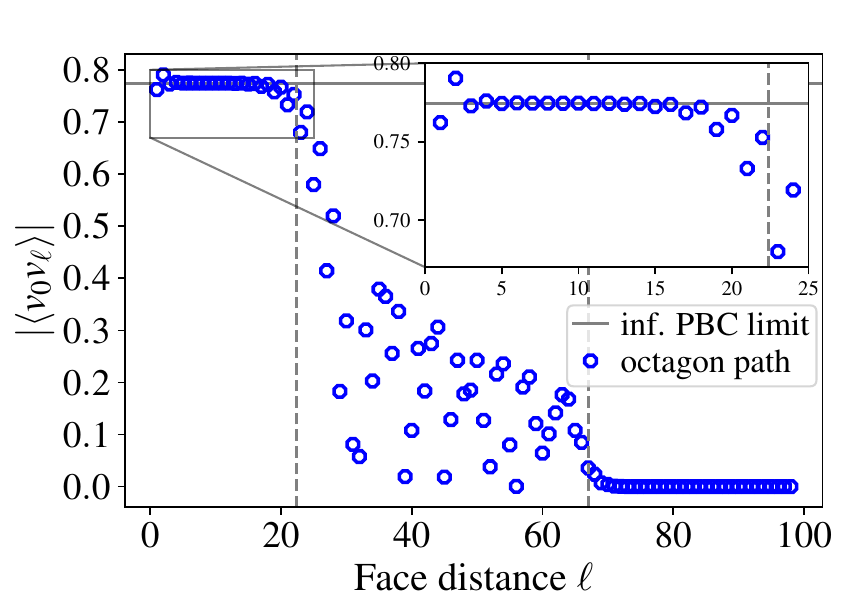}};
\node at (\rshift,0) {\includegraphics[height=\hth]{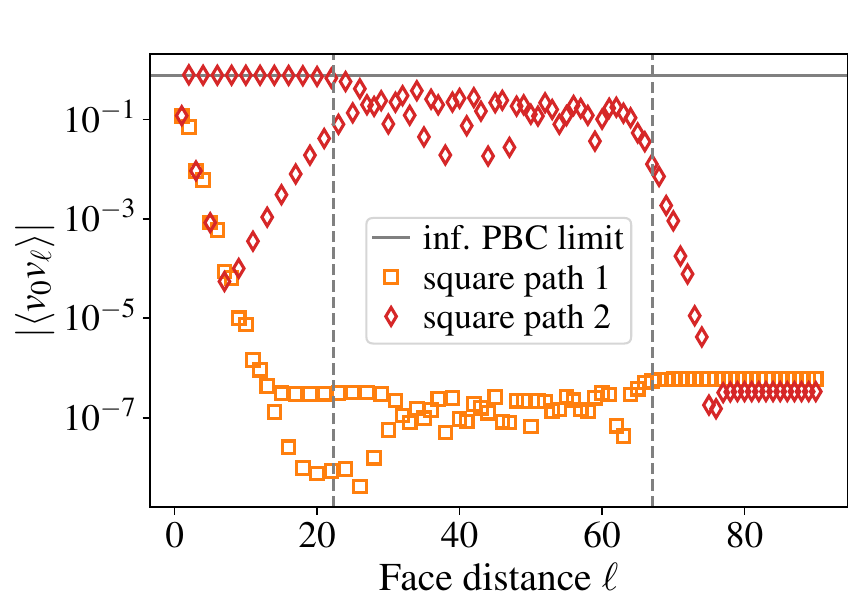}};
\node at (2*\rshift,0) {\includegraphics[height=\hth]{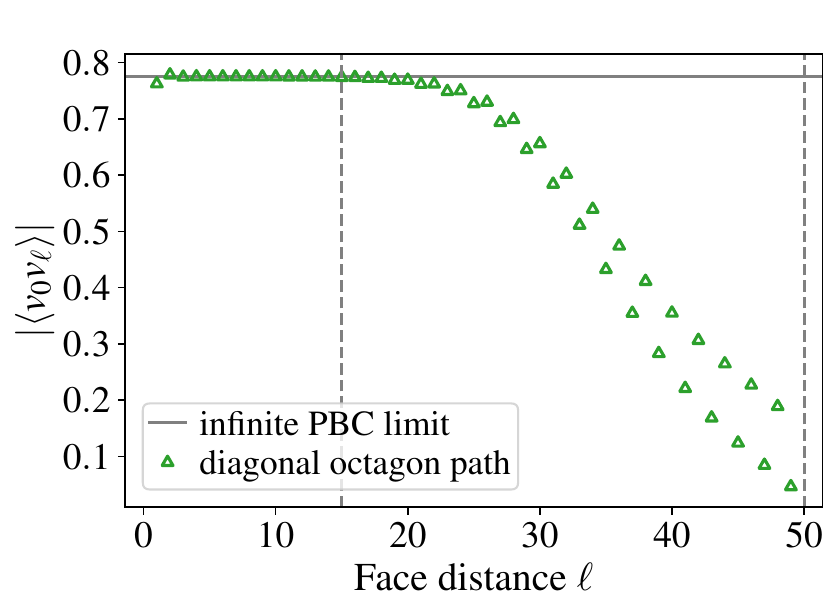}};
\def\lcoordx{2.5};
\def\lcoordy{2};
\begin{scope}[fontscale=-.5]
\node at (-\lcoordx,\lcoordy) {(a)};
\node[xshift={\rshift cm}] at (-\lcoordx,\lcoordy) {(b)};
\node[xshift={2*\rshift cm}] at (-\lcoordx,\lcoordy) {(c)};
\end{scope}
\end{tikzpicture}
\caption{
Vison correlators along the four considered paths in the square-octagon fortress with radius $R=100$, calculated using fraction-free LU (FFLU) decomposition. The vertical dashed gray lines show the limiting phase boundaries as before. The vison correlator behavior in the center phase region closely follows that of the rectangular boundary square-octagon lattice shown in \cref{fig:sqoct-vison}. In particular, the octagon and diagonal octagon paths (left-most and right-most plots respectively) show a constant vison correlator in the center phase region. The negative of the best fit slope (inverse correlation length) for the exponential decaying part of square path 1 (orange square marker) is 1.1 as calculated from $\ell=2$ to 12.
}
\label{fig:sqoctfortress-vison}
\end{figure*}

Now we describe the behavior of the vison correlator along the four paths shown in \cref{fig:sqoct-fortress_paths}.
Using \cref{eqn:vison-fast}, we calculate the vison correlators numerically and plot them in \cref{fig:sqoctfortress-vison}.
In the central region, the vison correlator along the octagon and diagonal octagon paths approaches a constant for long strings lying entirely within the region.
Moreover, this constant matches closely with the analytic value obtained for the square-octagon lattice with rectangular boundaries [\cref{eqn:sqoct-vison-limit0}].
In contrast, along square path 1 and square path 2, the vison correlator shows exponential decay with similar correlation lengths or displays oscillating behavior.

In summary, the behavior of both the dimer-dimer connected correlator and the vison correlator in the central region closely matches that on the square-octagon lattice with rectangular boundaries, both qualitatively and quantitatively.
This confirms that the central region is in the same phase as the square-octagon lattice with rectangular boundaries, as described in \cref{subsec:phase-description}.

\subsection{Nature of the phase}

As in \cref{subsec:phase-description}, it is useful to consider the diagonal dimers to identify the phases, since these paths should be closely related to \emph{height function} \cite{thurston1990conways,elkies1992alternating1} level lines, which exhibit distinct behavior in different phases (cf. Ref.~\cite[Fig. 2]{kenyon2006dimers}).

We recall the random dimer covering of the square-octagon fortress shown in \cref{fig:so-random-config}(a), where diagonal dimer are indicated in black. 
Their behavior is noticeably different across the phases.
In the central region, the configuration closely resembles that of the rectangular boundary case in \cref{fig:sor-random-config}---which is also expected to approximate the infinite-size limit---with diagonal dimers appearing almost exclusively in single-octagon loops.
In the intermediate phase, longer and more irregular chains of diagonal dimers begin to emerge.
In contrast, the frozen regions consist entirely of rigidly arranged diagonal dimer chains.
In terms of the height function, the center region corresponding to the ``smooth/gaseous" phase exhibits minimal fluctuations, with level lines forming only small, localized loops. 
In the critical or ``liquid" region, it is expected that the the height function fluctuations converge, in an appropriate scaling limit, to the Gaussian free field~\cite{kenyon2007limit,kenyon2008height,gorin2021lectures}, whose level lines are described by the random fractal curve $\mathrm{SLE}(4)$ \cite{schramm2009contour}.
In contrast, the frozen phase features a rigid, staggered arrangement, resulting in deterministic level lines for the height function.

\begin{figure}[htb]
\includegraphics[width=.48\textwidth]{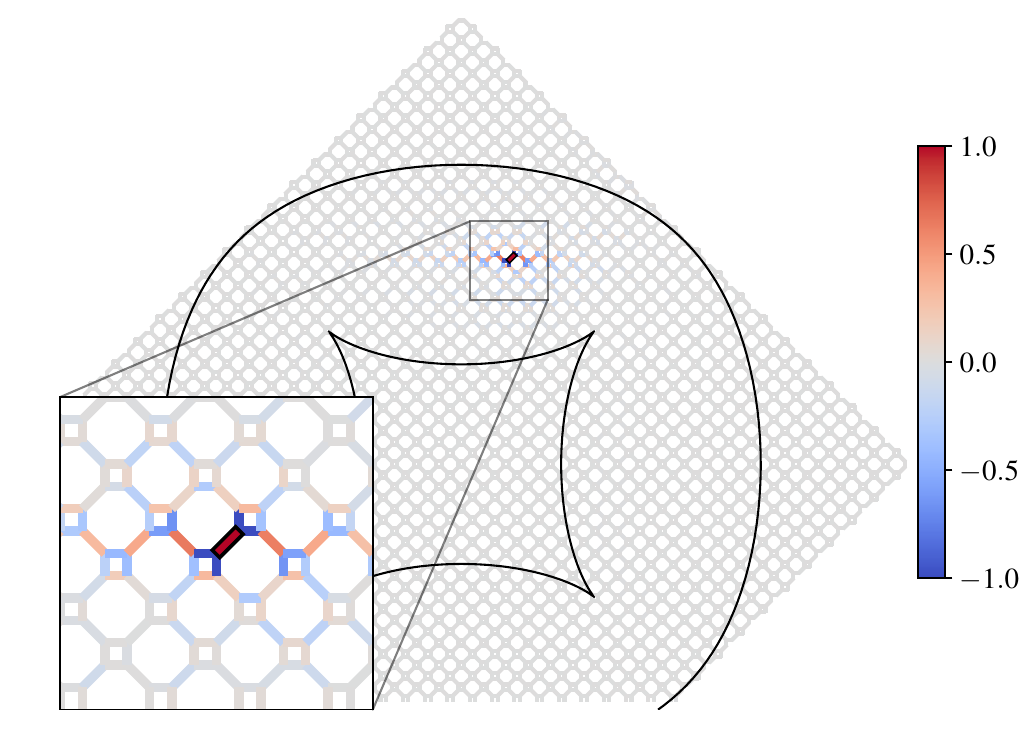} 
\caption{Conditional probability changes in the critical region of square octagon fortress $R=40$, using the same color scale as in \cref{fig:cond-prob}. As before, the value 0, which is gray, represents no change in the conditional probability from the usual (non-conditional) probability of that particular dimer, while red shows an increase for the conditional probability, and blue shows a decrease for the conditional probability.
There is a clear difference here compared to \cref{fig:cond-prob}. Instead of coming in a group of four diagonal edges, certain diagonal edges may now form longer horizontal or vertical (depending on location) ``strings''. 
}\label{fig:cond-prob-liquid}
\end{figure}

The contrast between loop and chain configurations is also evident in \cref{fig:cond-prob-liquid}, which displays the conditional dimer occupation probabilities in the critical region of the square-octagon fortress. 
Unlike the case of the square-octagon lattice with rectangular boundaries shown in \cref{fig:cond-prob}, where diagonal edges tend to form small loops, in the critical region, diagonal dimers more frequently appear as part of extended chains.

\section{Discussion}\label{sec:discusssion}

In this work, we presented two examples of quantum spin and quantum dimer models whose exact ground states have two or three spatially separated phases, challenging standard notions of the thermodynamic limit.
Crucially, this phase separation disappears when the same models are placed on domains with differently shaped boundaries.
The spin Hamiltonians we consider are defined on graphs of corner-sharing polygons and map to dimer models in their low-energy limit.
Our first example is a quantum dimer model defined on the Aztec diamond~\cite{elkies1992alternating1,elkies1992alternating2,jockusch1998random,kenyon1997local}.
At the Rokhsar--Kivelson~\cite{rokhsar1988superconductivity} point, we saw that in the ground state, dimers are frozen and have a staggered order in the corner regions outside the Arctic circle~\cite{jockusch1998random,kenyon1997local,propp2003generalized,kenyon2009lectures}, while dimers are in a critical phase with power-law decaying dimer-dimer correlators inside the Arctic circle.
The inner region resembles the critical dimer phase on the square lattice with rectangular boundaries originally studied by Kasteleyn~\cite{kasteleyn1961statistics}, and Temperley and Fisher~\cite{temperley1961dimer}.
Given the critical correlations inside the Arctic circle, it is likely that the observed phase separation might be fine-tuned to the RK point and would not survive small perturbations away from it.
This motivated the second example, a quantum dimer model on the square-octagon lattice.
On a square-shaped domain, Kasteleyn methods  show that the dimer-dimer correlators decay exponentially with distance along four different paths.
To probe if the system is in a quantum spin liquid phase, we numerically calculated the nonlocal vison correlator \textit{exactly} using Kasteleyn-type techniques described in \cref{subsec:vison}.
Along the octagon path, we were also able to calculate the infinite system size and infinite string length limit of the vison correlator analytically, and showed that it approaches a nonzero constant.
This implies the system is not in a $\mathbb{Z}_2$ quantum spin liquid phase, rather it is in an ordered phase.
We then commented on the order present in the system by considering the diagonal dimers and identified the system to be in a short-range entangled phase.
We then studied the same quantum dimer model and corresponding spin Hamiltonian on the square-octagon fortress.
Combining numerical calculations of the dimer-dimer and vison correlators with insights from the existing mathematical literature, we showed that at the RK point, the system separates into three distinct phases with different behaviors of the dimer-dimer correlations.
The regions near the corners are in an ordered phase with staggered dimers, while the innermost region is in the same short-range entangled phase as the square-octagon lattice on a square-shaped domain.
These regions are separated by a critical region and the boundaries of the three regions are given by the octic curve~\cite{propp2003generalized,difrancesco2014acoe}.
For the Kasteleyn-based calculations on the fortress, we encounter Kasteleyn matrices with condition numbers growing exponentially in the system size.
To get around this issue, we implement fraction-free LU (FFLU) \cite{lee1995fraction,nakos1997fraction,wilson1997determinant} decomposition to stably calculate the required inverses.
Since the innermost region on the fortress is gapped, we expect that the phase separation persists under small perturbations away from the RK point, likely providing an example of the quantum phase separation extending over a finite region of the phase diagram.

In \cref{subsec:vison}, we explained a method for calculating the vison correlator exactly and efficiently.
This approach applies to general planar graphs and can be used to check if a dimer model on a lattice is in a quantum spin liquid phase at the RK point.
Previously, vison correlators were primarily calculated using Monte Carlo methods, and the present method provides a significant advantage in both accuracy and speed.
This is particularly valuable for certain lattices with certain boundaries such as the square-octagon fortress, because it is known that the convergence of Monte Carlo simulation with natural ring exchanges is slow~\cite{bhakta2017sampling}.

It is often useful to generate a random dimer covering of a lattice.
In the mathematics literature, uniform random dimer coverings of the Aztec diamond and the square-octagon fortress are typically generated using the generalized domino shuffling algorithm \cite{elkies1992alternating2,propp2003generalized}, which is efficient and guaranteed to produce a dimer covering drawn uniformly at random. 
We point out that another method in the literature to produce random dimer coverings is via sampling from determinantal point processes, since Kenyon's formula [\cref{eqn:kenyon}] shows the inverse Kasteleyn matrix essentially gives the correlation kernel for the determinantal point processes formed by the dimers \cite{kenyon1997local,kenyon2009lectures}. 
There are efficient methods to generate random samples from such processes \cite{poulson2020high}, which have been applied to generate random samples of tilings of the Aztec diamond.
Since this method only requires knowing the inverse Kasteleyn matrix elements, it is easily adaptable to different lattices and boundary shapes. 
Implementing the determinantal point process sampling algorithm described in Ref.~\cite{poulson2020high}, in combination with the fraction-free LU methods from Refs.~\cite{lee1995fraction,nakos1997fraction} described in \cref{subsec:vison}, would provide a useful alternative to using Monte Carlo methods to generate random dimer coverings, particularly when the Monte Carlo convergence is slow.

We list below several interesting future directions:
\begin{enumerate}
    \item Study wavefunctions where dimer coverings appear with amplitudes determined by doubly periodic edge weights.
    Classical mixtures of such doubly periodic weighted dimer coverings have been studied extensively in the mathematical literature~\cite{berggren2023geometry,chhita2016domino,chhita2014coupling,borodin2023biased}.
    A natural question is whether one can construct an RK-like Hamiltonian whose ground state realizes these doubly periodic weights.
    \item Identify a lattice which shows phase separation for certain boundary shapes across a finite region of the phase diagram, with at least one region realizing a quantum spin liquid phase.
    \item Explicitly test the stability of phase separation of the quantum dimer model on the square-octagon fortress [\cref{eqn:hsos1-dimer}] away from the RK point using quantum Monte Carlo calculations, which has been used to study several other dimer models~\cite{syljuasen2006plaquette,yan2021widely,shannon2012quantum}. 
    \item The emergence of macroscopic regions with distinct behavior of correlators in dimer models is not unique to the Aztec diamond or the square-octagon fortress; similar phenomena are seen in lozenge tilings of a hexagon~\cite{gorin2021lectures,cohn2002shape} as well as with Aztec-like boundaries in other bipartite lattices. 
Investigating the phases of a quantum dimer model corresponding to these different lattices is another compelling direction.
\item From an experimental standpoint, Rydberg atom arrays offer a natural platform for realizing dimer models~\cite{semeghini2021probing,verresen2021,zeng2025quantum,shah2025quantum,glaetzle2014quantum,yan2022triangular}.
The Rydberg blockade mechanism can be used to impose the dimer constraint on the Kagome lattice~\cite{semeghini2021probing,verresen2021}, as well as the square and the triangular lattices~\cite{zeng2025quantum}.
Extending these approaches to implement the dimer model on the square-octagon lattice and to prepare the RK wavefunction is interesting.
\item Three-dimensional quantum dimer models are known to exhibit exotic phases, such as a $U(1)$ quantum spin liquid. 
Dimer models on the simple cubic and diamond lattices have been studied~\cite{huse2003coulomb,hermele2004pyrochlore,balasubramanian2024interplay}, and it is known that the RK wavefunction lies in a $U(1)$ quantum spin liquid phase.
An interesting question is whether phase separation can occur on a cubic lattice, perhaps with octahedral boundaries, where the regions near the corners may be in an ordered phase, while region inside a sphere lies in a $U(1)$ spin liquid phase.
This would be a three-dimensional analogue of the Aztec diamond and the Arctic circle.
\item The method for calculating the vison correlator described in \cref{subsec:vison}, can be extended to calculate many other important observables, such as the Wilson loops.
In particular, the ability to calculate Wilson loops exactly provides a valuable tool for probing confinement/deconfinement in dimer model.
Since these quantities decay exponentially with the area or the perimeter, it is crucial to calculate them to a high precision, something which is made possible by the technique in~\cref{subsec:vison}.
\end{enumerate}

\begin{acknowledgments}
The authors would like to thank Amol Aggarwal, Alexei Borodin, Sunil Chhita, and Gautam Nambiar for useful discussions.
This research was sponsored by the Army Research Office under Grant Number W911NF-23-1-0241 and Simons Foundation (VG), the National Science Foundation  QLCI grant OMA-2120757 (Jeet Shah and Jeremy Shuler), and Schwinger Foundation (LS).
\end{acknowledgments}

\FloatBarrier 

\bibliography{dimers}

\appendix

\section{Kasteleyn method on a torus}\label{app:torus}

As Kasteleyn showed in Ref.~\cite{kasteleyn1961statistics}, counting dimer configurations on graphs with periodic boundary conditions (PBC) (equivalently, on a torus) requires four distinct Kasteleyn matrices $\Kzz_n,\Kzo_n,\Koz_n$, and $\Koo_n$. 
In this section, we write the straightforward generalizations of the dimer-dimer and vison correlator formulas given in \cref{eqn:Kdimer,eqn:vison-fast} to the torus case, which are used in our analytical results.
In particular, we verify that in the infinite-size PBC limit, \cref{eqn:kenyon,eqn:vison-fast} hold with a single inverse function $\Ki^{-1}$ defined via contour integrals, such as in~\cref{eqn:k-inv}.

We begin by defining the four Kasteleyn matrices. 
In what follows, we continue to use the half-size Kasteleyn matrices $K_n$, defined using the bipartite graph structure, although for non-bipartite graphs, the full-size Kasteleyn matrices $\Kl_n$ must be used. 
The periodic graph $G_n$ can be represented as a planar graph $PG_n$ with additional edges connecting the boundary vertices to enforce periodic boundary conditions, thereby realizing the torus topology.
These added edges inherit the Kasteleyn orientation from the extended periodic planar graph.
The four Kasteleyn matrices $K^\ab_n$ are then defined on $PG_n$ with the usual Kasteleyn weighting, but differ in the weights assigned to the additional edges added to the boundary. 
For edges $(b,w)$ within $PG_n$, we have the usual definitions as in the planar case,
\begin{align}\label{eqn:Kab-planar}
\Kab_n(b,w)&=\begin{cases}
    1,&(b\to w)\text{ positive orientation}\\
    -1,& (b\to w)\text{ negative orientation}
\end{cases}.
\end{align}
However, for the additional edges $(b',w')$ that wrap around the graph to implement the torus topology, the Kasteleyn matrices differ~\cite{kasteleyn1961statistics}.
The first matrix, $\Kzz_n$, is the simplest to define: it retains the same edge weights as in the planar case, including for the wrap-around edges, using the prescription in \cref{eqn:Kab-planar}.
While this is the most natural extension, it assigns incorrect signs to certain classes of dimer configurations in which dimers wind around the boundaries \cite{kasteleyn1961statistics}.
To correct for this, the other three matrices $\Kzo_n$, $\Koz_n$, and $\Koo_n$ modify the signs on the wrap-around edges based on their winding direction.
Specifically, for edges $(b',w')$ that wrap around the torus horizontally or vertically, the sign of the edge weight is flipped according to: 
\begin{align}
\begin{aligned}
\Kzo_n(b',w')&=\begin{cases}
\Kzz_n(b',w'),&(b',w') \text{ horizontal}\\
-\Kzz_n(b',w'),&(b',w')\text{ vertical}
\end{cases},\\
\Koz_n(b',w')&=\begin{cases}
-\Kzz_n(b',w'),&(b',w')\text{ horizontal}\\
\Kzz_n(b',w'),&(b',w')\text{ vertical}
\end{cases},\\
\Koo_n(b',w')&=-\Kzz(b',w').
\end{aligned}
\end{align}
In other words, the Kasteleyn matrix $\Kab_n$ behaves as follows on the additional wrap-around edges: if $\alpha=1$, it flips the sign of horizontal wrap-around edges, and if $\beta=1$, it flips the sign of vertical wrap-around edges.
Although each individual Kasteleyn matrix $\Kab_n$ assigns incorrect signs to certain classes of dimer coverings involving dimers crossing the periodic boundary edges, Kasteleyn~\cite{kasteleyn1961statistics} showed that a linear combination of their Pfaffians yields the correct number of dimer coverings.
The number of dimer coverings (i.e., the partition function) on the torus is given by
\begin{align}
Z_n=\frac{1}{2}\left(-Z^\zz_n+Z^\zo_n+Z^\oz_n+Z^\oo_n\right),
\end{align}
where $Z^\ab_n:=\det\Kab_n$ for $\ab=\zz,\zo,\oz$, and $\oo$.
The relative signs in this expression account for the topological winding sectors and ensure that all dimer configurations are counted exactly once.

Using this counting method, all configurations are accounted for via four determinants instead of a single one.
For the dimer-dimer connected correlator $C_{\mu\nu}^d$ between two edges $\mu=(b_1,w_1)$ and $\nu=(b_2,w_2)$, we have \cite[\S8]{cohn2001variational},
\begin{widetext}
\begin{align}
C_{\mu\nu}^d=\langle d_\mu d_\nu\rangle-\langle d_\mu\rangle\langle  d_\nu\rangle&=\frac{\sum\limits_{\ab}\varepsilon_\ab\,\left|\left(\Kab_n\right)^{-1}(w_2,b_1)\left(\Kab_n\right)^{-1}(w_1,b_2)\right|\,\det\left[K_n^\ab\right]}{\sum\limits_{\ab}\varepsilon_\ab\det\left[\Kab_n\right]},
\end{align}
where $\varepsilon_\zz=-1$, $\varepsilon_\ab=1$ for the other three combinations of $\ab\in\{01,10,11\}$.
\end{widetext}
In the infinite-size limit $n\to\infty$, it is known (or can be verified) that the matrix elements $(\Kab)^{-1}(w,b)$ converge (at least along a subsequence, in the case of critical behavior) to a common limiting function $\Ki^{-1}(w,b)$, given by a contour integral representation~\cite[\S4]{kenyon2006dimers}, such as that in \cref{eqn:k-inv}.
This yields a simple expression for the dimer-dimer connected correlator in the infinite-volume limit (in the uniform edge weight/unweighted case, which includes all the graphs in the main text):
\begin{align}\label{eqn:dimer-torus}
|C_{\mu\nu}^d|&
=|\Ki^{-1}(w_2,b_1)\Ki^{-1}(w_1,b_2)|.
\end{align}

For the vison correlator, we have
\begin{align}
\langle(-1)^{\eta_E}\rangle &= \frac{ \sum_{\ab} \varepsilon_{\ab} \det\tilde K^\ab_n}{\sum_{\ab} \varepsilon_{\ab} \det K^\ab_n},
\end{align}
where $\tilde K^\ab_n=\Kab_n-\left(\Kab_n\right)'$, and the matrices $\tilde K^\ab_n$ and $\left(\Kab_n\right)'$ are defined as in~\cref{subsec:vison}. 
Using the block structure of $\left(\Kab_n\right)'$ and denoting by $(\cdot)_E$ the restriction to the $\ell \times \ell$ submatrix associated with the string $E$, we obtain 
\begin{widetext}
\begin{align}\label{eqn:vison-torus-expand}
\langle(-1)^{\eta_E}\rangle&=\frac{
\sum\limits_{\ab}\varepsilon_\ab\det\left[\Kab_n\right]\det\left[I_\ell-\left(\Kab_n\right)'_E\left((\Kab_n)^{-1}\right)_E\right]}{\sum\limits_{\ab}\varepsilon_\ab\det\left[\Kab_n\right]},
\end{align}
\end{widetext}
where the sum is over the four values of $\ab$, $\varepsilon_\zz=-1$, and $\varepsilon_\ab=+1$ for the other three $\ab$.
If the string $E$ does not wrap around the torus, as will be the case since we consider a finite-length string in the infinite-size limit, then all four matrices $\left(\Kab\right)'_E$ are identical.
In the limit $n\to\infty$, as before we know the matrix elements $\left(\Kab\right)^{-1}(w,b)$ all converge to the same value $\Ki^{-1}(w,b)$ given by contour integrals. 
Consequently, all determinants $\det\left[I_\ell-\left(\Kab_n\right)'_E\left((\Kab_n)^{-1}\right)_E\right]$ converge to the same value $\det\left[ I_\ell-\left(\Ki'\right)_E\left(\Ki^{-1}\right)_E \right]$, and \cref{eqn:vison-torus-expand} becomes
\begin{align}\label{eqn:vison-torus}
\langle(-1)^{\eta_E}\rangle&=\det\left[I_\ell-\left(\Ki'\right)_E\left(\Ki^{-1}\right)_E\right].
\end{align}

\section{Kasteleyn derivation for the infinite square-octagon lattice}\label{app:inf}

In this section we provide the details for deriving \cref{eqn:k-inv}, which gives the matrix elements of the inverse of the Kasteleyn matrix for an square-octagon lattice on an infinite torus.

\subsection{Full-size Kasteleyn matrices}\label{subsec:large}
For this derivation, we primarily work with the full-size Kasteleyn matrices $\Kl^\zz_n,\Kl^\zo_n,\Kl^\oz_n$, and $\Kl^\oo_n$, both for convenience and to better follow along with the method in Ref.~\cite{cohn2001variational}. 
These are the same type of matrices used when working with non-bipartite planar graphs.
Instead of starting with the half-size adjacency matrix $A$, whose rows are indexed by vertices in $B$ and columns by vertices in $W$, we consider the full-size $2N\times 2N$ adjacency matrix $\mathbb{A}$, with both rows and columns indexed by the full set of vertices $V$.
The entries of $\mathbb{A}$ are defined as $\mathbb{A}(v_i,v_j)=1$ if the vertices $v_i$ and $v_j$ are connected by an edge, and $0$ otherwise.
Similarly, we define a full-size $2N\times 2N$ signed adjacency matrix (i.e., the Kasteleyn matrix) $\Kl$ in the same way as before, using the Kasteleyn weighting and \cref{eqn:K-def}, applied to all vertex pairs $(v_i,v_j)$.
Kasteleyn~\cite{kasteleyn1963dimer,kasteleyn1967graph} showed that the number of dimer coverings is given by the \emph{Pfaffian} of the skew-symmetric matrix $\Kl$:
\begin{align}
\begin{aligned}
\#\{\text{dimer}&\text{ coverings}\}=|\Pf(\Kl)|\\
&= \frac{1}{2^NN!}\left|\sum_{\sigma\in S_{2N}}\operatorname{sgn}(\sigma)\prod_{i=1}^N\Kl\left(\sigma(2i-1),\sigma(2i)\right)\right|.
\end{aligned}
\end{align}
For any skew-symmetric $\Kl$, we have the identity $|\Pf(\Kl)|=\sqrt{\det \Kl}$, so the number of dimer coverings can equivalently be expressed in terms of the determinant.

In the case where the graph $G$ is bipartite, indexing the vertices such that all black vertices $B$ precede the white vertices $W$ gives the large Kasteleyn matrix $\Kl$ the block structure
\begin{align}\label{eqn:K-block}
\Kl=\begin{pmatrix}0&K\\-K^\dagger&0\end{pmatrix},
\end{align}
where $K$ denotes the half-size Kasteleyn matrix.
Throughout, we use $K$ to refer to this smaller half-size Kasteleyn matrix and reserve the notation $\Kl$ for the ``large-size'' Kasteleyn matrix. 
With this block structure, the Pfaffian and determinant are related by $\Pf(\Kl)^2=\det \Kl= |\det K|^2$, thereby recovering the determinant formula \cref{eqn:detk} for bipartite graphs.
Note that in this formulation, the matrix $K$ has rows indexed by black vertices $b_i$ and columns indexed by white vertices $w_i$, so the inverse matrix $K^{-1}$ has rows indexed by $w_i$  and columns by $b_i$.

\subsection{Square-octagon partition function}

\begin{figure}[htb]
\begin{tikzpicture}[scale=1.3]
\begin{scope}[decoration={
    markings,
    mark=at position 0.5 with {\arrow{Stealth}}}
    ] 
\foreach \col in {0,1,2,3,4}{
\foreach \row in {0,-1}{
\draw[postaction={decorate},yshift={\row cm}] (\col-1,0)--(\col,0);
}
}
\draw[postaction={decorate}](0,-1)--++(0,1) node[midway,left] {$b$};
\draw[postaction={decorate}](1,1)--++(0,-1) node[midway,left] {$b$};
\draw[postaction={decorate}](2,0)--++(0,1) node[midway,right] {$b$};
\draw[postaction={decorate}](1,-1)--(1,-2) node[midway,left] {$b$};
\draw[postaction={decorate}](2,-2)--(2,-1) node[midway,right] {$b$};
\draw[postaction={decorate}](3,0)--(3,-1) node[midway,right] {$b$};
\end{scope}
\begin{scope}[xshift=.5cm]
\foreach \b in {-1,1,3}{
\node[above] at (\b,0) {$b$};
\node[below] at (\b,-1) {$b$};
}
\foreach \a in {-1,1}{
\node[above] at (\a+1,0) {$a$};
\node[below] at (\a+1,-1) {$a$};
}
\end{scope}
\def\rad{.08}
\foreach \col in {0,1}
{
\draw[fill=black,xshift={2*\col cm}] (0,0) circle (\rad);
\draw[xshift={2*\col cm},yshift=-1cm,fill=white] (0,0) circle (\rad);
\draw[xshift={2*\col cm},fill=white] (1,0) circle (\rad);
\draw[fill=black, xshift={2*\col cm},yshift=-1cm] (1,0) circle (\rad);
}
\begin{scope}[yshift=.7mm]
\node[above] at (0,0) {$v_1$};
\node[above] at (3,0) {$v_7$};
\node[above] at (2,-1) {$v_8$};
\node[above] at (1,-1) {$v_4$};
\end{scope}
\begin{scope}[yshift=-.7mm]
\node[below] at (3,-1) {$v_3$};
\node[below] at (0,-1) {$v_5$};
\node[below] at (1,0) {$v_6$};
\node[below] at (2,0) {$v_2$};
\end{scope}
\node [left] at (-1,0) {$z^{-j}v_7$};
\node[left] at (-1,-1) {$z^{-j}v_3$};
\node[right] at (4,0) {$z^jv_1$};
\node[right] at (4,-1) {$z^jv_5$};
\node[above] at (1,1) {$z^kv_4$};
\node[above] at (2,1) {$z^kv_8$};
\node[below] at (1,-2) {$z^{-k}v_6$};
\node[below] at (2,-2) {$z^{-k}v_2$};
\end{tikzpicture}
\caption{Fundamental cell of the square-octagon lattice, drawn as a subgraph of a square lattice by ``flattening'' the octagons. Values of a vector in the $z^j$ eigenspace of $T_{(4,0)}$ and the $z^k$ eigenspace of $T_{(0,2)}$. The vector is determined by its values $v_1,\ldots,v_8$ in a fundamental domain.}\label{fig:sqoct-jk}
\end{figure}
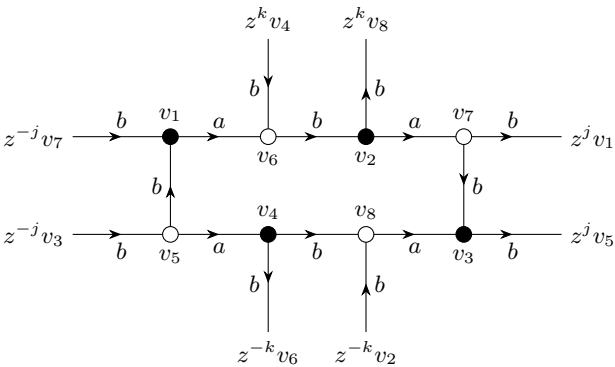

Because it can be useful for calculating quantities such as edge probabilities, we place edge weights $a$ on the octagon (non-square) edges and $b$ on the square edges, as depicted in \cref{fig:sqoct-jk}. 
Ultimately, we will set $a=b=1$ to recover the uniform (unweighted) case studied in the main text.
To clarify the horizontal and vertical translation of fundamental cells, we depict the octagons in a compressed form in \cref{fig:sqoct-jk}, allowing the lattice to be viewed as a subset of $\Z^2$.
Recall that $G_n$ denotes the PBC graph formed by applying $n$ horizontal and $n$ vertical translations of the fundamental cell.
Viewed as a subset of $\Z^2$, it is a $2n\times 4n$ (height $\times$ width) subset of $\Z^2$ with edges kept according to the square-octagon lattice pattern. 
The partition function $Z_n$ with edge weights $a$ and $b$ is defined as 
\begin{align*}
Z_n=\sum_{\substack{\{\text{dimer}\text{ coverings}\}}}a^{N_a}b^{N_b},
\end{align*}
where $N_a$ and $N_b$ denote the number of dimers on edges with weights $a$ and $b$ respectively in each dimer covering.

Recall we will primarily work with the large Kasteleyn matrices $\Kl^\zz_n,\Kl^\zo_n,\Kl^\oz_n$, and $\Kl^\oo_n$, for convenience and to better follow along with the method in Ref.~\cite{cohn2001variational}. 
These matrices are skew-symmetric, satisfying $\left(\Kl^\ab_n\right)^\dagger=-\Kl^\ab_n$, and have the block structure written in \cref{eqn:K-block}, with the $\Kl^\ab$ defined in \cref{app:torus}.
These full-size matrices each have dimensions $(8n^2)\times(8n^2)$, with rows and columns indexed by all vertices of the graph $G_n$.
Although complex phases are sometimes used in defining Kasteleyn matrices, we restrict ourselves to real phases (such as signs), since this is important for numerical calculations involving exact integer arithmetic.

Recall that we label each vertex in $G_n$ by coordinates $(x,y,t)$, where $(x,y)$ denotes the fundamental cell translation, and $t\in\{1,\ldots,8\}$ specifies the internal vertex label within the cell, corresponding to positions $v_1,\ldots,v_8$. 
With this indexing, the matrix $\Kl^\zz_n$ is given by
\begin{multline}
\langle x,y,t|\Kl^\zz_n|j,k,s\rangle=\\
\begin{cases}
\varepsilon_e\,a,&e=\left((x,y,t)\to (j,k,s)\right)\text{ has weight $a$}\\
\varepsilon_e\, b,&e=\left((x,y,t)\to (j,k,s)\right)\text{ has weight $b$}\\
0,&\left((x,y,t)\to (j,k,s)\right)\text{ is not an edge}
\end{cases},
\end{multline}
where the orientation $\varepsilon_e\in\{\pm1\}$ of an edge $e$ is defined to be $+1$ if it agrees with the Kasteleyn orientation (i.e., the arrow directions) indicated in \cref{fig:sqoct-fc,fig:sqoct-jk}, and $-1$ if it opposes it.
The other matrices $\Kl^\ab_n$ are defined analogously, except that the signs are flipped on certain edges that wrap around the torus, as described earlier in \cref{app:torus}.
Then we have \cite{kasteleyn1961statistics}
\begin{multline}
Z_n=\frac{1}{2}\bigg(-\sqrt{\det \Kl_n^{00}}+\sqrt{\det \Kl_n^{01}}\\+\sqrt{\det \Kl_n^{10}}+\sqrt{\det \Kl_n^{11}}\bigg).
\end{multline}
To compute $\det \Kl_n^{\ab}$ and $\left(\Kl_n^{\ab}\right)^{-1}$, we block-diagonalize the matrices $\Kl_n^{\ab}$ using a Fourier transform, following for example the method in Ref.~\cite[\S7]{cohn2001variational}. 
For brevity, we provide full details only for the $\Kl^\zz_n$ case, since as discussed in Refs.~\cite[\S7]{cohn2001variational} and \cite[\S4]{kenyon2006dimers}, it can be verified that the entries of $(\Kl^\ab_n)^{-1}$ converge, in the infinite-size limit, to the same contour integral expressions for all choices of $\ab$.
The matrix $\Kl^\zz_n$ is especially convenient to work with since it commutes with the translation operators $T_{(4,0)}$ and $T_{(0,2)}$, where $T_{(m,n)}$ denotes a shift in the $x$-direction by $m$ and in the $y$ direction by $n$. 
The translation operator $T_{(4,0)}$ is a shift by one fundamental domain in the $x$-direction, while $T_{(0,2)}$ is a shift by one fundamental domain in the $y$-direction.
The eigenspaces of these shift operators can be easily determined. 
Since $T_{(4,0)}^n=I$ and $T_{(0,2)}^n=I$, the eigenvalues of these operators are $n$th roots of unity, i.e., powers of $z:=e^{2\pi i/n}$.
For a fixed eigenvalue $z^j$ of $T_{(4,0)}$, the corresponding eigenspace has dimension $4(2n)=8n$, and an eigenvector is determined by its values on four adjacent columns, which each have length $2n$.
Similarly, for an eigenvalue $z^k$ of $T_{(0,2)}$, the eigenspace has dimension $8n$ and an eigenvector is determined by its values on two adjacent rows, which each have length $4n$.
A vector in the $z^j$-eigenspace of $T_{(4,0)}$ and the $z^k$-eigenspace of $T_{(0,2)}$ is fully determined by its values in a $2\times 4$ block of the lattice, which can be taken to be a fundamental domain.

Following Ref.~\cite{cohn2001variational}, define the vector $e_{j,k}^{(s)}$ by
\begin{align*}
\langle x,y,t|e_{j,k}^{(s)}\rangle =\frac{1}{n}\begin{cases}z^{jx+ky},&t=s\\0,&\text{otherwise}\end{cases}.
\end{align*}
The vector $e_{j,k}^{(s)}$ is a $z^j$-eigenvector of $T_{(4,0)}$ and a $z^k$-eigenvector of $T_{(0,2)}$. It selects a single vertex $s$ in a fundamental domain and translates it across all of $G_n$, with the appropriate eigenvalue prefactors. Since there are $n^2$ translates of a fundamental domain vertex $s$ [one for each $(x,y)$], the vector $e_{j,k}^{(s)}$ is properly normalized (in $\ell^2$ norm).  
Let $S$ be the $8n^2\times 8n^2$ matrix whose columns are the $e_{jk}^{(s)}$ for $s\in\{1,\ldots,8\}$ and $(j,k)\in\{0,\ldots,n-1\}^2$, ordered so that vectors with the same $(j,k)$ appear consecutively,
\[
\left(e_{jk}^{(1)},\ldots,e_{jk}^{(8)}\right)_{jk}.
\]
One can verify that $S$ is unitary, and that it block-diagonalizes $\Kl^\zz_n$ as follows:
\begin{align}\label{eqn:sks}
S^{-1} \Kl^\zz_n S&=\begin{pmatrix}
    B_{0,0} &&&\\
    & B_{0,1} &&\\
    &&\ddots &\\
    &&&B_{n-1,n-1}
\end{pmatrix},
\end{align}
where each block
\begin{align}
B_{j,k}&=\begin{pmatrix}0&B_{jk,1}\\-B_{jk,1}^\dagger&0\end{pmatrix}
\end{align}
is an $8\times 8$ matrix, indexed by the vertex labels $v_1,\ldots,v_8$ as shown in \cref{fig:sqoct-jk}, and describes the action of $\Kl^\zz_n$ on the intersection of the $z^j$ eigenspace of $T_{(4,0)}$ and the $z^k$ eigenspace of $T_{(0,2)}$.

Using \cref{fig:sqoct-jk}, we see that
\begin{align}\label{eqn:bjk}
B_{jk,1}&=\begin{pmatrix}
-b&a&-bz^{-j}&0\\
0&-b&a&bz^k\\
bz^j&0&-b&-a\\
-a&bz^{-k}&0&b
\end{pmatrix}.
\end{align}
Due to the block-diagonal structure of $S^{-1}\Kl^\zz_nS$, we obtain the determinant of $\Kl^\zz_n$ as the product:
\begin{align*}
\det \Kl^\zz_n&=\prod_{j,k=0}^{n-1}|\det B_{jk,1}|^2\\
&=\prod_{j,k=0}^{n-1}\left|-a^4-4b^4+a^2b^2(z^j+z^{-j}+z^k+z^{-k})\right|^2,\numberthis\label{eqn:det-prod}
\end{align*}
where $z=e^{2\pi i/n}$. 
Recall that the full partition function $Z_n$ involves a linear combination of four such determinants, as in \cref{eqn:Z-torus}.
The formulas for the other three Kasteleyn-type matrices are similar, though they may commute with a composition of a translation and negation of certain entries, rather than just the translations. In the $n\to\infty$ limit all quantities we are interested in will be the same for any of the four terms \cite{cohn2001variational,kenyon2006dimers}.

Taking a logarithm or derivative of $\sqrt{\det \Kl^\zz_n}$ converts the product over $j,k$ in \cref{eqn:det-prod} into a sum over $j,k$.
In the limit $n\to\infty$, the discrete sums over $j$ and $k$ become Riemann sums, which approximate double integrals over the torus.
Specifically, terms involving $z^j=e^{2\pi ij/n}$ are replaced by continuous variables $e^{2\pi i t}$ for $t\in[0,1)$, leading to contour integrals or integrals over the torus $[0,1)^2$ in the thermodynamic limit. 
For example, in the uniform case $a=b=1$, we have
\begin{align}\label{eqn:Z}
Z_n^\ab=\prod_{z^n=(-1)^\alpha}\prod_{w^n=(-1)^\beta}(-5+z+z^{-1}+w+w^{-1}),
\end{align}
where the products are taken over $n$th roots of unity $z$ and $w$ satisfying the appropriate twisted boundary conditions determined by $\ab$. 
Using a Riemann sum approximation, this product becomes a double integral in the limit $n \to \infty$, yielding the limiting free energy per unit cell:
\begin{multline}\label{eqn:free-energy}
\lim_{n\to\infty}\frac{1}{n^2}\log Z_n=\\\frac{1}{(2\pi i)^2}\oint_{S^1}\frac{dz}{z}\oint_{S^1}\frac{dw}{w}\,\log|-5+z+z^{-1}+w+w^{-1}|.
\end{multline}
The function $P(z,w)=-5+z+z^{-1}+w+w^{-1}$ is known as the \emph{characteristic polynomial} of the graph \cite{kenyon2006dimers}. 
It can be easily calculated from the fundamental cell and determines many key properties of the model, including whether the dimer-dimer correlators have exponential decay or power-law decay \cite{kenyon2006dimers}.

\subsection{Dimer occupation probabilities}\label{appsub:dimerprob}

In the limit $n\to\infty$, the above calculations give the expected dimer occupation number on the diagonal edges (edges with weight $a$ in \cref{fig:sqoct-jk}) as: 
\begin{align}
\E[e_a]&=\lim_{n\to\infty}\frac{a}{4n^2Z_n}\frac{\partial Z_n}{\partial a}\nonumber\\
=\frac{1}{4\pi^2}&\int_0^{2\pi}\int_0^{2\pi}\frac{-a^4+a^2b^2(\cos t+\cos s)}{-a^4-4b^4+2a^2b^2(\cos t+\cos s)}\,dt\,ds.
\end{align}
For uniform edge weights $a=b=1$, this simplifies to:
\begin{align}\label{eqn:edge-prob-oct}
\E[e_a]&=\frac{1}{4\pi^2}\int_0^{2\pi}\int_0^{2\pi}\frac{-1+\cos t+\cos s}{-5+2(\cos t+\cos s)}\,dt\,ds\nonumber\\
&=0.118925\text{ (to six decimal places).}
\end{align}
The expected dimer occupation number for the square edges (edges with weight $b$ in \cref{fig:sqoct-jk}) is then
\begin{align}\label{eqn:edge-prob-square}
\E[e_b]&=\frac{1-\E[e_a]}{2}=0.440537\text{ (to six decimal places).}
\end{align}

\subsection{Inverse Kasteleyn matrix}\label{subsec:Kinv-details}
To determine the entries of the inverse Kasteleyn matrix, we return to the block-diagonal form of $S^{-1}\Kl^\zz_nS$ as described in \cref{eqn:sks}. 
Taking the inverse yields 
\begin{align}
\left(\Kl^\zz_n\right)^{-1}&=S\begin{pmatrix}
    B_{0,0}^{-1} &&&\\
    & B_{0,1}^{-1} &&\\
    &&\ddots &\\
    &&&B_{n-1,n-1}^{-1}
\end{pmatrix}
S^{-1},
\end{align}
where each $B_{jk}^{-1}$ has the block form
\begin{align}\label{eqn:Binverse}
B_{jk}^{-1}&=\begin{pmatrix}0&\left(-B_{jk,1}^\dagger\right)^{-1}\\B_{jk,1}^{-1}&0\end{pmatrix},
\end{align}
where $B_{jk,1}$ is defined in~\cref{eqn:bjk}.
The inverse of $B_{jk,1}$ can be explicitly calculated as:
\begin{widetext}
\begin{multline}
B_{jk,1}^{-1}=\frac{1}{-a^4-4b^4+a^2b^2(z^j+z^{-j}+z^k+z^{-k})}\\
\times\left(\begin{array}{cccc} 2 b^3-a^2 b z^{-k} & a b^2-a b^2 z^{-j-k} & a^2 b-2 b^3 z^{-j} & a^3-a b^2 z^{-j}-a b^2 z^k \\ -a^3+a b^2 z^j+a b^2 z^k & 2 b^3-a^2 b z^{-j} & a b^2-a b^2 z^{k-j} & a^2 b-2 b^3 z^k \\ 2 b^3 z^j-a^2 b & -a^3+a b^2 z^j+a b^2 z^{-k} & 2 b^3-a^2 b z^k & a b^2-a b^2 z^{j+k} \\ a b^2-a b^2 z^{j-k} & a^2 b-2 b^3 z^{-k} & a^3-a b^2 z^{-j}-a b^2 z^{-k} & a^2 b z^j-2 b^3 \\\end{array}\right).
\end{multline}
Noting that
$\langle w,z,s|S|j,k,t\rangle=e^{2\pi i(jw+kz)}\delta_{t,s}/n$, we obtain the inverse matrix entries as
\begin{align}
\langle w,z,s|\left(\Kl^\zz_n\right)^{-1}|x,y,t\rangle
= \sum_{j,k=0}^{n-1}\frac{1}{n^2}e^{2\pi i(jw+jz)/n}e^{-2\pi i(jx+ky)/n}\langle t|B_{jk}^{-1}|s\rangle.
\end{align}
\end{widetext}
Similar formulas apply to the other matrices $\Kl^\ab$ for $\ab = \zo, \oz, \zz$. 
In the limit $n\to\infty$, the entries of all $\Kl^\ab_n$ converge to the same contour integral expressions \cite[\S4]{kenyon2006dimers} via a Riemann sum approximation.
With $2\pi ij/n$ and $2\pi ik/n$ becoming the continuous variables $k_x$ and $k_y$ respectively, this yields \cref{eqn:k-inv}. 
The expressions in \cref{eqn:numerator} are just the entries of $B_{jk,1}^{-1}$ in the special case $a=b=1$ and without the prefactor $1/P(z,w)$. 
Note that for \cref{eqn:numerator}, we determine the row and column indices of the table and of $B_{jk,1}^{-1}$ by looking at its placement in the block structure of the larger skew-symmetric matrix $B_{jk}^{-1}$ in \cref{eqn:Binverse}, which has rows and columns indexed by $1,2,\ldots,8$.

To simplify the double integrals appearing in \cref{eqn:k-inv}, we consider the following transformation for generic $j,k\in\Z$:
\begin{widetext}
\begin{align*}
\frac{1}{4\pi^2}\int_0^{2\pi}\int_0^{2\pi}\frac{e^{ij\theta}e^{ik\phi}}{-5+2\cos\theta+2\cos\phi}\,d\theta\,d\phi&=\frac{1}{(2\pi i)^2}\oint_{S^1}\oint_{S^1}\frac{z^jw^k}{-5+z+z^{-1}+w+w^{-1}}\,\frac{dz}{z}\,\frac{dw}{w}\\
&=\frac{1}{(2\pi i)^2}\oint_{S^1}\frac{z^j\,dz}{z}\oint_{S^1}\frac{w^k\,dw}{\left[w-\alpha(z)\right]\left[w-\beta(z)\right]},\numberthis\label{eqn:contour-int}
\end{align*}
where the functions $\alpha(z)$ and $\beta(z)$ are given by:
\begin{align}
\alpha(z),\beta(z)&=\frac{1}{2}\left(5-z^{-1}- z\mp\sqrt{(-5+z^{-1}+z)^2-4}\right).
\end{align}
For $|z|=1$, the functions $\alpha(z)$ and $\beta(z)$ are real-valued, and one can verify that $|\alpha(z)|<1$ and $|\beta(z)|>1$. 
Therefore, in the contour integral \cref{eqn:contour-int}, if $k\ge0$, the only pole inside the unit circle is at  $w=\alpha(z)$, giving the single-residue contribution:
\begin{align*}
\frac{1}{4\pi^2}\int_0^{2\pi}\int_0^{2\pi}\frac{e^{ij\theta}e^{i|k|\phi}}{-5+2\cos\theta+2\cos\phi}\,d\theta\,d\phi
&=\frac{1}{2\pi i}\oint_{S^1}\frac{z^j\alpha(z)^{|k|}\,dz}{z\left[\alpha(z)-\beta(z)\right]}\\
&=-\frac{1}{2\pi}\int_0^{2\pi}\frac{e^{ij\theta}\left[\frac{1}{2}\left(5-2\cos\theta-\sqrt{(5-2\cos\theta)^2-4}\right)\right]^{|k|}}{\sqrt{(5-2\cos\theta)^2-4}}\,d\theta.\numberthis\label{eqn:single-int}
\end{align*}
\end{widetext}
If $k<0$ in the integral on the left hand side of \cref{eqn:contour-int}, we can first substitute $\phi\mapsto-\phi$ in the starting integral of \cref{eqn:single-int} to take $k>0$.
Note this agrees intuitively, since in an infinite-size limit or on a torus it is symmetric to shift the fundamental cell by $k$ or $-k$.

\section{Numerical details}\label{sec:numdetails}

\begin{figure}[htb]
\includegraphics[width=.33\textwidth]{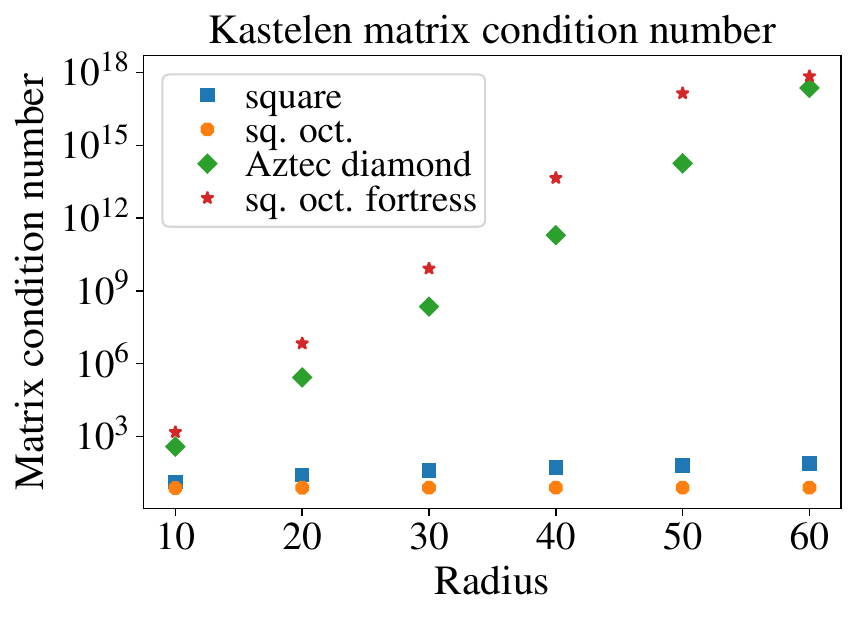}
\caption{Condition number of the Kasteleyn matrices for four models as a function of the system size. While the square boundary condition Kasteleyn matrices have a very small condition number (of order $10^1$ or $10^2$ in the plot), the Aztec diamond and fortress boundary conditions lead to exponentially large condition numbers, indicating that the standard floating point LU or determinant calculations can be entirely inaccurate for these matrices.
}\label{fig:cond-number}
\end{figure}

In this section, we briefly discuss numerical details specific to the Aztec diamond and the square-octagon fortress.
In contrast to the square or square-octagon lattices with square boundary conditions, it turns out that the Kasteleyn matrices for the Aztec diamond and square-octagon fortress have large condition numbers $\kappa:=\|K\|\|K^{-1}\|$, which may exceed $10^{16}$ even for small system sizes (see \cref{fig:cond-number}). 
This implies that standard inversion techniques and LU decomposition are numerically unreliable: for every power of 10 in the condition number, one may lose roughly that many digits of precision when solving linear systems~\cite{burden2015numerical}.
In practice, attempting to use the standard floating point LU decomposition can give vison correlator values greater than one which is impossible and indicates numerical inaccuracies.
Similarly, using floating point LU to calculate dimer densities for large fortresses can give clear inaccuracies and non-symmetric values despite the horizontal and vertical reflection symmetries of the fortress.
Numerical instabilities involving $K^{-1}$ for the Aztec diamond were also observed in Ref.~\cite{poulson2020high}.

To circumvent these issues, we implement a fraction-free LU decomposition as described in \cref{subsec:vison}, based on the sparse version of the Bareiss algorithm introduced in Ref.~\cite{lee1995fraction} (see also Ref.~\cite{nakos1997fraction}).
Although our implementation is significantly slower than standard floating point LU methods, it uses exact integer arithmetic---handling integers of magnitude $>10^{1700}$ for the plots in \cref{sec:sqoct-fortress}---and produces the exact rational values of the submatrix $(K^{-1})_E$.

\section{Correspondence between lattices}
\label{sec:correspondence}

Here, we explain briefly the correspondence between the weighted Aztec diamond and the square-octagon fortress.
It is well-established in the mathematical literature that the unweighted square-octagon fortress corresponds to a particular assignment of edge weights of the Aztec diamond~\cite{propp2003generalized,difrancesco2014acoe}.

While we consider the fortress with uniform weights since it can be formulated clearly in terms of spin Hamiltonians, many results in the mathematical literature deal with the weighted Aztec diamond model instead.
The $T$-system procedure, described in Ref.~\cite{difrancesco2014acoe}, establishes an equivalence between various weighted Aztec diamonds and weighted square-octagon fortresses.
The correspondence holds up to general patterns of \textit{face}-weighting, where each face $F$ is assigned a weight $W_F$. 
This induces an edge weight $w_e$ for each edge $e$ given by the product of the face weights of the two faces adjoining the edge $e$. 
Note that while all face-weightings are edge weightings, the converse is not true.
The partition function for the Aztec diamond of radius $k$ is given in terms of face weights $W_{f}$, and in terms of edge weights $w_e$:
\begin{align}
\begin{aligned}
Z_k &= \sum_{\text{configs.}} \prod_{\text{faces } f} \left(W_f\right)^{D_f - 1} \\
 &= \left( \prod_{\text{faces } f} \frac{1}{W_f} \right) \sum_{\text{configs.}} \prod_{\text{edges } e} (w_e)^{D_e},
 \end{aligned}
\end{align}
where $D_f$ is the sum of the dimer occupation numbers of the four edges surrounding a face $f$, and $D_e$ is the dimer occupation number of a single edge $e$. 

For the square-octagon fortress, the definition of the partition function has to be modified compared to the simple square lattice, due to faces having different numbers of edges surrounding them:
\begin{align}
\begin{aligned}
 Z_{k} &= \sum_{\text{configs.}} \prod_{\text{faces } f} (W_f)^{D_f + 1 - \frac{v_f}{2}} \\
  &=\left( \prod_{\text{faces } f} {W_f}^{1 - \frac{v_f}{2}} \right) \sum_{\text{configs.}} \prod_{\text{edges } e} (w_e)^{D_e},
\end{aligned}
\end{align}
where $v_f$ represents the number of vertices (4 or 8) surrounding the face designated $f$, and $w_e$ is defined in the same way it was previously. 

The correspondence provides, in our case, a set of face weights on the square Aztec diamond whose Arctic curves \cite{difrancesco2014acoe} and whose partition function are identical to that of a square-octagon fortress with another set of weights.
Details of the process for obtaining this correspondence are given in Ref.~\cite{difrancesco2022tsnd}, where the square-octagon lattice is considered a particular case of a more general type of lattice consisting of square, hexagon, and octagon plaquettes. 
The method treats these lattices as an alteration of the basic square grid, such that the correspondence between plaquettes is one-to-one. 
In this case, the alteration contracts the four octagon-octagon edges emanating from each square into single points.

We will only state the relevant final results here. 
For a square-octagon fortress where we assign a face weight $w_s$ to all squares and a face weight $1$ to all octagons, or equivalently, a weight of $1$ to all diagonal edges and a weight of $w_s$ to all horizontal and vertical edges, the corresponding face weights for the Aztec diamond that has the same partition function and same Arctic curve are given as:
\begin{equation}\label{eqn:correspondence}
   \begin{cases}
      W_{i,j} = 1, & i+j \equiv 1 \mod 2 \\
      W_{i,j} = w_s, & i \equiv 0 \mod 2, j \equiv 0 \mod 2 \\
      W_{i,j} = \frac{1}{2w_s}, & i \equiv 1 \mod 2, j \equiv 1 \mod 2,
   \end{cases}
\end{equation}
where $(i,j)$ represent coordinates on a two-dimensional grid that are assigned to each face. Since these weights depend only on the faces' coordinates modulo 2, this weight assignment is called ``2-periodic". 
The weights of faces are conventionally denoted by $a, b, c, d$ whose coordinates are (even, odd), (odd, even), (even, even), and (odd, odd) respectively. These parameters are employed in the formula ~\cite[\S3.3]{difrancesco2014acoe} for the Arctic octic curve (as used in \cref{sec:octic}) of the square-octagon fortress, and additional formulae for properties of the weighted Aztec diamond.

\end{document}